\documentclass[aps,prd,floatfix,nofootinbib,showpacs,twocolumn,10pt]{revtex4-1}

\usepackage{amssymb}
\usepackage[intlimits]{amsmath}
\usepackage{amsfonts}
\usepackage{dsfont}
\usepackage{subfigure}
\usepackage[usenames,dvipsnames]{color}
\usepackage{colortbl}
\usepackage{array}
\usepackage{morefloats}

\definecolor{webgreen}{rgb}{0,0.75,0}
\definecolor{webred}{rgb}{0.75,0,0}
\definecolor{webblue}{rgb}{0,0,0.75}
\definecolor{darkblue}{rgb}{0,0,0.6}
\definecolor{dunkelgrau}{rgb}{0.8,0.8,0.8}
\definecolor{darkgray}{rgb}{0.5,0.5,0.5}
\definecolor{lgray}{rgb}{0.95,0.95,0.95}
\definecolor{lgreen}{rgb}{0.95,1.00,0.90}
\definecolor{lblue}{rgb}{0.9,0.95,1.00}
\definecolor{lred}{rgb}{1.00,0.92,0.85}
\definecolor{shadecolor}{rgb}{1.00,0.92,0.82}

\usepackage[colorlinks=true,linkcolor=darkblue,citecolor=darkblue,urlcolor=darkblue]{hyperref}
\usepackage{graphicx}

\usepackage{natbib}
\usepackage{multirow}
\usepackage{bbm}

        \usepackage[vcentermath]{youngtab}

      \def\llongrightarrow{
      \relbar\joinrel\relbar\joinrel\relbar\joinrel\rightarrow}
      \def\longlongrightarrow{
      \relbar\joinrel\relbar\joinrel\relbar\joinrel\relbar\joinrel\rightarrow}

      \def\p{\partial}

      \newcommand{\mA}{\mathcal{A}}
      \newcommand{\mS}{\mathcal{S}}
      \newcommand{\mD}{\mathcal{D}}
      \newcommand{\mT}{\mathcal{T}}

      \newcommand{\gray}[1]{\color{darkgray} #1 }

\begin{document}

\title{Four-point functions and the permutation group $S_4$}

\author{Gernot Eichmann}
\author{Christian S. Fischer}
\author{Walter Heupel}
\affiliation{Institut f\"ur Theoretische Physik, Justus-Liebig--Universit\"at Giessen, 35392 Giessen, Germany.}

\begin{abstract}
Four-point functions are at the heart of many interesting physical processes. A prime example is
the light-by-light scattering amplitude, which plays an important role in the calculation of hadronic
contributions to the anomalous magnetic moment of the muon. In the calculation of such quantities
one faces the challenge of finding a suitable and well-behaved basis of tensor structures in coordinate
and/or momentum space. Provided all (or many) of the external legs represent similar particle content,
a powerful tool to construct and organize such bases is the permutation group $S_4$. We introduce an
efficient notation for dealing with the irreducible multiplets of $S_4$, and we highlight the
merits of this treatment by exemplifying four-point functions with gauge-boson legs such as the
four-gluon vertex and the light-by-light scattering amplitude. The multiplet analysis is also
useful for isolating the important kinematic regions and the dynamical singularity content of such
amplitudes. Our analysis serves as a basis for future efficient calculations of these and similar objects.
\end{abstract}

\maketitle

\section{Introduction}

        In the study of the structure of elementary and composite particles one often faces the situation that the
        physical properties of interest are encoded in higher $n-$point functions. These appear in different contexts.
        For example, hadron properties are experimentally extracted from nucleon-lepton or $NN$ scattering, $\pi\pi$
        scattering, Compton scattering or pion electroproduction amplitudes, which are all four-point functions. The
        structure of electrons is inferred from Compton scattering and even the structure of photons can be measured
        in principle via light-by-light (LbL) scattering on the photon mass shells. The corresponding off-shell photon
        four-point function, although not directly measurable, is a particularly interesting case since it receives
        hadronic corrections that play an important role in the muon $(g-2)$ puzzle~\cite{Jegerlehner:2009ry}. On a
        technical level, the photon four-point function is challenging because of the gauge boson nature of its
        external legs, but it is also subject to potentially extensive simplifications due to the corresponding
        Bose symmetry.

        Higher $n-$point functions are also important at the level of quarks and gluons: hadrons appear as poles in the
        $q\bar{q}$ four-point and $qqq$ six-point correlation functions, whose residues define the $q\bar{q}$ and $qqq$
        Bethe-Salpeter vertex functions, and their properties are linked with the fundamental quark-gluon, three-gluon
        and four-gluon vertices. With respect to the recent experimental discoveries of a variety of potential exotic
        states (XYZ-states) in the heavy quark region, a further interesting quantity is the $qq\bar{q}\bar{q}$
        eight-point function, with tetraquark amplitudes as residues on the corresponding pole positions. Tetraquark
        amplitudes are five-point functions with four (anti-)quark legs and display symmetries similar to the corresponding
        four-point functions.

        It is desirable to develop the technology to deal with such quantities efficiently and make the most possible
        use of the underlying symmetries. A generic $n-$point function depends on $n-1$ independent momenta and has
        the form
      \begin{equation}
      \begin{split}
          \Gamma^{\mu\nu\dots}(p_1, \dots p_n) &= \sum_i^N f_i(\dots)\,\tau^{\mu\nu\dots}_{i}(p_1, \dots p_n)
      \end{split}
      \end{equation}
        in momentum space, modulo potential flavor and color factors, where $\mu$, $\nu$ denote Lorentz and/or Dirac
        indices. It can be decomposed into $N$ Dirac-Lorentz tensors $\tau_i^{\mu\nu\dots}$ with corresponding dressing
        functions $f_i$. The $f_i$ depend on $M=n(n-1)/2-m$ independent Lorentz invariants $p_1^2$, $p_2^2$,
        $p_1\cdot p_2 \dots$ that constitute the phase space, where $m$ is the number of legs that are on-shell.
        For example, in linear covariant gauges one finds
        \begin{itemize}
        \item $(N,M)=(14,3)$ for the three-gluon vertex~\cite{Ball:1980ax},
        \item $(N,M)=(136,6)$ for the four-gluon vertex and, without implementing gauge invariance,
        also for the LbL amplitude (see Sec.~\ref{sec:basis-counting}),
        \item $(N,M)=(64,5)$ for a nucleon Bethe-Salpeter amplitude~\cite{Eichmann:2009qa}
        and (128,5) for that of a $\Delta-$baryon~\cite{SanchisAlepuz:2011jn},
        \item and $(N,M)=(256,9)$ for a tetraquark Bethe-Salpeter amplitude~\cite{tetraquark}.
        \end{itemize}
        Clearly, for a growing number of external legs this becomes unmanageable and poses a challenge for
        theoretical approaches. For example, in the non-perturbative region of QCD the fundamental two-point functions
        are reasonably well understood by now, partially owing to their relatively simple structure.
        Progress in determining three-point functions is underway, see e.g.
        \cite{Blum:2014gna,Eichmann:2014xya,Braun:2014ata,Aguilar:2014lha}
        and references therein. Concerning four- and higher $n-$point functions, however, only exploratory
        studies exist so far \cite{Kellermann:2008iw,Cyrol:2014kca,Binosi:2014kka}. Within QED, an important and prominent
        example is the photon four-point function, which received an increasing amount of interest recently
        in connection with the problem of the anomalous magnetic moment of the muon. Its hadronic corrections
        account for a sizable contribution to the total error budget of the theoretical determination of
        $(g-2)_\mu$. With efforts for a substantial decrease of the experimental error underway
        \cite{LeeRoberts:2011zz,Iinuma:2011zz} various theoretical approaches to capture the rich physics of
        the photon four-point function on a quantitative level have been explored
        \cite{Goecke:2012qm,Pauk:2014rfa,Blum:2014oka,Dorokhov:2014iva,Colangelo:2014dfa,Colangelo:2014pva,Eichmann:2014ooa,Benayoun:2014tra}.
        Clearly, a suitable and well-organized basis of tensor structures of this object will greatly facilitate
        these studies.

        In general, the organizing principles for a basis of tensor structures of an $n$-point function are gauge
        invariance (if gauge-boson legs are involved), momentum counting and permutation-group symmetries.
        Correlation functions that are subject to gauge-invariance constraints can be split into a `gauge' part
        and a transverse contribution, where the gauge part is often used as a zeroth-order approximation to
        the full result. For gauge-\textit{invariant} quantities the gauge part vanishes, which complicates
        matters further because the transverse part is also subject to analyticity constraints. The momentum
        counting entails that tensor structures with higher momentum powers are less important, and in practice
        it is often sufficient to restrict oneself to the simplest momentum-independent tensors to obtain a
        reasonable approximation of the full $n$-point function. Finally, permutation-group symmetries are
        useful because they allow one to arrange both tensor structures and momentum variables into multiplets
        and thereby isolate the most relevant momentum dependence of the dressing functions $f_i$.

   		While our main motivation for the present paper is the muon $g-2$ problem, we will not repeat the discussion
    	of the underlying physics problems here. Instead, we will focus on technical aspects: we develop the
    	terminology to deal with permutation-group symmetries, and we apply it to investigate the phase space in
    	four-point amplitudes and construct appropriate tensor bases for them. To some extent our results are also
    	applicable to other systems where permutation-group symmetries play a role, such as the four-gluon vertex
    	or the tetraquark Bethe-Salpeter amplitude, and in principle they can be generalized to higher $n-$point
    	functions.

    	The paper is organized as follows.
    	We exemplify the basic problems in constructing gauge-invariant tensor bases in Sec.~\ref{sec:2photon}
    	for a scalar two-photon current.
    	In Sec.~\ref{sec:s3} we illustrate our multiplet notation for the permutation group $S_3$ and its
		application to baryon flavor wave functions.
    	We subsequently generalize it to the permutation group $S_4$ in Sec.~\ref{sec:bose}, where we discuss
		the color structure of the four-gluon vertex as an example.
    	In Sec.~\ref{sec:phase-space} we apply the terminology to examine the phase space in four-point functions,
    	and we finally construct appropriate tensor bases for vector four-point functions in Sec.~\ref{sec:tensorbasis}.
    	We summarize and conclude in Sec.~\ref{sec:summary}.

          \begin{figure}[t]
                    \begin{center}
                    \includegraphics[scale=0.17]{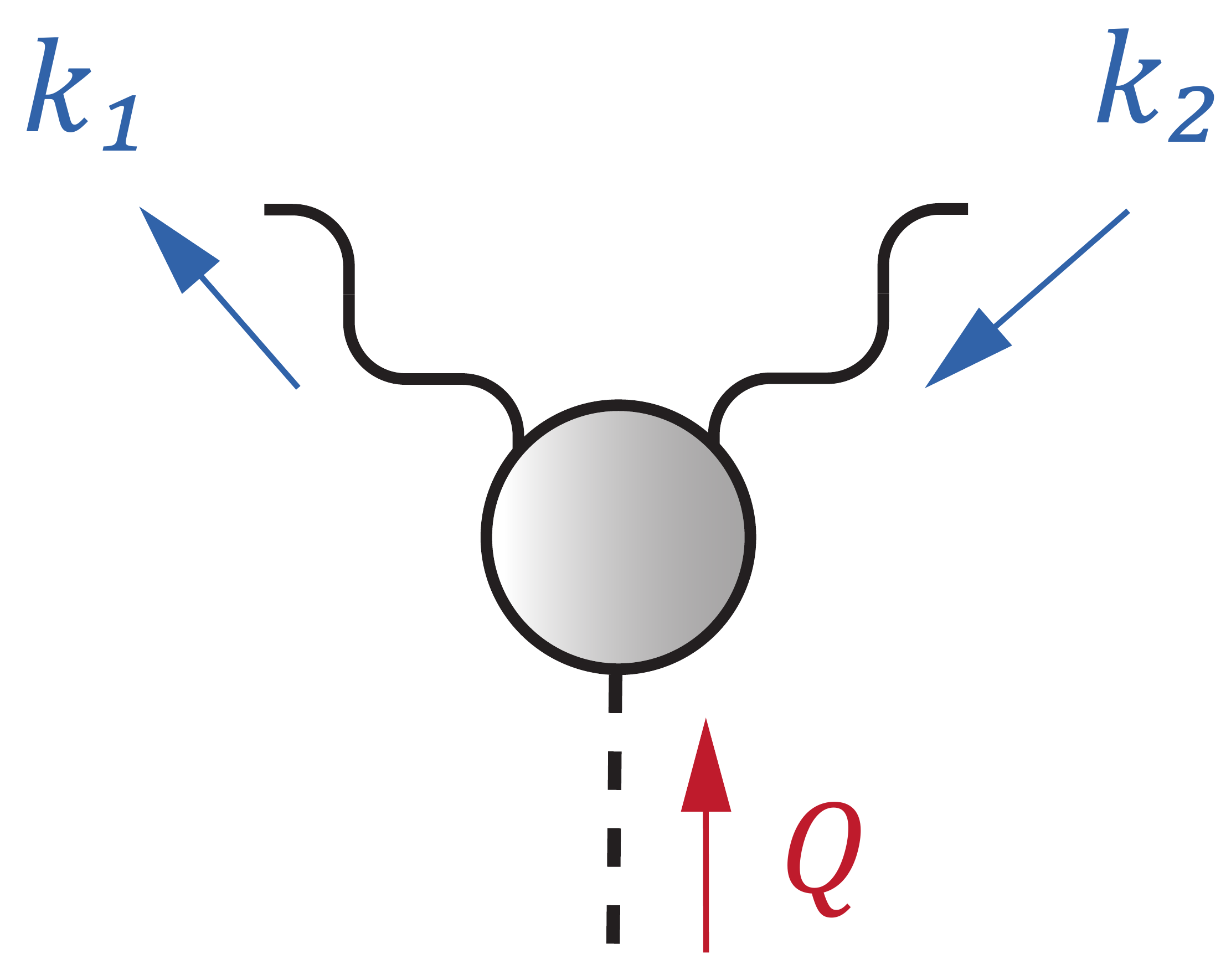}
                    \caption{ Kinematics for a scalar two-photon current.
                                            }\label{fig:scalar-current}
                    \end{center}
        \end{figure}

        \section{Example: scalar two-photon current}\label{sec:2photon}

            Ultimately our goal is to establish tensor bases for vector four-point functions that implement
            the constraints from Bose symmetry, gauge invariance and analyticity. The template for this is
            the analogous case of Compton scattering that has been investigated in
            Refs.~\cite{Bardeen:1969aw,Tarrach:1975tu,Drechsel:1996ag,Drechsel:1997xv,L'vov:2001fz,Drechsel:2002ar,Gorchtein:2009wz}.
            We will discuss four-point functions in Sec.~\ref{sec:tensorbasis}, but
            to illustrate the main points we first consider a simpler system,
            namely a two-photon current with scalar ($0^{++}$) quantum numbers.

            The current $\Gamma^{\mu\nu}(k,Q)$ depends on two independent momenta (see Fig.~\ref{fig:scalar-current});
            we denote the outgoing and incoming photon momenta by
            \begin{equation}
                k_1 = k+\frac{Q}{2}\,, \qquad
                k_2 = k-\frac{Q}{2}\,,
            \end{equation}
            where $Q=k_1-k_2$ is the total bound-state momentum and $k=(k_1+k_2)/2$ is the average photon momentum.
            Electromagnetic gauge invariance entails that $\Gamma^{\mu\nu}(k,Q)$ is transverse with respect to
            $k_1^\mu$ and $k_2^\nu$:
            \begin{equation}\label{2gamma-transversality}
                k_1^\mu\, \Gamma^{\mu\nu} = 0, \qquad k_2^\nu\, \Gamma^{\mu\nu}=0\,.
            \end{equation}
            Since the current has two photon legs, transversality and analyticity require it to be at least quadratic in the photon momenta.

            The most general decomposition of the current consists of five tensor structures,
            which we can express in terms of the photon momenta $k_1$ and $k_2$:
            \begin{equation}\label{scalar-current-1}
            \begin{split}
                \Gamma^{\mu\nu}= \sum_{i=1}^5 f_i\,\tau_i^{\mu\nu}
                & =  f_1\, \delta^{\mu\nu}
                +f_2\, k_2^\mu \,k_1^\nu \\[-4mm]
                & +f_3\, k_1^\mu \,k_2^\nu
                 +f_4\,(k_1^\mu\,k_1^\nu + k_2^\mu\,k_2^\nu) \\
                &+f_5\, \omega\,(k_1^\mu\,k_1^\nu - k_2^\mu\,k_2^\nu)\,.
            \end{split}
            \end{equation}
            The dressing functions or form factors $f_i(k^2,\omega,Q^2)$ depend on the Lorentz invariants $k^2$, $Q^2$ and $\omega=k\cdot Q$, or equivalently
            \begin{equation} \renewcommand{\arraystretch}{1.2}
            \begin{split}
                \left\{ \begin{array}{c} k_1^2 \\ k_2^2 \end{array}\right\} &=  k^2+\frac{Q^2}{4} \pm k\cdot Q = \eta_+ \pm \omega, \\
                k_1\cdot k_2 &= k^2 - \frac{Q^2}{4} = \eta_-.
            \end{split}\label{defeta}
            \end{equation}
            Bose symmetry entails $\Gamma^{\mu\nu}(k,Q) = \Gamma^{\nu\mu}(-k,Q)$. The Lorentz invariants $k^2$ and $Q^2$ are symmetric whereas $\omega$ is antisymmetric.
            To ensure Bose symmetry at the level of the basis elements, the tensor structure for $f_5$ includes an angular prefactor $\omega$
            so that all form factors $f_i$ are even in $\omega$, so they depend on $k^2$, $Q^2$ and $\omega^2$.
            In other words, all basis elements and all form factors are now \textit{singlets} under the permutation group $S_2$.
            The consequences of analyticity are not yet manifest in this basis;
            the requirement we will need is that the $f_i$ are nonsingular in any kinematic limit.

            Instead of working out Eqs.~\eqref{2gamma-transversality} it is more convenient in practice to equate the
            transverse projection of the current with the current itself:
            \begin{equation}\label{2gamma-transversality-2}
                T^{\mu\alpha}_1\,\Gamma^{\alpha\beta}\,T^{\beta\nu}_2 \stackrel{!}{=} \Gamma^{\mu\nu}\,.
            \end{equation}
            Here,
            $T^{\mu\nu}_i = \delta^{\mu\nu} - k_i^\mu \,k_i^\nu/k_i^2$ is a transverse projector with respect to photon $i$.
            Only the tensor structures for $f_1$ and $f_2$ in Eq.~\eqref{scalar-current-1} survive the projection (we will call
            them `\textit{transverse survivors}' in the following) whereas the others are longitudinal.
            This leads to three conditions for the $f_i$ whose solution is
            \begin{equation} \label{scalar-current-transversality-relations}
            \begin{split}
                f_1 &= -\eta_- f_2 + (\eta_+^2-\omega^2)\,f_5, \\
                f_3 &= \eta_-f_5, \\
                f_4 &= -\eta_+ f_5\,,
            \end{split}
            \end{equation}
            with $\eta_\pm$ defined in Eqs.~(\ref{defeta}).
            Here we exploited the constraint that the $f_i$ should be nonsingular:
            had we solved for any other set of three form factors,
            we would have introduced denominators that can become singular, contrary to our initial assumption.
            Note also that the Bose symmetry of the $f_i$ was crucial: without the antisymmetric variable $\omega$
            in the last row of Eq.~\eqref{scalar-current-1} the resulting factors of $\omega$ would obscure the correct solution.
            Hence, to solve the transversality constraints we should start from a basis made of  permutation-group singlets.

            When substituting the solution into Eq.~\eqref{scalar-current-1}, the result for the transverse current becomes
            \begin{equation} \label{scalar-current-temp}
            \begin{split}
                \Gamma^{\mu\nu} &= f_2\,\big[\tau_2-\eta_- \tau_1\big]^{\mu\nu} \\
                                     &+ f_5\,\big[\tau_5 + (\eta_+^2-\omega^2)\,\tau_1 + \eta_- \tau_3 - \eta_+ \tau_4\big]^{\mu\nu}  \,,
            \end{split}
            \end{equation}
            These two new tensor structures do not look particularly enlightening, but they can be cast in a compact form
            if we define
            \begin{equation}\label{transverse-proj-safe}
                t^{\mu\nu}_{ij} = k_i\cdot k_j\,\delta^{\mu\nu} - k_j^\mu\,k_i^\nu\,.
            \end{equation}
            This quantity is transverse with respect to $k_1^\mu$ and $k_2^\nu$ without introducing a pole,
            and for $i=j$ it becomes proportional to the usual transverse projector: $t^{\mu\nu}_{ii} = k_i^2\,T^{\mu\nu}_i$ with $T^{\mu\nu}_i$ defined below Eq.~\eqref{2gamma-transversality-2}.
            The resulting current in Eq.~\eqref{scalar-current-temp} can now be written as
            \begin{equation}\label{scalar-current-temp2}
                \Gamma^{\mu\nu} = -f_2\,t_{12}^{\mu\nu} + f_5\,t_{11}^{\mu\alpha}\,t_{22}^{\alpha\nu}\,,
            \end{equation}
            from where the transversality and analyticity properties can be read off directly.\footnote{Following Tarrach's procedure for Compton scattering~\cite{Tarrach:1975tu},
            we would write instead of Eq.~\eqref{2gamma-transversality-2}:
           \begin{equation} \label{2gamma-transversality-3}
                T^{\mu\alpha}_{12}\,\Gamma^{\alpha\beta}\,T^{\beta\nu}_{12} \stackrel{!}{=} \Gamma^{\mu\nu}\,, \qquad T_{ij}^{\mu\nu} = \frac{t^{\mu\nu}_{ij}}{k_i\cdot k_j}\,,
            \end{equation}
            which leads to the same result.}
            The two transverse structures exhibit the correct power counting.
            Since the current can be expressed entirely in terms of photon momenta, the photon momentum powers are equivalent to their mass dimension:
            $t_{12}^{\mu\nu}$ has mass dimension two and $t_{11}^{\mu\alpha}\,t_{22}^{\alpha\nu}$ has dimension four.
            The transverse tensors read explicitly:
            \begin{equation}
            \begin{split}
                t_{12}^{\mu\nu} &= k_1\cdot k_2\,\delta^{\mu\nu} - k_2^\mu\,k_1^\nu, \\[2mm]
                t_{11}^{\mu\alpha}\,t_{22}^{\alpha\nu} &= (k_1^2\,\delta^{\mu\alpha} - k_1^\mu\,k_1^\alpha)\,(k_2^2\,\delta^{\alpha\nu}-k_2^\alpha\,k_2^\nu) \\
                                                       &= k_1^2\,k_2^2\,\delta^{\mu\nu} -k_2^2\,k_1^\mu\,k_1^\nu - k_1^2\,k_2^\mu\,k_2^\nu \\
                                                       & \quad + k_1\cdot k_2\,k_1^\mu\,k_2^\nu\,.
            \end{split}
            \end{equation}
            Their form factors $f_2$ and $f_5$ are free of kinematic singularities and zeros and become constant
            in any kinematic limit of vanishing photon momenta.
            Hence, their only possible singularities are of dynamical origin.
            The basis~\eqref{scalar-current-temp2} is therefore \textit{`minimal'}.

            This simple example highlights the main features that we will again encounter in the discussion of the LbL amplitude.
            What if the two-photon current happened to be \textit{not} transverse, for example as a consequence of an \textit{incomplete} dynamical calculation
            that violates gauge invariance?
            We can add again the tensor structures for those coefficients in Eq.~\eqref{scalar-current-transversality-relations} that we found to be linearly dependent:
            \begin{equation} \label{scalar-current-full}
            \begin{split}
                \Gamma^{\mu\nu}  &=  g_1\, \delta^{\mu\nu} + g_3\, k_1^\mu \,k_2^\nu  + g_4\,(k_1^\mu\,k_1^\nu + k_2^\mu\,k_2^\nu) \\
                & - f_2\,t_{12}^{\mu\nu} + f_5\,t_{11}^{\mu\alpha}\,t_{22}^{\alpha\nu}\,.
            \end{split}
            \end{equation}
            Violations of gauge invariance will then become manifest in nonzero functions $g_1$, $g_3$ and $g_4$.
            If we apply another transverse projection to Eq.~\eqref{scalar-current-full} and express the result in the `physical' basis $t_{12}^{\mu\nu}$ and $t_{11}^{\mu\alpha}\,t_{22}^{\alpha\nu}$,
            these gauge artifacts will produce unphysical singularities in the two-photon current,
            \begin{equation}\label{trans-proj-art-sing}
                T^{\mu\alpha}_1\,\Gamma^{\alpha\beta}\,T^{\beta\nu}_2 = - f_2\,t_{12}^{\mu\nu} + \left[ f_5 + \frac{g_1}{k_1^2\,k_2^2}\right]t_{11}^{\mu\alpha}\,t_{22}^{\alpha\nu}\,,
            \end{equation}
            and therefore invalidate the extraction of form factors in the limits where the photon momenta vanish.
            Ultimately, in a fully gauge-invariant calculation all $g_i$ must cancel,
            so instead of projecting transversely we could simply \textit{omit} the $g_i$ terms in Eq.~\eqref{scalar-current-full}.
            The form factors $f_2$, $f_5$ of the transverse
            remainder would then still provide a sensible estimate of the result.

            Eq.~\eqref{scalar-current-full}
            is the general decomposition of a scalar-vector-vector amplitude in terms of a 'gauge part' and a transverse part that implements
            the analyticity constraints. It also applies to a scalar glueball amplitude \cite{Meyers:2012ka,Sanchis-Alepuz:2015hma},
            and analogous constructions are well-known for
            the quark-photon and quark-gluon vertex, the three-gluon vertex etc.
            In those cases, the gauge parts are nonzero and constrained
            by Ward-Takahashi or Slavnov-Taylor identities, and
            they are not artifacts but usually even the dominant contributions.
            For example, the gauge part in the quark-photon vertex is the Ball-Chiu vertex~\cite{Ball:1980ay} that is known to give a large contribution
            to hadronic form factors. Only for gauge-invariant amplitudes the gauge parts are forced to be zero,
            and in combination with the power counting this leads to the problem of unphysical singularities if gauge invariance is not respected.

            The example also provides us with another, more direct way to arrive at Eq.~\eqref{scalar-current-full}.
            Since the transverse projection of Eq.~\eqref{scalar-current-1} leaves only two survivors, it is clear that there can be only two transverse tensor structures.
            Using the definition~\eqref{transverse-proj-safe}, we could have immediately written down
            the two lowest-dimensional transverse tensors that are permutation-group singlets and free of kinematic singularities:
            $t_{12}^{\mu\nu}$ and $t_{11}^{\mu\alpha}\,t_{22}^{\alpha\nu}$. When expressed in terms of the generic basis via Eq.~\eqref{scalar-current-temp},
            the former depends on $\tau_2$ and the latter on $\tau_5$. These are the only dependencies that do not involve any kinematic prefactors.
            Therefore, the only kinematically safe option is to eliminate the contributions from $\tau_2$ and $\tau_5$ in Eq.~\eqref{scalar-current-1} in favor of the new transverse tensors,
            thereby leaving $\tau_1$, $\tau_3$ and $\tau_4$ as the gauge part.
            The result is identical to Eq.~\eqref{scalar-current-full}.

            To facilitate the following discussion, we will refer to the two types of bases in Eq.~\eqref{scalar-current-1} and Eq.~\eqref{scalar-current-full} as 'type I' and 'type II', respectively:
            \begin{equation}
                \Gamma = \underbrace{\Gamma_\text{sur} + \Gamma_L}_{\text{type I}} = \underbrace{\Gamma_\text{gauge} + \Gamma_T}_{\text{type II}} \,.
            \end{equation}
            The type-I basis consists of a `survivor' and a longitudinal part and the type-II basis of a gauge and a transverse part.
            The transverse survivors depend upon the type of projection: Eqs.~\eqref{2gamma-transversality-2} and~\eqref{2gamma-transversality-3} will produce different survivors,
            but their total number is the same.
            Although the gauge parts are usually called `longitudinal', neither $\Gamma_\text{sur}$ nor $\Gamma_\text{gauge}$ are
            fully transverse or longitudinal. A transverse or longitudinal projection will rather result in the following structure:
            \begin{equation}
            \begin{split}
                T(\Gamma)  &= T(\Gamma_\text{sur}) = T(\Gamma_\text{gauge}) + \Gamma_T\,, \\
                L(\Gamma)  &= L(\Gamma_\text{sur}) + L(\Gamma_L) = L(\Gamma_\text{gauge})\,.
            \end{split}
            \end{equation}
            Whereas the basis dimensions are identical,
            \begin{equation}
                \text{dim}(\Gamma_\text{sur}) = \text{dim}(\Gamma_T) , \quad
                \text{dim}(\Gamma_\text{gauge}) = \text{dim}(\Gamma_L),
            \end{equation}
            there is no one-to-one relation between their dressing functions,
            which is also apparent in the two-photon current example of Eqs.~\eqref{scalar-current-1} and \eqref{scalar-current-full}.

            The example also illustrates another generic problem: type-I bases are straightforward to write down, whereas the construction of type-II bases, in which
            the underlying physics becomes more transparent, is much more difficult.
            QCD vertices that are contracted with transverse gluon propagators in Landau gauge have the form $T(\Gamma)$ with $\Gamma_\text{gauge}\neq 0$,
            hence type-I bases are usually sufficient for their analyses.\footnote{In the case of the quark-photon vertex, Eq.~(75) in Ref.~\cite{Eichmann:2012mp} constitutes
            the type-I basis and Eq.~(89), together with the Ball-Chiu vertex of Eq.~(72), the type-II basis. The same applies to the quark-gluon vertex except that
            $\Gamma_\text{gauge}$  follows from a Slavnov-Taylor identity that contains the quark-ghost kernel.
            For the three-gluon vertex, Table V in Ref.~\cite{Eichmann:2014xya}
            is of type I and Eq.~(66) is the type-II basis.}
            In the case of gauge-invariant photon amplitudes only $\Gamma_T$ survives, which makes it mandatory to find an appropriate basis of type II.

            The program in applying these findings to the photon or gluon four-point functions is then the following:
            \begin{itemize}
            \item Work out the irreducible representations of the permutation group $S_4$.
            \item Cast the Lorentz-invariant momentum variables into $S_4$ multiplets and work out the structure of the phase space, the
                  relevant kinematic limits, etc.
            \item For the photon four-point function, write down a generic Lorentz tensor basis of \mbox{type I} and cast it into $S_4$ singlets.
                  For the four-gluon vertex, arrange also the color structures into $S_4$ multiplets and combine them together with the Lorentz tensors into $S_4$ singlets.
            \item Apply the transversality constraints and derive a type-II basis which is made of $S_4$ singlets.
                  It should be the sum of a gauge and a transverse part, where the gauge part must vanish
                  if the amplitude is gauge invariant.
            \end{itemize}
            Since the construction of permutation-group multiplets
            requires a certain amount of formalism, we will first discuss the simpler case of the permutation group $S_3$
            and postpone its application to the group $S_4$ to Sec.~\ref{sec:bose}.

\section{Permutation group $S_3$} \label{sec:s3}

\subsection{Multiplets}\label{sec:s3-multiplets}

\begin{figure}[t]
                    \begin{center}
                    \includegraphics[scale=0.25]{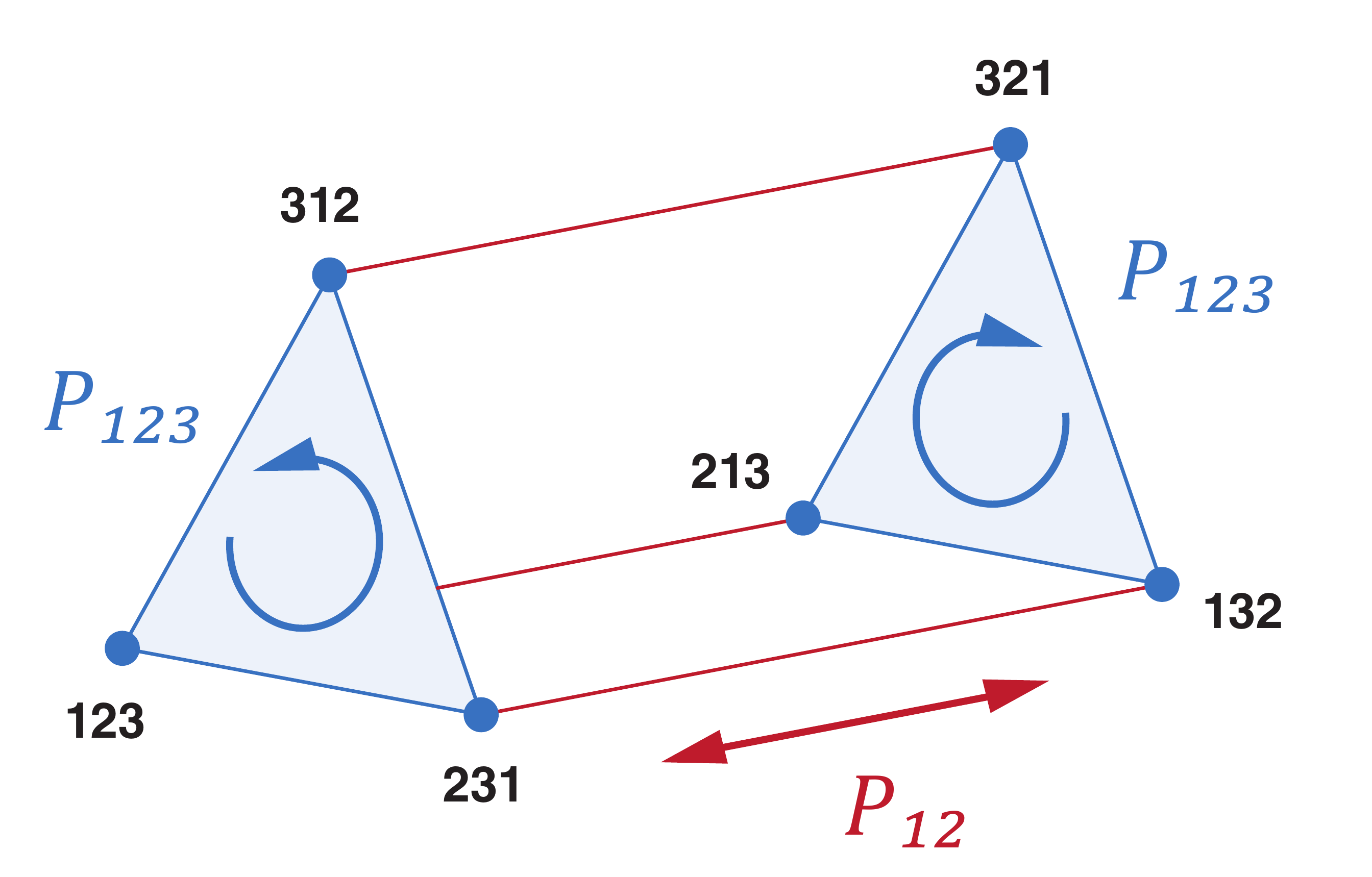}
                    \caption{Cayley graph for the permutation group $S_3$. Any permutation can be reconstructed from
                                     a transposition $P_{12}$ and a cycle $P_{123}$.
                                      }\label{fig:cayley-s3}
                    \end{center}
            \end{figure}

        The permutation group $S_3$ consists of $3!=6$ elements.
        The group manifold is visualized by the Cayley graph in Fig.~\ref{fig:cayley-s3}:
        any permutation of an object $f_{123}$ can be reconstructed from  a transposition $P_{12}$ and a cyclic permutation $P_{123}$.
        The former interchanges the indices $1\leftrightarrow 2$ and the latter
        is a cyclic permutation $1\rightarrow 2$, $2\rightarrow 3$, $3\rightarrow 1$. The group elements acting on $f_{123}$ are given by
        \begin{equation}\label{perm-reconstruct}  \renewcommand{\arraystretch}{1.3}
        \begin{array}{l}
           1\,, \\ P_{13}\,P_{12} = P_{123}\,, \\ P_{23}\,P_{12} = P_{123}^2\,,
        \end{array}\qquad\quad
        \begin{array}{l}
           P_{12}\,, \\ P_{23} = P_{12}\,P_{123}\,, \\ P_{13} = P_{12}\,P_{123}^2
        \end{array}
        \end{equation}
        and can be visualized by paths along the Cayley graph, starting from the lower left corner.

        The six permutations of $f_{123}$ can be rearranged into a symmetric singlet $\mS$, an antisymmetric singlet (`antisinglet') $\mA$, and two doublets $\mD_j$ ($j=1,2$)
        whose explicit form can be chosen as\footnote{Our notation is based on Ref.~\cite{Carimalo:1992ia}; see also~\cite{Eichmann:2011vu,Eichmann:2014xya} for applications to the baryon Faddeev amplitude and the three-gluon vertex.}
        \begin{equation}\label{singlets-doublets}
        \begin{split}
            \mathcal{S} &= \psi_1^+ + \psi_2^+ + \psi_3^+ \,, \\
            \mathcal{A} &= \psi_1^- + \psi_2^- + \psi_3^- \,, \\[2mm]
            \mathcal{D}_1
             &= \left[ \begin{array}{c}
                                    \psi_2^--\psi_3^- \\
                                    -\tfrac{1}{\sqrt{3}} \left(  \psi_2^+ + \psi_3^+ -2\psi_1^+ \right) \end{array}\right] , \\
            \mathcal{D}_2
             &= \left[ \begin{array}{c}
                                    \tfrac{1}{\sqrt{3}} \left( \psi_2^- + \psi_3^- -2\psi_1^- \right) \\
                                    \psi_2^+-\psi_3^+ \end{array}\right] ,
        \end{split}
        \end{equation}
        where
        \begin{equation}\label{psi-1-3}
            \psi_1^\pm = P_\pm f_{123}\,, \quad
            \psi_2^\pm = P_\pm f_{231}\,, \quad
            \psi_3^\pm = P_\pm f_{312}
        \end{equation}
        and $P_\pm = 1 \pm P_{12}$.
        They transform under the irreducible representations of $S_3$, which correspond to the following Young diagrams:
        \begin{equation*}
             \begin{array}{c @{\qquad\quad}  c @{\qquad\quad\;\;}  c}
             \mS &  \mD_j &  \mA \\[2mm]
             \scriptsize\yng(3) &
             \scriptsize\yng(2,1) &
             \scriptsize\yng(1,1,1)
             \end{array}
        \end{equation*}
        The multiplets do not mix under any permutation and thereby generate three invariant subspaces:
        \begin{equation}\label{permutation-tf-s3} \renewcommand{\arraystretch}{1.1}
        \begin{split}
           P_{12}\left\{ \begin{array}{c} \mathcal{S} \\ \mathcal{A} \\ \mathcal{D}_j \end{array}\right\} &=
                 \left\{ \begin{array}{c} \mathcal{S} \\ -\mathcal{A} \\ \mathsf{M}_{12}^\mathsf{T}\,\mathcal{D}_j \end{array}\right\} , \\
           P_{123}\left\{ \begin{array}{c} \mathcal{S} \\ \mathcal{A} \\ \mathcal{D}_j \end{array}\right\} &=
                 \left\{ \begin{array}{c} \mathcal{S} \\ \mathcal{A} \\ \mathsf{M}_{123}^\mathsf{T}\,\mathcal{D}_j \end{array}\right\},
        \renewcommand{\arraystretch}{1.0}
        \end{split}
        \end{equation}
        with the two-dimensional orthogonal representation matrices ($\mathsf{M^T}$ denotes the matrix transpose)
        \begin{equation}\label{perm-rep-matrices-s3}
            \mathsf{M}_{12} = \left(\begin{array}{rr} -1 & \, 0 \\  0 & 1 \end{array}\right), \quad
            \mathsf{M}_{123} = \frac{1}{2}\left(\begin{array}{cc} -1 &  \sqrt{3} \\  -\sqrt{3} & -1 \end{array}\right).
        \end{equation}
        $\mathcal{S}$ is totally symmetric and invariant under any permutation. $\mathcal{A}$ is totally antisymmetric
        under exchange of any two entries and therefore it picks up a minus sign under a transposition.
        Both subspaces are one-dimensional.
        The doublets form a two-dimensional subspace;
        they transform under the same matrix representations $\mathsf{M}_{12}$ and $\mathsf{M}_{123}$ from
        where all other ones can be reconstructed via Eq.~\eqref{perm-reconstruct}, e.g.: $P_{23} \rightarrow (\mathsf{M}_{12}\,\mathsf{M}_{123})^\mathsf{T}$.
        Their upper (lower) components are antisymmetric (symmetric) under transpositions $P_{12}$ and we denote them by
        \begin{equation}  \renewcommand{\arraystretch}{1.0}
            \mD_j = \left[ \begin{array}{c} a_j \\ s_j \end{array}\right].
        \end{equation}

        In practical applications the tensor products of $S_3$ multiplets are often of interest.
        Given two sets of singlets $\mS$, $\mS'$, antisinglets $\mA$, $\mA'$ and doublets $\mD = [ a, s ]$, $\mD' = [a',s']$,
        there are 16 possible combinations
        \begin{equation}
           \{ \mS, \, \mA, \, a, \, s \} \times \{ \mS', \, \mA', \, a', \, s' \}
        \end{equation}
        which we can again arrange into multiplets. Of course
        the products of two singlets ($\mathcal{S} \,\mathcal{S'}$) or antisinglets ($\mathcal{A}\, \mathcal{A}'$) must be singlets.
        The inner product $\mD\cdot\mD'$ of two doublets is also a singlet and invariant under any permutation
        because the representation matrices $\mathsf{M} \in \{ \mathsf{M}_{12},\,\mathsf{M}_{123}\}$ are orthogonal:
        \begin{equation}
            (\mathsf{M}^\mathsf{T} \mathcal{D})\cdot(\mathsf{M}\,\mathcal{D}') = \mathcal{D}^\mathsf{T}\mathsf{M} \mathsf{M}^\mathsf{T} \mathcal{D}' = \mathcal{D}\cdot\mathcal{D}'\,.
        \end{equation}
        Therefore, there are three possibilities
        for constructing singlets in the product space:
        \begin{equation}\label{s3-prod-s}
            \mS \,\mS', \qquad \mA \,\mA', \qquad \mathcal{D}\cdot\mathcal{D}' =aa'+ss'\,.
        \end{equation}
        Antisinglets are obtained from
        \begin{equation}\label{s3-prod-a}
            \mS \,\mA', \qquad \mA\,\mS', \qquad \mathcal{D} \wedge \mathcal{D'} := as'-sa'\,,
        \end{equation}
        where we defined an antisymmetric wedge product,
        and doublets are formed by
        \begin{equation} \label{s3-prod-d}
            \begin{array}{c} \mS \,\mD', \\[1mm] \mS' \mD, \end{array} \quad
            \begin{array}{c} \mA\,(\varepsilon\,\mD') \,, \\[1mm] \mA'\,(\varepsilon\,\mD)\,, \end{array}\quad
            \mathcal{D} \ast \mathcal{D}' := \left[ \begin{array}{c} as'+sa' \\ aa'-ss' \end{array}\right],
        \end{equation}
        where
        \begin{equation}\label{perm-product-doublets-2-s3}
           \varepsilon=\left(\begin{array}{r@{\quad}c} 0 & 1 \\ -1 & 0 \end{array}\right) \;\; \Rightarrow \;\; \varepsilon\,\mathcal{D} = \left[\begin{array}{c} s \\ -a\end{array}\right].
        \end{equation}
        These relations are complete because they cover all $16$ possible combinations.
        One can verify them explicitly using Eq.~\eqref{permutation-tf-s3}:
        the resulting singlets stay invariant under permutations, the antisinglets pick up a minus sign for odd permutations, and the doublets
        transform under $\mathsf{M}_{12}$ and $\mathsf{M}_{123}$.

            \begin{figure}[t]
                    \begin{center}
                    \includegraphics[scale=0.25]{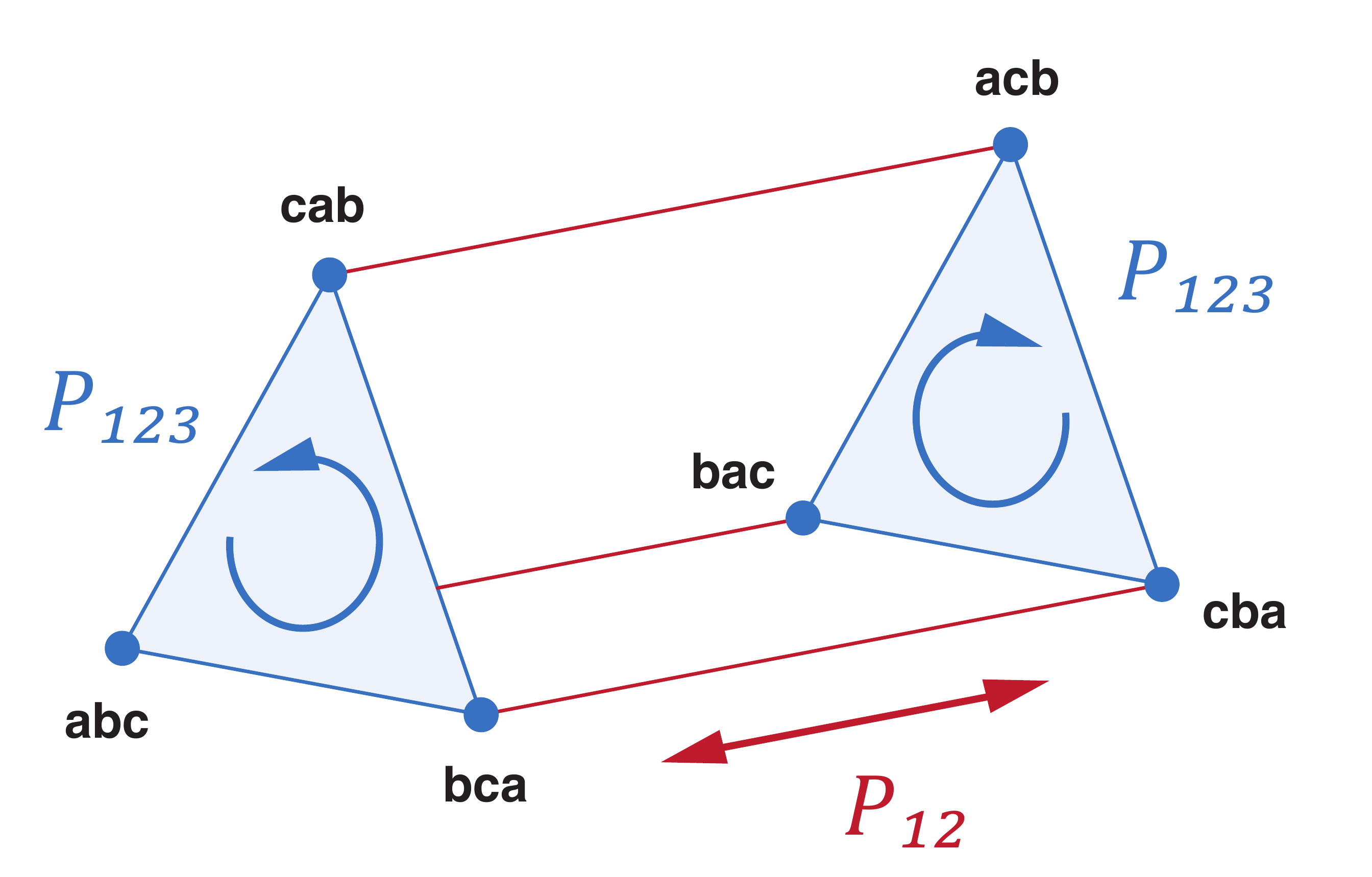}
                    \caption{Cayley graph for the permutation group $S_3$, with permutations of positions rather than indices.
                                      }\label{fig:cayley-s3-3}
                    \end{center}
            \end{figure}

        \subsection{Example: baryon flavor wave functions}

        Let us illustrate these properties using the example of the well-known $SU(3)_f$ flavor wave functions for baryons.
        Their textbook construction can be somewhat involved but with the help of the relations above
        we can derive them in a few lines.

        The $SU(3)_f$ irreducible representations that are constructed from the flavor vectors $u$, $d$, $s$ in the fundamental representation
        via \mbox{$\textbf{3}\otimes \textbf{3} \otimes \textbf{3} =$} \mbox{$\mathbf{1} \oplus \textbf{8} \oplus \textbf{8} \oplus \textbf{10}$}
        differ by their symmetry, so we must classify them into simultaneous irreducible representations of the permutation group $S_3$.
        In this case we would like the group elements in Eq.~\eqref{perm-reconstruct} to
        interchange the \textit{positions} of the objects $a, b, c \,\in \, \{u, d, s\}$ in a tensor product $abc=a \otimes b \otimes c$ rather than the indices of $f_{123}$.
        This requires slight modifications: in Fig.~\ref{fig:cayley-s3} we have to replace
        \begin{equation} \renewcommand{\arraystretch}{1.3}
           \begin{array}{c}  123 \rightarrow abc \\ 231 \rightarrow bca \\ 312 \rightarrow cab \end{array}\qquad
           \begin{array}{c}  213 \rightarrow bac \\ 132 \rightarrow cba \\ 321 \rightarrow acb \end{array}
        \end{equation}
        and hence $P_{13} \leftrightarrow P_{23}$ in Eq.~\eqref{perm-reconstruct}:
        \begin{equation}\label{perm-reconstruct-2}  \renewcommand{\arraystretch}{1.3}
        \begin{array}{l}
           1\,, \\ P_{23}\,P_{12} = P_{123}\,, \\ P_{13}\,P_{12} = P_{123}^2\,,
        \end{array}\qquad\quad
        \begin{array}{l}
           P_{12}\,, \\ P_{13} = P_{12}\,P_{123}\,, \\ P_{23} = P_{12}\,P_{123}^2\,.
        \end{array}
        \end{equation}
        All other relations remain the same as long as the permutations are understood to act on positions rather than indices.

        To construct the flavor wave function for a baryon with flavor content $uud$,
        take $a=b=u$ and $c=d$ and apply Eq.~\eqref{psi-1-3} to the `permutation-group seed' $uud$:
        \begin{equation}\label{uud-1}
            \psi_1^+=uud\,, \quad
            \psi_1^-=0\,, \quad
            \psi_2^\pm = \pm \psi_3^\pm = \frac{ud\pm du}{2}\,u\,.
        \end{equation}
        Up to prefactors one obtains
        \begin{equation}\label{uud-2}
        \begin{split}
            \mathcal{S} &= uud+udu+duu\,,  \\
            \mathcal{D}_1 &=   \left[ \begin{array}{c} udu-duu \\ -\tfrac{1}{\sqrt{3}}\left(udu+duu-2uud\right) \end{array}\right]
        \end{split}
        \end{equation}
        together with $\mathcal{D}_2=\mathcal{A}= 0$.
        Apart from overall normalization, $\mathcal{S}$ is the flavor wave function of the $\Delta^+$ and $\mathcal{D}_1$ that of the proton.
        Had we started from $ddu$ instead of $uud$, we would have obtained the $\Delta^0$ and the neutron (replace $u\leftrightarrow d$ above).
        The combination $uuu$ returns only a singlet ($\Delta^{++}$), and from $uds$ we get everything: $\mathcal{S}$, $\mathcal{A}$ and two doublets.
        Taking into account all 10 combinations with different flavor content ($uuu$, $ddd$, $sss$, $uud$, $uus$, $ddu$, $dds$, $ssu$, $ssd$, $uds$),
        we arrive at the result in Table~\ref{tab:baryons-flavor}:
        \begin{itemize}
  \setlength{\itemsep}{3pt}
  \setlength{\parskip}{0pt}
  \setlength{\parsep}{0pt}
        \item ten symmetric singlets, which form the flavor decuplet with $\Delta$, $\Sigma$, $\Xi$ and $\Omega$,
        \item eight doublets that form the flavor octet, including proton, neutron, $\Sigma$, $\Xi$ and $\Lambda$,
        \item and one antisymmetric singlet from $uds$, the flavor singlet for $\Lambda$.
        \end{itemize}
        It is simple to extend the construction to $SU(4)_f$, which produces
        20 singlets, 20 doublets and four antisinglets ($\mathbf{4} \otimes \mathbf{4} \otimes \mathbf{4} = \mathbf{20}_S \oplus \mathbf{20}_{M_A} \oplus \mathbf{20}_{M_S} \oplus \mathbf{4}_A$),
        or restrict it to $SU(2)_f$ ($\mathbf{2} \otimes \mathbf{2} \otimes \mathbf{2} = \mathbf{4}_S \oplus \mathbf{2}_{M_A} \oplus \mathbf{2}_{M_S}$).

         \renewcommand{\arraystretch}{1.2}

             \begin{table}[t]

                \begin{center}
                \footnotesize
                \begin{tabular}{     @{ \;} l @{ \;} | @{ \;\;} l  @{\;\;} l @{\;\;} l @{\;\;} l @{ \;} | @{ \;\;}  l @{\;\;} l @{\;\;} l @{ \;} | @{ \;\;} l @{\;\;} l @{ \;} | @{ \;\;} l @{ \;}  }   \rule{0mm}{0.2cm}

                                            &  $uuu$ & $uud$ & $ddu$ & $ddd$
                                            &  $uus$ & $uds$ & $dds$
                                            &  $ssu$ & $ssd$
                                            &  $sss$  \\[1mm] \hline\hline  \rule{0mm}{0.4cm}

                                              \!\!\!   $\mathcal{S}$
                                            &  $\Delta^{++}$ & $\Delta^{+}$ & $\Delta^{0}$ & $\Delta^{-}$
                                            &  $\Sigma^{+}$ & $\Sigma^{0}$ & $\Sigma^{-}$
                                            &  $\Xi^{0}$ & $\Xi^{-}$
                                            &  $\Omega^-$  \\[0.5mm] \hline   \rule{0mm}{0.4cm}

                                              \!\!\!\!   $\mathcal{D}_1$
                                            &    & $p$ & $n$ &
                                            &  $\Sigma^{+}$ & $\Sigma^{0}$ & $\Sigma^{-}$
                                            &  $\Xi^{0}$ & $\Xi^{-}$
                                            &    \\

                                                $\mathcal{D}_2$
                                            &    &   &   &
                                            &    & $\Lambda^0$ &
                                            &  &
                                            &     \\[1mm] \hline   \rule{0mm}{0.4cm}

                                               \!\!\! $\mathcal{A}$
                                            & &&&
                                            & & $\Lambda^0$ &
                                            &  &
                                            &

                \end{tabular}
                \end{center}

               \caption{$SU(3)_f$ flavor wave functions for baryons.}
               \label{tab:baryons-flavor}

        \end{table}

             \begin{table}[t]

                \begin{center}

                \begin{tabular}{    @{\;\;}c@{\;\;} |  @{\;\;}c@{\;\;}  |  @{\;\;}c@{\;\;}  }

                                     \multicolumn{3}{c}{\normalsize{$\mathcal{S}_\text{spin-flavor}: \; 56$}}         \\[1mm] \hline  \rule{0mm}{0.4cm}

                                   \!\!\!\!   $\mathcal{D}_s\cdot\mathcal{D}_f$  &  $(2,8)$  & $16$  \\
                                             $\mathcal{S}_s\, \mathcal{S}_f$      &  $(4,10)$ & $40$  \\[1mm] \hline

                                   \multicolumn{3}{c}{} \\[-1mm]
                                   \multicolumn{3}{c}{\normalsize{$\mathcal{A}_\text{spin-flavor}: \; 20$}}         \\[1mm] \hline   \rule{0mm}{0.4cm}

                                   \!\!\!\!   $\mathcal{D}_s\wedge\mathcal{D}_f$  &  $(2,8)$  & $16$  \\
                                              $\mathcal{S}_s\, \mathcal{A}_f$      &  $(4,1)$ & $4$    \\[1mm] \hline

                \end{tabular}\qquad
                \begin{tabular}{    @{\;\;}c@{\;\;}  |   @{\;\;}c@{\;\;}  |  @{\;\;}c@{\;\;}  }

                                   \multicolumn{3}{c}{\normalsize{$\mathcal{D}_\text{spin-flavor}: \; 70$}}         \\[1mm] \hline   \rule{0mm}{0.4cm}

                                   \!\!\!\!   $\mathcal{D}_s\ast\mathcal{D}_f$  &  $(2,8)$  & $16$  \\
                                              $\mathcal{S}_s \,\mathcal{D}_f$      &  $(4,8)$ & $32$   \\
                                              $\mathcal{D}_s\,\mathcal{S}_f $  &  $(2,10)$  & $20$  \\
                                              $(\varepsilon\,\mathcal{D}_s)\,\mathcal{A}_f$      &  $(2,1)$ & $2$   \\[1mm] \hline

                \end{tabular}

                \end{center}

               \caption{Spin-flavor wave functions from Eqs.~(\ref{s3-prod-s}--\ref{s3-prod-d}). The brackets count the number of multiplets:
                        the combination of two spin doublets $\mD_s$ and eight flavor doublets $\mD_f$ gives 16 independent spin-flavor singlets, etc.}
               \label{tab:Hosc}

        \end{table}

         \renewcommand{\arraystretch}{1.0}

        The notation is especially convenient for constructing product wave functions.
        Take the full (Bethe-Salpeter) wave function of a baryon,
        \begin{equation}\label{baryon-wf-all}
            \psi = \text{Dirac} \times \text{Flavor} \times \text{Color}\,,
        \end{equation}
        which must be totally antisymmetric under exchange of any two quarks.
        'Dirac' is here a shorthand for the full spatial-spin (or momentum-spin) contribution that transforms
        under the Poincar\'e group.
        The color wave function $\mA_c = \varepsilon_{ijk}$ is totally antisymmetric.
        The flavor wave functions form multiplets $\mS_f$, $\mD_f$ and $\mA_f$. Hence, the only combinations that produce a fully symmetric Dirac-flavor part are given by
        \begin{equation}\renewcommand{\arraystretch}{1.2}
            \mathcal{A}_\text{total} \; = \;  \left\{ \begin{array}{rl} (\mathcal{D}_D \cdot \mathcal{D}_f)\,\mathcal{A}_c &\quad (\text{octet})\,,\\
                                                                         (\mathcal{S}_D \, \mathcal{S}_f )\,\mathcal{A}_c &\quad  (\text{decuplet})\,, \\
                                                                         (\mathcal{A}_D \, \mathcal{A}_f )\,\mathcal{A}_c&\quad (\text{singlet}) \,.
                                                                         \end{array}\right. \label{spin-color-flavor-delta}
        \renewcommand{\arraystretch}{1.0}
        \end{equation}
        In the nonrelativistic quark model, the Dirac parts are the direct products of $O(3)$ orbital and $SU(2)$ spin wave functions.
        The spin wave functions are identical to Eq.~\eqref{uud-2} and the first four columns in Table~\ref{tab:baryons-flavor}
        if we replace $u$ by $\uparrow$ and $d$ by $\downarrow$: there are two doublets $\mathcal{D}_s$ with spin $S=\tfrac{1}{2}$
        and four singlets $\mathcal{S}_s$ with $S=\tfrac{3}{2}$.
        For orbital ground states ($L=0$ $\Rightarrow$ $P=+$, $J=S$) the  orbital wave functions are spatially symmetric, so the only possible combinations are
        \begin{equation}\renewcommand{\arraystretch}{1.2}
            \mathcal{A}_\text{total} \; = \;  \left\{ \begin{array}{rl} (\mathcal{D}_s \cdot \mathcal{D}_f)\,\mathcal{A}_c &\quad (J=\tfrac{1}{2}^+, \; \text{octet})\,,\\
                                                                         (\mathcal{S}_s \, \mathcal{S}_f )\,\mathcal{A}_c &\quad  (J=\tfrac{3}{2}^+, \; \text{decuplet})   \,
                                                                         \end{array}\right.
        \end{equation}
        whereas the flavor-singlet baryon $\Lambda^0$ does not appear in an orbital ground state.

        Combining the $SU(2)$ spin multiplets $\mathcal{D}_s$, $\mathcal{S}_s$ with the $SU(3)_f$ flavor multiplets $\mathcal{D}_f$, $\mathcal{S}_f$ and $\mathcal{A}_f$
        yields the standard $SU(6)$ quark-model classification. The product wave functions can be read off directly from Table~\ref{tab:Hosc}:
        there are 56 singlets from Eq.~\eqref{s3-prod-s} which are relevant for orbital ground states,
        20 antisinglets from Eq.~\eqref{s3-prod-a}, and 70 doublets from Eq.~\eqref{s3-prod-d}.

\section{Permutation group $S_4$} \label{sec:bose}

        \renewcommand{\arraystretch}{1.0}

\subsection{Multiplets}\label{sec:multiplets}

\begin{figure}[!t]
                  \begin{center}
                  \includegraphics[scale=0.40]{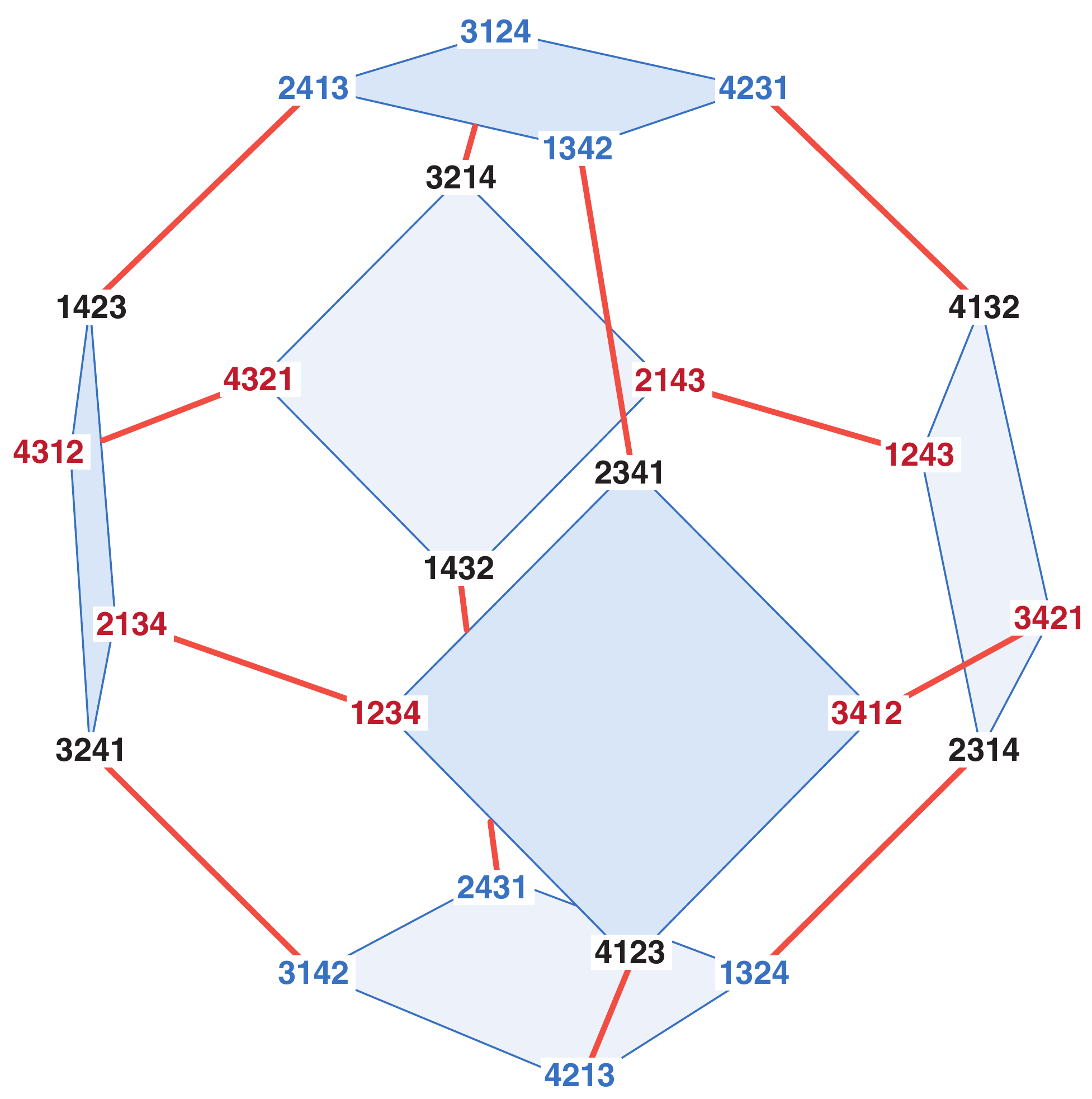}
                  \caption{Cayley graph for the group $S_4$. The elements in the vertical center (red) are the $t-$channel permutations,
                           the ones at the top and bottom (blue) are the $u-$channel permutations, and the remaining ones are the $s-$channel permutations.
                  }\label{fig:cayley}
                  \end{center}
      \end{figure}

        We will now generalize our notation from the last section to the permutation group $S_4$.
        The group consists of $4!=24$ elements. Each permutation of an object $f_{1234}$ can
        be reconstructed from two group elements, a transposition $P_{12}$ and a 4-cycle $P_{1234}$. The former
        interchanges the indices $1\leftrightarrow 2$ and the latter
        is a cyclic permutation $1\rightarrow 2$, $2\rightarrow 3$, $3\rightarrow 4$, $4\rightarrow 1$. For example, one has
        \begin{equation}\label{P23-P34}
        \begin{split}
            P_{23} &= (P_{1234})^2\,P_{12}\,P_{1234}\,P_{12}\,, \\
            P_{34} &= P_{12}\,(P_{1234})^2\,P_{12}\,(P_{1234})^2\,P_{12}\,.
        \end{split}
        \end{equation}
        The Cayley graph in Fig.~\ref{fig:cayley}
        represents the group manifold as a geometric structure made of squares and hexagons.
        The squares contain the elements that are connected to each other by 4-cycles, for example
        \begin{equation}
            f_{1234} \, \stackrel{P_{1234}}{\llongrightarrow} \,
            f_{2341} \, \stackrel{P_{1234}}{\llongrightarrow} \,
            f_{3412} \, \stackrel{P_{1234}}{\llongrightarrow} \,
            f_{4123} \,.
        \end{equation}
        In total there are six squares with four elements each, which
        are connected to each other by transpositions $P_{12}$.
        The two permutation chains in Eq.~\eqref{P23-P34} then correspond to paths along the Cayley diagram.
        Instead of $P_{12}$ and $P_{1234}$ it is also common to span the group by the transpositions $P_{12}$, $P_{23}$ and $P_{34}$,
        which leads to a similar Cayley graph.

        One can rearrange the 24 permutations in multiplets
        that transform under the irreducible representations of $S_4$. The respective Young diagrams are
        \begin{equation*}
             \begin{array}{c @{\qquad} c @{\qquad} c @{\qquad\;\;} c @{\qquad\;\;} c}
             \mS & \mT^+_i & \mD_j & \mT^-_i & \mA \\[2mm]
             \scriptsize\yng(4) &
             \scriptsize\yng(3,1) &
             \scriptsize\yng(2,2) &
             \scriptsize\yng(2,1,1) &
             \scriptsize\yng(1,1,1,1)
             \end{array}
        \end{equation*}
        As for the case of $S_3$, $\mS$ is a  singlet and $\mA$ an antisinglet, and
        the doublets $\mD_j$ ($j=1,2$) form a two-dimensional irreducible subspace.
        The triplets $\mT^+_i$ and antitriplets $\mT^-_i$ ($i=1,2,3$) are new: they transform under inequivalent irreducible representations
        and thereby form two different three-dimensional subspaces. We denote their elements by
        \begin{equation}\label{S4-multiplets-generic}
            \mT^\pm_i = \left[ \begin{array}{c} u^\pm_i \\ v^\pm_i \\ w^\pm_i \end{array}\right]\,.
        \end{equation}

        For the explicit construction of the multiplets it is helpful to group the 24 permutations of an element $f_{1234}$
        into the following three subclasses:
        \renewcommand{\arraystretch}{1.2}
        \begin{equation*}
        f^{(1)} =
        \left[ \begin{array}{c}
            f_{1234}\\
            f_{3412} \\
            f_{2143} \\
            f_{4321}
        \end{array}\right]\!,\;\;
        f^{(2)} =
        \left[ \begin{array}{c}
            f_{2314}  \\
            f_{4132}  \\
            f_{1423} \\
            f_{3241}
        \end{array}\right]\!,\;\;
        f^{(3)} =
        \left[ \begin{array}{c}
            f_{3124} \\
            f_{1342} \\
            f_{4213} \\
            f_{2431}
        \end{array}\right]\!,
        \end{equation*}
        together with the transpositions
        \begin{equation}
            P_{12}\,f^{(1)}, \quad P_{12}\,f^{(2)}, \quad P_{12}\,f^{(3)}.
        \end{equation}
        We label the column vectors with $i=1,2,3$ and denote the column index by $k=1\dots 4$.
        Together with the transpositions $P_{12}$, each column defines a closed path in Fig.~\ref{fig:cayley} that contains eight elements.

        A similar classification is that in terms of $t-$, $s-$ and $u-$channel permutations:
        the $t-$channel permutations $f^{(1)}$, $P_{12}\,f^{(1)}$ are the eight elements in the vertical center of the Cayley graph where $\{12\}$ and $\{34\}$ are grouped,
        the $u-$channel permutations $f^{(3)}$, $P_{12}\,f^{(2)}$ are those at the top and the bottom with $\{31\}\{24\}$,
        and the $s-$channel permutations $f^{(2)}$, $P_{12}\,f^{(3)}$ contain the remaining eight elements with $\{23\}\{14\}$.
        Going from the $t-$channel to the $s-$ and $u-$channels amounts to
        \begin{equation}
            f_{1234} \, \stackrel{P_{12}\,P_{23}}{\longlongrightarrow} \, f_{2314}\,, \qquad
            f_{1234} \, \stackrel{P_{23}\,P_{12}}{\longlongrightarrow} \, f_{3124} \,.
        \end{equation}

        Now take the sum of the entries in each column and (anti-)symmetrize with  $P_\pm = 1 \pm P_{12}$:
        \begin{equation}\label{perm-001}
            \psi_i^\pm = P_\pm \sum_{k=1}^4 f^{(i)}_k\,, \qquad i=1,2,3\,.
        \end{equation}
        The resulting six combinations constitute the singlet, antisinglet and the two doublets
        which have the same form as the multiplets in $S_3$:
        \renewcommand{\arraystretch}{1.2}
        \begin{equation}\label{perm-multiplets}
        \begin{split}
            \mathcal{S} &= \left(\psi_1 + \psi_2 + \psi_3\right)^+ \,, \\
            \mathcal{A} &= \left(\psi_1  + \psi_2 + \psi_3\right)^- \,, \\[2mm]
            \mathcal{D}_1
             &= \left[ \begin{array}{c}
                                    \left(\psi_2-\psi_3\right)^- \\
                                    -\tfrac{1}{\sqrt{3}} \left(  \psi_2 + \psi_3 -2\psi_1 \right)^+ \end{array}\right] ,\\
            \mathcal{D}_2
            & = \left[ \begin{array}{c}
                                    \tfrac{1}{\sqrt{3}} \left( \psi_2 + \psi_3 -2\psi_1 \right)^- \\
                                    \left(\psi_2-\psi_3\right)^+ \end{array}\right].
        \end{split}
        \end{equation}
        The remaining 18 elements are reserved for the triplets and antitriplets. If we generalize Eq.~\eqref{perm-001} to
        \begin{equation}\label{triplet-derivation-1}
        \begin{split}
               a_i^\lambda &= P_\lambda\left(f^{(i)}_1-f^{(i)}_2+f^{(i)}_3-f^{(i)}_4\right),   \\
               b_i^\lambda &= P_\lambda\left(f^{(i)}_1-f^{(i)}_2-f^{(i)}_3+f^{(i)}_4\right),   \\
               c_i^\lambda &= P_\lambda\left(f^{(i)}_1+f^{(i)}_2-f^{(i)}_3-f^{(i)}_4\right),
        \end{split}
        \end{equation}
        with $\lambda=\pm$, and define
        \renewcommand{\arraystretch}{1.4}
        \begin{equation}
        \begin{array}{rl}
            (\phi_1)_1^\lambda \!\!&= \lambda \,a_1^\lambda,\\
            (\phi_2)_1^\lambda \!\!&= \lambda \,b_2^\lambda, \\
            (\phi_3)_1^\lambda \!\!&= \lambda \,c_3^\lambda,
        \end{array}\;\;
        \begin{array}{rl}
            (\phi_1)_2^\lambda \!\!&= -a_2^\lambda,\\
            (\phi_2)_2^\lambda \!\!&= -c_1^\lambda, \\
            (\phi_3)_2^\lambda \!\!&= -b_3^\lambda ,
        \end{array}\;\;
        \begin{array}{rl}
            (\phi_1)_3^\lambda    \!\!&= a_3^\lambda,\\
            (\phi_2)_3^\lambda \!\!&= c_2^\lambda, \\
            (\phi_3)_3^\lambda \!\!&= b_1^\lambda,
        \end{array}
        \end{equation}
        then we can cast the triplets and antitriplets in a common form:
        \renewcommand{\arraystretch}{1.2}
        \begin{equation}\label{perm-triplets}
            \mathcal{T}_i^\pm
             = \left[ \begin{array}{c}
                                    \sqrt{\tfrac{2}{3}}\left( \phi_1+\phi_2+\phi_3\right)^\pm_i  \\
                                    \tfrac{1}{\sqrt{3}}\left( \phi_2+\phi_3-2\phi_1\right)^\pm_i \\
                                    \left(\phi_2-\phi_3\right)^\mp_i
                                    \end{array}\right].
        \end{equation}

        \renewcommand{\arraystretch}{1.4}
        The transformation laws for any permutation can be reconstructed from
        \begin{equation}\label{permutation-tf-s4} \renewcommand{\arraystretch}{1.2}
        \begin{split}
           P_{12}\left\{ \begin{array}{c} \mS \\ \mA \\ \mD_j \\[1mm] \mT_i^\lambda\end{array}\right\} &=
                 \left\{ \begin{array}{c} \mS \\ -\mA \\ \mathsf{M}_{12}^\mathsf{T}\,\mD_j \\[1mm] \lambda\, \mathsf{H}_{12}^\mathsf{T}\,\mT_i^\lambda \end{array}\right\} , \\
           P_{1234}\left\{ \begin{array}{c} \mS \\ \mA \\ \mD_j \\[1mm] \mT_i^\lambda\end{array}\right\} &=
                 \left\{ \begin{array}{c} \mS \\ -\mA \\ \mathsf{M}_{1234}^\mathsf{T}\,\mD_j \\[1mm] \lambda\, \mathsf{H}_{1234}^\mathsf{T}\,\mT_i^\lambda\end{array}\right\},
        \renewcommand{\arraystretch}{1.0}
        \end{split}
        \end{equation}
        with the two-dimensional representation matrices
        \renewcommand{\arraystretch}{1.0}
        \begin{equation}\label{doublet-rep-matrices}
           \mathsf{M}_{12} = \left(\begin{array}{rr} -1 & \, 0 \\ 0 & \ 1 \end{array}\right), \qquad
           \mathsf{M}_{1234} = \frac{1}{2}\left(\begin{array}{cc} 1 & \sqrt{3} \\ \sqrt{3} & -1 \end{array}\right)
        \end{equation}
        and the three-dimensional matrices
        \begin{equation}\label{triplet-rep-matrices}
        \begin{split}
           \mathsf{H}_{12} &= \left(\begin{array}{rrr} 1 & \, 0  &   0 \\ 0 & 1 & 0 \\ 0 & 0 & -1 \end{array}\right), \\
           \mathsf{H}_{1234}  &= -\frac{1}{6}\left(\begin{array}{ccc} 2 &  4\sqrt{2}  &   0 \\ -2\sqrt{2} & 1 & 3\sqrt{3} \\ 2\sqrt{6} & -\sqrt{3} & 3 \end{array}\right).
        \end{split}
        \end{equation}
        The representation matrices for the $\mathcal{T}^-_i$ differ
        by a minus sign from those of the $\mathcal{T}^+_i$, and it is
        not possible to rearrange the entries $\{u,v,w\}$
        to obtain a common transformation law; i.e., the triplets and antitriplets transform under inequivalent representations.\footnote{An
        equivalent form of the representation matrices can be found in Table 3.2 of Ref.~\cite{vanBeveren:1998}.
        They follow if one writes instead of Eq.~\eqref{S4-multiplets-generic}:
        \renewcommand{\arraystretch}{1.0}
        \begin{equation*}
            \mathcal{T}^+_i = \left[ \begin{array}{c} u \\ -v \\ w \end{array}\right]^+_i, \quad
            \mathcal{T}^-_i = \left[ \begin{array}{c} w \\ v \\ u \end{array}\right]^-_i, \quad
            \mathcal{D}_j = \left[ \begin{array}{c} s \\ a \end{array}\right]_j,
        \end{equation*}
        however at the price that Eqs.~(\ref{perm-triplets}--\ref{permutation-tf-s4}) and the notation for the product representations become less compact.
        The group characters (the traces over the representation matrices) are necessarily the same.}

\subsection{Product representations}\label{sec:product-representations}

        \renewcommand{\arraystretch}{1.0}

        The next step is to find all possible
        products of elementary multiplets ($\mathcal{S}$, $\mathcal{A}$, $\mathcal{D}_j$ and $\mathcal{T}_i^\pm$) that also satisfy
        the transformation laws in Eq.~\eqref{permutation-tf-s4}:

        \medskip
        (i)\,
        \textbf{Singlets} are obtained from the combinations
        \begin{equation}\label{perm-product-singlets}
            \mathcal{S}\,\mathcal{S}\,, \quad \mathcal{A}\,\mathcal{A}\,, \quad \mathcal{D}_i\cdot\mathcal{D}_j\,, \quad \mathcal{T}_i^\pm\cdot\mathcal{T}_j^\pm\,,
        \end{equation}
        where $(\cdot)$ is the usual dot product for vectors:
        \begin{equation}
        \begin{split}
            \mathcal{D}\cdot\mathcal{D}' &:=aa'+ss' \,, \\
            \mathcal{T}\cdot\mathcal{T}' &:=uu'+vv'+ww'\,.
        \end{split}
        \end{equation}
        The singlet property follows from the orthogonality of the representation matrices.
        Similarly, the combinations
        \begin{equation}\label{perm-product-antisinglets}
            \mathcal{S}\,\mathcal{A}\,, \quad \mathcal{D}_i\wedge\mathcal{D}_j\,, \quad \mathcal{T}_i^\pm\cdot\mathcal{T}_j^\mp
        \end{equation}
        produce \textbf{antisinglets}, where we used the antisymmetric product $\mathcal{D} \wedge \mathcal{D'} = as'-sa'$.

        \medskip
        (ii)\,
        \textbf{Doublets} can be constructed in various ways. $\mathcal{S}\,\mathcal{D}$ and $\mathcal{A}\,(\varepsilon\,\mathcal{D})$ are doublets;
        $\varepsilon$ was defined in Eq.~\eqref{perm-product-doublets-2-s3}.
        The combination of two doublets and two (anti-)triplets also produces doublets:
        \begin{equation}\label{perm-product-doublets}
            \mathcal{D}_i \ast \mathcal{D}_j\,, \quad
            \mathcal{T}_i^\pm \ast \mathcal{T}_j^\pm\,, \quad
            \varepsilon\left(\mathcal{T}_i^\pm \ast\mathcal{T}_j^\mp\right),
        \end{equation}
        where the ($\ast$) operation has the form
        \begin{align}
           \mathcal{D} \ast \mathcal{D}'&:= \left[ \begin{array}{c} as'+sa' \\ aa'-ss' \end{array}\right], \\
           \mathcal{T} \ast \mathcal{T}'&:= \left[ \begin{array}{c} vw'+wv'+\sqrt{2}\,(uw'+wu') \\ ww'-vv'+\sqrt{2}\,(uv'+vu') \end{array}\right].
        \end{align}

        \medskip
        (iii)\, \textbf{Triplets} $\mathcal{T}^+$ are obtained from
        \begin{equation}\label{perm-product-triplets}
            \mathcal{S}\,\mathcal{T}^+ , \quad
            \mathcal{A}\,\mathcal{T}^- , \quad
            \begin{array}{l} \mathcal{T}^+\vee \mathcal{D}\,, \\\mathcal{T}^- \wedge \mathcal{D}\,,\end{array}\quad
            \begin{array}{l} \mathcal{T}^\pm\vee \mathcal{T}^\pm\,,\\ \mathcal{T}^\pm\wedge \mathcal{T}^\mp \end{array}
        \end{equation}
        and \textbf{antitriplets} $\mathcal{T}^-$ from
        \begin{equation}\label{perm-product-antitriplets}
            \mathcal{S}\,\mathcal{T}^-, \quad
            \mathcal{A}\,\mathcal{T}^+, \quad
            \begin{array}{l} \mathcal{T}^+ \wedge \mathcal{D}\,,\\ \mathcal{T}^-\vee \mathcal{D}\end{array}\quad
            \begin{array}{l} \mathcal{T}^\pm\wedge \mathcal{T}^\pm\,,\\ \mathcal{T}^\pm\vee \mathcal{T}^\mp\,.\end{array}
        \end{equation}
        We defined the `wedge' and `vee' products for the triplets and antitriplets as
        \begin{align}
            \mathcal{T}\wedge\mathcal{T}' &:= \left[ \begin{array}{c} vw'-wv' \\ wu'-uw' \\ uv'-vu' \end{array}\right], \\
            \mathcal{T}\vee\mathcal{T}' &:= \left[ \begin{array}{c} vv'+ww'-2uu' \\ uv'+vu'+\sqrt{2}\,(vv'-ww') \\ uw'+wu'-\sqrt{2}\,(vw'+wv') \end{array}\right],
        \end{align}
        and for the combination of (anti-)triplet and doublet:
        \begin{align}
            \mathcal{T}\wedge\mathcal{D} &:= \left[ \begin{array}{c} va-ws \\ ua-\tfrac{1}{\sqrt{2}}\,(va+ws) \\ -us-\tfrac{1}{\sqrt{2}}\,(vs-wa) \end{array}\right], \\
            \mathcal{T}\vee\mathcal{D} &:= \left[ \begin{array}{c} vs+wa \\ us-\tfrac{1}{\sqrt{2}}\,(vs-wa) \\ ua+\tfrac{1}{\sqrt{2}}\,(va+ws) \end{array}\right].
        \end{align}
        The wedge product for the triplets is the usual vector product; as such
        it satisfies the identity
        \begin{equation}
            \mT_i\cdot (\mT_j\wedge\mT_k)  = \mT_j\cdot(\mT_k\wedge\mT_i)\,,
        \end{equation}
        where $\{ijk\}$ is a cyclic permutation of $\{123\}$. It turns out that the identity
        also holds for the vee and star products,
        \begin{equation}
        \begin{split}
            \mT_i\cdot (\mT_j\vee\mT_k)  &= \mT_j\cdot(\mT_k\vee\mT_i)\,,\\
            \mD_i\cdot (\mD_j\ast\mD_k)  &= \mD_j\cdot(\mD_k\ast\mD_i)\,,
        \end{split}
        \end{equation}
        and there is a similar relation for the combination of triplet and doublet:
        \begin{equation}
        \begin{split}
            \mT_1\cdot (\mT_2\wedge\mD)  &= \mT_2\cdot (\mT_1\wedge\mD) \,,\\
            \mT_1\cdot (\mT_2\vee\mD)  &= \mT_2\cdot (\mT_1\vee\mD)\,.
        \end{split}
        \end{equation}
        Finally, we mention the identities
        \begin{align}
           &\sum_{\{ijk\}} (\mD_i\wedge\mD_j)\,\mD_k = 0, \\
           &\sum_{\{ijkl\}} (-1)^P \left[ \mT_i\cdot(\mT_j\wedge\mT_k)\right] \mT_l=0,
        \end{align}
        where the sums are over cyclic permutations and $P=\pm$ for even/odd permutations.
        They are useful for expanding a doublet in a basis defined by two other doublets, or
        a triplet in a basis defined by three triplets.

    \renewcommand{\arraystretch}{1.8}

        \begin{table*}[t]

                \begin{center}
                \begin{tabular}{    @{\;\;\;} l @{\;\;\;}  || @{\;\;\;} c @{\;\;\;} |  @{\;\;\;} c @{\;\;\;}  |  @{\;\;\;} c @{\;\;\;} ||  @{\;\;\;} c @{\;\;\;} |  @{\;\;\;} c @{\;\;\;} | @{\;\;\;} c @{\;\;\;} | @{\;\;\;} c @{\;\;\;}       }

                                           & $(2,0)$         & $(1,1)$           & $(0,2)$          & $(3,0)$                  & $(2,1)$                  & $(1,2)$                 & $(0,3)$                    \\   \hline\hline

                             Singlet       & $\mD\cdot\mD$   &                   & $\mT\cdot\mT$    & $\mD\cdot(\mD\ast\mD)$   &                          & $\mD\cdot(\mT\ast\mT)$  & $\mT\cdot(\mT\vee\mT)$     \\[1mm] \hline
                             Doublet       & $\mD\ast\mD$    &                   & $\mT\ast\mT$     &                          &                          & $\mD\ast(\mT\ast\mT)$   &                            \\[-1.5mm]
                                           &                 &                   &                  & \gray{$(\mD\cdot\mD)\,\mD$} &                       & \gray{$(\mT\cdot\mT)\,\mD$} &                        \\[1mm] \hline
                             Triplet       &                 & $\mT\vee\mD$      & $\mT\vee\mT$     &                          & $\mT\vee(\mD\ast\mD)$    & $\mT\vee(\mT\vee\mD)$   & $\mT\vee(\mT\vee\mT)$       \\[-1.5mm]
                                           &                 &                   &                  &                          & \gray{$(\mD\cdot\mD)\,\mT$} &                      & \gray{$(\mT\cdot\mT)\,\mT$}  \\[1mm] \hline
                             Antitriplet   &                 & $\mT\wedge\mD$    &                  &                          & $\mT\wedge(\mD\ast\mD)$  & $\mT\wedge(\mT\vee\mD)$ & $\mT\wedge(\mT\vee\mT)$    \\[1mm] \hline
                             Antisinglet   &                 &                   &                  & $\mD\wedge(\mD\ast\mD)$  &                          & $\mD\wedge(\mT\ast\mT)$ &

                \end{tabular}
                \end{center}

               \caption{         Products of a doublet $\mD$ and a triplet $\mT$ according to the rules~(\ref{perm-product-singlets}--\ref{perm-product-antitriplets})
                                 up to cubic terms. The brackets in the top row count the
                                 number of multiplets; e.g., $(2,1)$ refers to two doublets and one triplet in the product.
                                 The elements in gray are trivially obtained by multiplying $\mD$ and $\mT$ with singlets.  }
               \label{tab:momentum-multiplets-0}

        \end{table*}

    \renewcommand{\arraystretch}{1.0}

        The list defined by Eqs.~(\ref{perm-product-singlets}--\ref{perm-product-antitriplets}) is exhaustive.
        Checking the transformation properties of these quantities by hand can become tedious, but they are simple to implement in a computer algebra system such as \texttt{Mathematica}~\cite{mma}.
        For example, suppose we want to find all possible products of one triplet $\mathcal{T}^+$ with itself.
        The resulting six combinations $u^2$, $v^2$, $w^2$, $uv$, $uw$, $vw$ can be rearranged into a symmetric singlet,
        a doublet, and a triplet: 
        \begin{equation}\label{triplet-operations}
        \begin{split}
           \mathcal{T}^+\cdot\mathcal{T}^+ &=u^2+v^2+w^2\,, \\
           \mathcal{T}^+ \ast \mathcal{T}^+&= \left[ \begin{array}{c} 2w\,(v+\sqrt{2}\,u) \\ w^2-v^2+2\sqrt{2}\,uv \end{array}\right], \\
           \mathcal{T}^+\vee\mathcal{T}^+ &= \left[ \begin{array}{c} v^2+w^2-2u^2 \\ 2uv+\sqrt{2}\,(v^2-w^2) \\ 2w\,(u-\sqrt{2}\,v) \end{array}\right].
        \end{split}
        \end{equation}
        It is not possible to form an antisinglet, and
        the antitriplet vanishes because $\mathcal{T}^+\wedge\mathcal{T}^+=0$.

        Table~\ref{tab:momentum-multiplets-0} collects all possible multiplets that can be systematically constructed from a doublet $\mD$ and a triplet $\mT^+ = \mT$ up to cubic terms.
        This will become relevant in Sec.~\ref{sec:phase-space} because the six Lorentz invariants in a four-point function can be arranged
        into a singlet, a doublet and a triplet, which can generate further product multiplets with higher mass dimensions.

\subsection{Example: four-gluon vertex}\label{sec:color}

      As an application, let us recast the color factors of the four-gluon vertex~\cite{Pascual:1980yu} in the multiplet notation.
      There are four types of color structures that can be used as seed elements
      for obtaining all further multiplets:
      \begin{equation}
      \begin{split}
         A_{abcd} &= \delta_{ab}\,\delta_{cd}\,, \\
         B_{abcd} &= f_{abe}\,f_{cde}\,, \\
         C_{abcd} &= d_{abe}\,d_{cde}\,, \\
         D_{abcd} &= f_{abe}\,d_{cde}\,.
      \end{split}
      \end{equation}
      Here, $f_{abc}$ and $d_{abc}$ are the antisymmetric and symmetric structure constants of $SU(N)$, respectively,
      \begin{equation}
      \begin{split}
          f_{abc} &= -2i\,\text{Tr}\,([\mathsf{t}_a, \mathsf{t}_b]\,\mathsf{t}_c)\,, \\
          d_{abc} &= 2\,\text{Tr}\,(\{\mathsf{t}_a, \mathsf{t}_b\}\,\mathsf{t}_c)\,,
      \end{split}
      \end{equation}
      and $\mathsf{t}_a$ are the $SU(N)$ generators in the fundamental representation.

      We specialize to $SU(3)$, where $\mathsf{t}_a = \lambda_a/2$ and $\lambda_a$ are the Gell-Mann matrices.
      For the seed $A_{abcd} = \delta_{ab}\,\delta_{cd}$, the three column vectors in Eq.~\eqref{perm-001} simply become
      \begin{equation} \renewcommand{\arraystretch}{1.3}
         \begin{array}{rl}
         f^{(1)} &= \delta_{ab}\,\delta_{cd}\,\mathsf{V}, \\
         f^{(2)} &= \delta_{bc}\,\delta_{ad}\,\mathsf{V}, \\
         f^{(3)} &= \delta_{ca}\,\delta_{bd}\,\mathsf{V},
         \end{array} \qquad \renewcommand{\arraystretch}{1.0}
         \mathsf{V} = \left[\begin{array}{c} 1\\1\\1\\1\end{array}\right],
      \end{equation}
      and the transposition $P_{12}$ has the effect
      \begin{equation*}
         P_{12}\,f^{(1)} = f^{(1)}, \quad
         P_{12}\,f^{(2)} = f^{(3)}, \quad
         P_{12}\,f^{(3)} = f^{(2)}.
      \end{equation*}
      Since all entries in $\mathsf{V}$ are identical, Eq.~\eqref{triplet-derivation-1} can only produce zeros
      and therefore the triplets and antitriplets vanish. The $\psi_i^\pm$ of Eq.~\eqref{perm-001} become
      \begin{equation}
      \begin{split}
             \psi_1^+ &= 8\,\delta_{ab}\,\delta_{cd}\,,\\
             \psi_1^- &= 0\,, \\
             \psi_2^+ = \psi_3^+ &= 4\,(\delta_{bc}\,\delta_{ad} + \delta_{ca}\,\delta_{bd})\,, \\
             \psi_2^- = -\psi_3^- &= 4\,(\delta_{bc}\,\delta_{ad} - \delta_{ca}\,\delta_{bd})\,,
      \end{split}
      \end{equation}
      and hence only the singlet and the doublet $\mD_1$ survives:
      \begin{equation}\label{color-dd}
      \begin{split}
          \tfrac{1}{8}\,\mS(A) &= \delta_{ab}\,\delta_{cd} + \delta_{bc}\,\delta_{ad} + \delta_{ca}\,\delta_{bd}, \\
          \tfrac{1}{8}\,\mD_1(A) &= \left[\begin{array}{c} \delta_{bc}\,\delta_{ad} - \delta_{ca}\,\delta_{bd} \\ -\tfrac{1}{\sqrt{3}}\,( \delta_{bc}\,\delta_{ad} + \delta_{ca}\,\delta_{bd} -2 \delta_{ab}\,\delta_{cd} ) \end{array}\right].
      \end{split}
      \end{equation}
      The derivation for the seed $B_{abcd} = f_{abe}\,f_{cde}$ is identical except that the transpositions $P_{12}$ produce
      minus signs due to the antisymmetry of the structure constants. The resulting antisinglet vanishes because of the Jacobi identity,
      \begin{equation}
          \tfrac{1}{8}\,\mA(B) = f_{abe}\,f_{cde} + f_{bce}\,f_{ade} + f_{cae}\,f_{bde} = 0,
      \end{equation}
      and the remaining doublet is
      \begin{equation*}
          \tfrac{1}{8}\,\mD_2(B) = \left[\begin{array}{c} \tfrac{1}{\sqrt{3}}\,(f_{bce}\,f_{ade} + f_{cae}\,f_{bde} -2 f_{abe}\,f_{cde}) \\ f_{bce}\,f_{ade} - f_{cae}\,f_{bde} \end{array}\right].
      \end{equation*}
      The seed $C_{abcd} = d_{abe}\,d_{cde}$ produces another singlet and doublet $\mD_1$, but they linearly depend on the ones above:
      \begin{equation}
      \begin{split}
          \mS(C) &= \tfrac{1}{3}\,\mS(A), \\
          \mD_1(C) &= -\tfrac{2}{3}\,\mD_1(A) + \tfrac{1}{\sqrt{3}}\,\mD_2(B)\,,
      \end{split}
      \end{equation}
      which is a consequence of identities that can be found in Ref.~\cite{Pascual:1980yu}.
      Finally, the seed $D_{abcd} = f_{abe}\,d_{cde}$ produces one independent antitriplet and nothing else:
      \begin{equation*}
         \tfrac{1}{8}\,\mT_3^-(D) =\left[\begin{array}{c} \tfrac{1}{\sqrt{6}}\,(f_{abe}\,d_{cde} + f_{bce}\,d_{ade} + f_{cae}\,d_{bde}) \\
                                                             \tfrac{1}{\sqrt{3}}\,(f_{bde}\,d_{cae} - f_{ade}\,d_{bde})  \\
                                                             f_{cde}\,d_{abe} \end{array}\right],
      \end{equation*}
      with $\mT_2^-(D) = -\mT_3^-(D)$.
      Therefore, the four-gluon vertex in $SU(3)$ has eight independent color structures in total (we attach a subscript $c$ for color):
        \begin{equation}
            \mS_c=\mS(A), \quad
            \begin{array}{rl} \mD_c^{(1)} &=\mD_1(A), \\[1mm] \mD_c^{(2)}&=\mD_2(B),  \end{array}\quad \mT_c^-=\mT_3^-(D)\,.
        \end{equation}
      Ignoring the $d_{abc}$ symbols, they reduce to five.

        Now suppose we want to combine them also with the Lorentz structures.
        The four-gluon vertex is Bose-symmetric and therefore the products of Lorentz tensors, color factors, and momentum-dependent dressing functions must form symmetric singlets.
        If we restrict ourselves to the subset
        of momentum-independent Lorentz structures, then only one seed element $\delta^{\mu\nu}\,\delta^{\rho\sigma}$ contributes.
        It generates a singlet $\mS_L$ and a doublet $\mD_L$ (with subscript $L$ for Lorentz) which have the same form as in Eq.~\eqref{color-dd}.
        Consequently, there are three possible
        Lorentz-color singlets in the product space:
        \begin{equation}
            \mS_L\,\mS_c, \qquad
            \mD_L\cdot\mD_c^{(1)}, \qquad
            \mD_L\cdot\mD_c^{(2)}\,.
        \end{equation}
        The last element is the tree-level structure of the four-gluon vertex because it can be written as
        \begin{equation}
        \begin{split}
            -\tfrac{\sqrt{3}}{128}\,\mD_\text{L}\cdot\mD_\text{C}^{(2)} &= f_{abe}\,f_{cde}\,(\delta^{\nu\rho}\,\delta^{\mu\sigma} - \delta^{\rho\mu}\,\delta^{\nu\sigma}) \\
                                                          & +f_{bce}\,f_{ade}\,(\delta^{\rho\mu}\,\delta^{\nu\sigma} - \delta^{\mu\nu}\,\delta^{\rho\sigma}) \\
                                                          & +f_{cae}\,f_{bde}\,(\delta^{\mu\nu}\,\delta^{\rho\sigma} - \delta^{\nu\rho}\,\delta^{\mu\sigma}) \,.
        \end{split}
        \end{equation}
        If we include Lorentz tensors with higher mass dimension,
        or if we allow the momentum-dependent dressing functions to form multiplets other than singlets,
        there will be many more possible combinations. A more detailed discussion of the Lorentz tensor basis will follow in Sec.~\ref{sec:tensorbasis}.

\section{Phase space in four-point functions} \label{sec:phase-space}

         We proceed by applying the permutation-group technique to the phase space in four-point functions.
         As we discussed in Sec.~\ref{sec:2photon}, in a Bose-symmetric and minimal basis all form factors are free of kinematic singularities or zeros
         and their momentum evolution is governed purely by the dynamics of the system. A generic offshell four-point function has a rich
         phase space because its form factors depend on six Lorentz invariants. Therefore, the goal of this section is
         to isolate the relevant kinematic regions where dynamical singularities occur, and to identify possible `scaling variables'
         that simplify the description because they carry the main momentum dependence.

\subsection{Definitions} \label{sec:definitions}

          \begin{figure}[t]
                    \begin{center}
                    \includegraphics[scale=0.18]{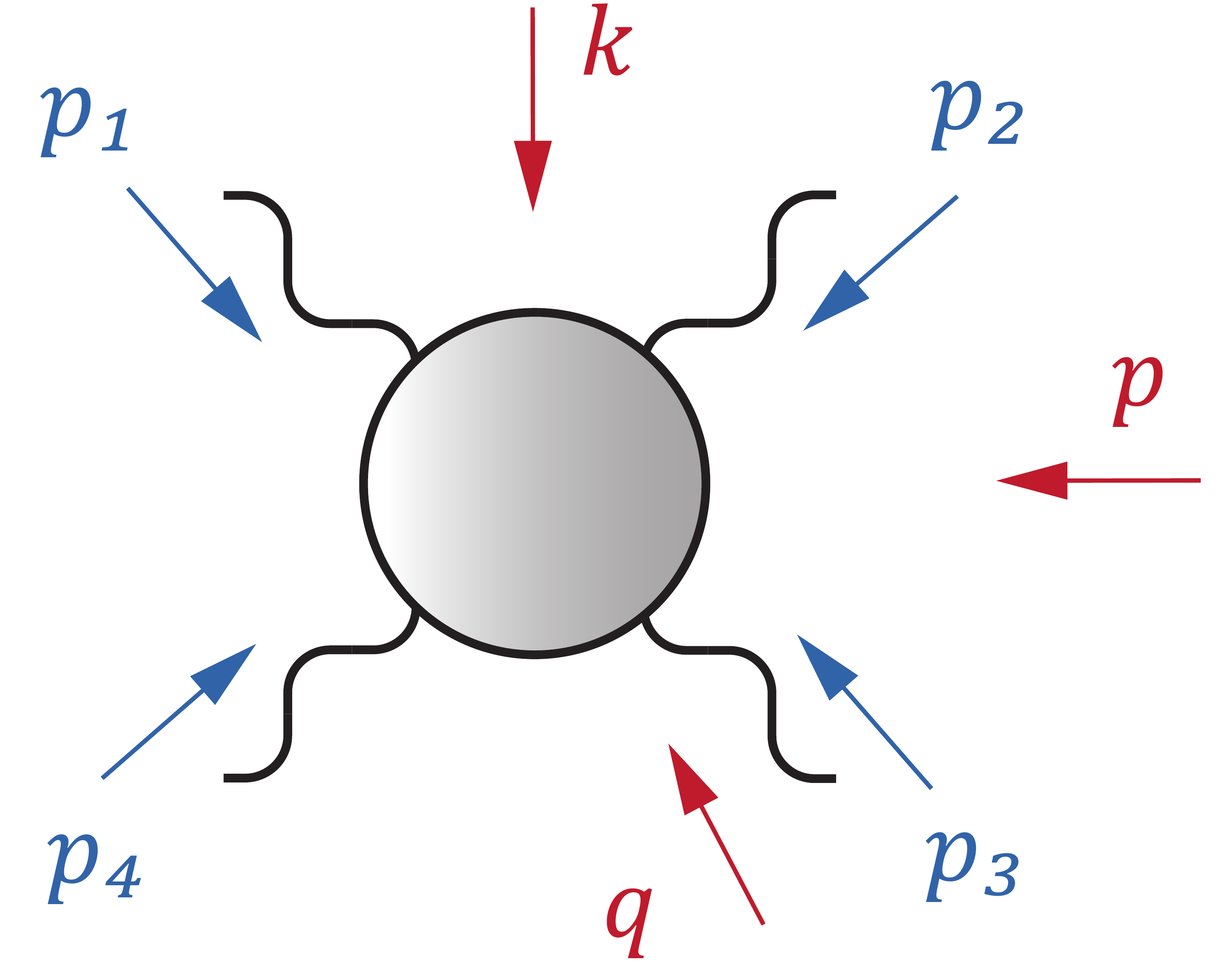}
                    \caption{ Kinematics in the photon four-point function.
                                            }\label{fig:kinematics}
                    \end{center}
        \end{figure}

         A generic vector four-point amplitude has the form
         \begin{equation}\label{4PA-decomposition-0}
             \mathcal{M}^{\mu\nu\rho\sigma}(p_1 \dots p_4) = \sum_{i=1}^N f_i(\dots)\,\tau_i^{\mu\nu\rho\sigma}(p_1 \dots p_4)\,,
         \end{equation}
         which is illustrated in Fig.~\ref{fig:kinematics} for the special case of four photons.
         It depends on four momenta $p_1$, $p_2$, $p_3$, $p_4$, of which only three are
         independent due to momentum conservation $p_1+p_2+p_3+p_4=0$.
         Since the Lorentz indices are not relevant for the discussion in this section,
         the analysis applies not only to the hadronic LbL amplitude but also to the four-gluon vertex or any other symmetric four-point function.
         $\mathcal{M}$ has a decomposition into $N$ tensor basis elements with $N$ Lorentz-invariant dressing functions $f_i$. They depend on $4\cdot3/2=6$
         independent Lorentz invariants which we will specify below.

         Bose symmetry entails that all 24 permutations of $\mathcal{M}$ (interchanges of momentum labels and Lorentz indices) are identical:
         \begin{equation*}
             \mathcal{M}^{\mu\nu\rho\sigma}(p_1,p_2,p_3,p_4) = \mathcal{M}^{\nu\mu\rho\sigma}(p_2,p_1,p_3,p_4) = \dots
         \end{equation*}
         We can group them into three classes with eight permutations each, as discussed earlier:
         $t$-channel $\{12\}\{34\}$, $s$-channel $\{23\}\{14\}$ and $u$-channel $\{31\}\{24\}$.
         Therefore, the simplest choice of independent momenta are the
         $t-$, $s-$, and $u-$channel Mandelstam momenta:
         \begin{equation}\label{Mandelstam-momenta}
         \begin{split}
             k &= p_1+p_2 = -p_3-p_4\,, \\
             p &= p_2+p_3 = -p_1-p_4\,, \\
             q &= p_3+p_1 = -p_2-p_4\,,
         \end{split}
         \end{equation}
         with the inverse relations
        \renewcommand{\arraystretch}{1.2}
         \begin{equation}\label{momenta-inverse}
         \begin{array}{rl}
             p_1 &= \tfrac{1}{2}\,(q-p+k)\,, \\
             p_3 &= \tfrac{1}{2}\,(q+p-k)\,,
         \end{array}\quad
         \begin{array}{rl}
             p_2 &= -\tfrac{1}{2}\,(q-p-k)\,, \\
             p_4 &= -\tfrac{1}{2}\,(q+p+k)\,.
         \end{array}
         \end{equation}
         Going from the $t-$channel to the
         $s-$ and $u-$channels then amounts to cyclic permutations
         \begin{alignat}{4}\label{3-permutations}
            t-\text{channel} : \quad & p_1, \;  p_2, \; p_3, \; p_4  && \quad \Leftrightarrow \quad &&  p, \; q, \; k\,,    \nonumber \\
            s-\text{channel} : \quad & p_2, \;  p_3, \; p_1, \; p_4  && \quad \Leftrightarrow \quad &&  q, \; k, \; p\,,      \\
            u-\text{channel} : \quad & p_3, \;  p_1, \; p_2, \; p_4  && \quad \Leftrightarrow \quad &&  k, \; p, \; q\,.    \nonumber
         \end{alignat}

 \subsection{Phase space}

        \renewcommand{\arraystretch}{1.0}

         In the next step we would like to develop a geometrical understanding of the phase space,
         in particular of the spacelike (`Euclidean') region. In the context of the LbL scattering amplitude, this is the integration domain
         that contributes to the muon $g-2$ value, whereas in the four-gluon vertex it is the domain that is mapped into itself in a Dyson-Schwinger equation.
         The $f_i$ in Eq.~\eqref{4PA-decomposition-0} depend on six Lorentz invariants that constitute the phase space:
         from the momenta above one can form
         the three Mandelstam variables $p^2$, $q^2$, $k^2$ and the angular variables
         \begin{equation}\label{angular-variables}
             \omega_1=q\cdot k, \qquad
             \omega_2=p\cdot k, \qquad
             \omega_3=p\cdot q\,.
         \end{equation}
         The Mandelstam variables are convenient for discussing two-photon intermediate states that appear in the LbL amplitude, as we shall see below.

         To expose the Bose symmetry of the phase space, we would like to arrange these variables
         into permutation-group multiplets.
         The simplest strategy is to first arrange the four-momenta $p$, $q$, $k$ themselves into multiplets.
         If we use the $t-$channel momentum $k=p_1+p_2$ as permutation-group seed  $f_{1234}$ and follow the steps in Eqs.~(\ref{perm-001}--\ref{perm-triplets}),
         the only nonvanishing multiplet turns out to be the triplet
         \begin{equation}\label{momentum-triplet}
             \mathcal{T}^+ =  \frac{1}{2}\left[\begin{array}{c} \tfrac{1}{\sqrt{3}}\,(p+q+k) \\ \tfrac{1}{\sqrt{6}}\,(p+q-2k) \\ \tfrac{1}{\sqrt{2}}\,(q-p) \end{array}\right].
         \end{equation}
         These are just the three independent Jacobi momenta of the system.
         Applying
         the three product operations in Eq.~\eqref{triplet-operations} to $\mathcal{T}^+$, where each operation
         is simultaneously a scalar product of four-momenta, yields
         six Lorentz invariants that form a singlet, a doublet and a triplet.
         The singlet is given by
         \begin{equation}\label{S0-def}
             \mS_0 = \mathcal{T}^+\cdot\mathcal{T}^+ =\frac{p^2+q^2+k^2}{4} \,,
         \end{equation}
         whereas the doublet is
         \begin{equation}\label{doublet-def}         \renewcommand{\arraystretch}{1.0}
             \mD_0 = \mS_0 \left[ \begin{array}{c} a \\ s \end{array}\right] = \frac{1}{4}\left[ \begin{array}{c} \sqrt{3}\,(q^2-p^2) \\[0.5mm] p^2+q^2-2k^2\end{array}\right]
         \end{equation}
         and the triplet has the form
         \begin{equation}\label{triplet-def}
             \mT_0 = \mS_0 \left[ \begin{array}{c} u \\ v \\ w \end{array}\right] = \frac{1}{4}\left[ \begin{array}{c} -2\,(\omega_1+\omega_2+\omega_3) \\[0.5mm] -\sqrt{2}\,(\omega_1+\omega_2-2\omega_3) \\[0.5mm] \sqrt{6}\,(\omega_1-\omega_2)\end{array}\right].
         \end{equation}
        We pulled out factors of $\mS_0$ to remove the mass dimension of the doublet and triplet variables, so that only
        $\mS_0 \in \mathds{R}_+$ carries a dimension.
        The advantage is that the doublet and triplet phase spaces $\{a,s\}$ and $\{u,v,w\}$ can be discussed independently of $\mS_0$ (otherwise they would scale with $\mS_0$).

        In Sec.~\ref{sec:tensorbasis} we will express the photon four-point function in a fully symmetric basis,
        so that the corresponding dressing functions must be also symmetric and can only depend on singlet Lorentz invariants.
        The six lowest-dimensional singlets constructed from $\mS_0$, $\mD_0$ and $\mT_0$ are those in the first row of Table~\ref{tab:momentum-multiplets-0}:
         \begin{equation}\label{perm-group-invariants-LI}  \renewcommand{\arraystretch}{1.3}
         \begin{array}{rl}
            \mathcal{S}_1 &= \mathcal{D}_0\cdot\mathcal{D}_0\,, \\
            \mathcal{S}_3 &= \mathcal{T}_0\cdot\mathcal{T}_0\,,
         \end{array}\quad
         \begin{array}{rl}
            \mathcal{S}_2 &= \mathcal{D}_0\cdot (\mathcal{D}_0 \ast\mathcal{D}_0)\,, \\
            \mathcal{S}_4 &= \mathcal{D}_0\cdot (\mathcal{T}_0 \ast\mathcal{T}_0)\,, \\
            \mathcal{S}_5 &= \mathcal{T}_0\cdot (\mathcal{T}_0 \vee\mathcal{T}_0)\,,
         \end{array}
         \end{equation}
        together with $\mS_0$ itself.
        Since $\mS_0$ has the lowest mass dimension, it is reasonable to assume that it also carries the main momentum dependence, whereas the dependencies in the remaining
         variables are weaker.

         What is the domain for $\{a,s\}$ and $\{u,v,w\}$ in the spacelike region?
         To find out, we express the three Mandelstam momenta in hyperspherical coordinates (using Euclidean conventions):
         \begin{equation}\label{phasespace-hyperspherical-co}
         \begin{split}
             k = &\sqrt{k^2}\left[\begin{array}{c} 0\\0\\0\\1\end{array}\right], \quad
             p = \sqrt{p^2}\left[\begin{array}{c} 0 \\ 0 \\ \sqrt{1-z^2} \\ z \end{array}\right], \\
             &q = \sqrt{q^2}\left[\begin{array}{l} \, 0 \\ \sqrt{1-{z'}^2}\,\sqrt{1-y^2} \\ \sqrt{1-{z'}^2}\,y \\ \, z' \end{array}\right] ,
         \end{split}
         \end{equation}
         The radial variables $k^2$, $p^2$ and $q^2$ are all real and positive,
         whereas the variables $z$, $z'$ and $y$ are the cosines of hyperspherical angles:
         \begin{equation}\label{spacelike}
              p^2, \,q^2, \,k^2 \;\in \; \mathds{R}_+, \qquad
              z,\,z',\,y \;\in \;[-1,1]\,.
         \end{equation}
         It is straightforward to derive their relations with the Lorentz-invariants in Eq.~\eqref{angular-variables}.
         It turns out that $\{a,s\}$ form the interior of a triangle and $\{u,v,w\}$ that of a tetrahedron, which we will discuss next.

\subsection{Doublet} \label{sec:phase-space-doublet}

\begin{figure}[t]
                    \begin{center}
                    \includegraphics[scale=0.18]{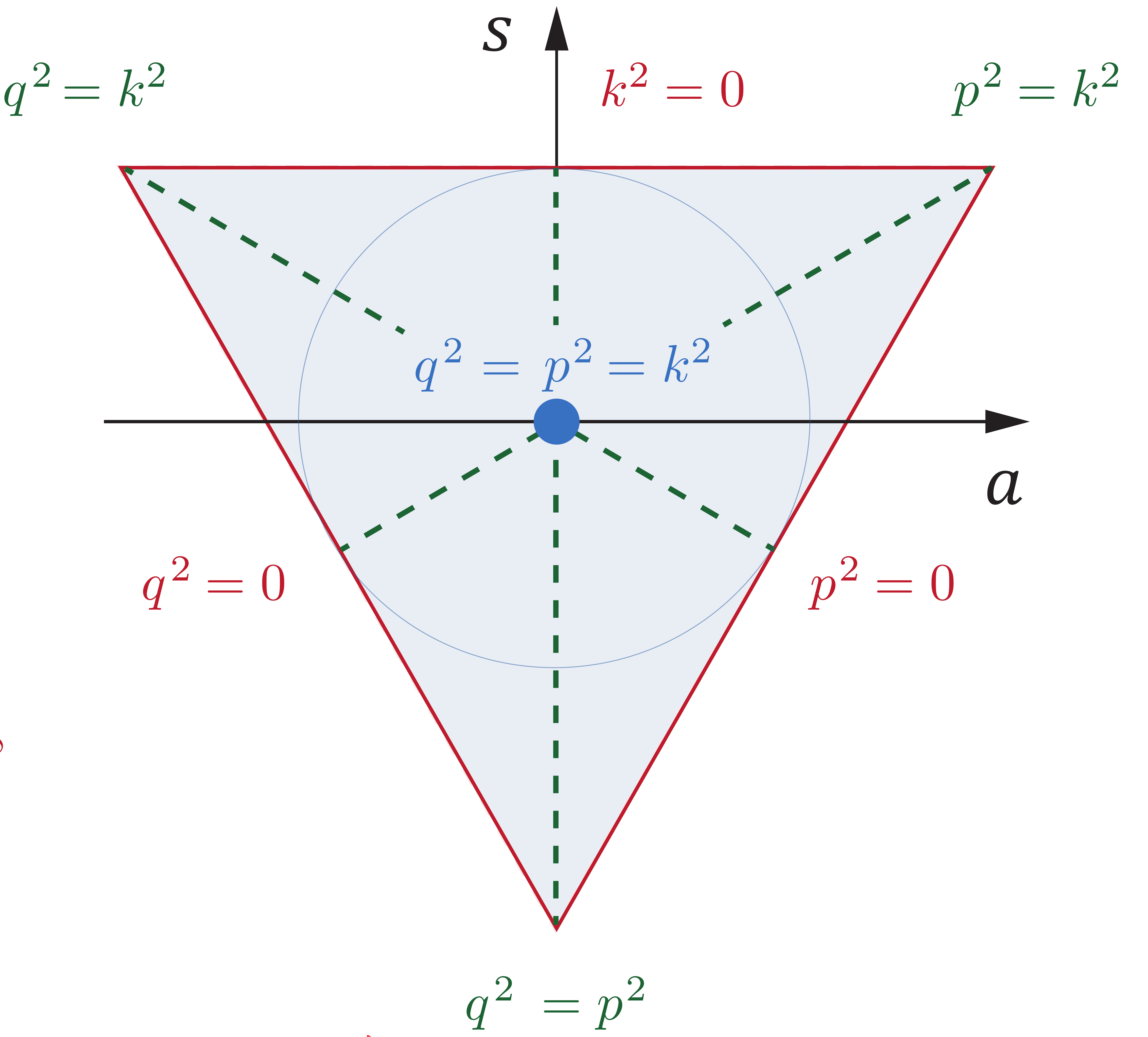}
                    \caption{ Doublet phase space for the photon four-point function in the $(a,s)$ plane at a slice of fixed $\mathcal{S}$.
                                            }\label{fig:phasespace}
                    \end{center}
        \end{figure}

        The doublet phase space is the Mandelstam plane spanned by the variables $a$ and $s$ defined in Eq.~\eqref{doublet-def}. It encodes the relations between
        $p^2$, $q^2$ and $k^2$ which are illustrated in Fig.~\ref{fig:phasespace}.
        The spacelike region defined by Eq.~\eqref{spacelike} forms an equilateral triangle of side length $2\sqrt{3}$, enclosed by the lines
        $p^2=0$, $q^2=0$ and $k^2=0$.
        The three corners
        \begin{equation} \renewcommand{\arraystretch}{1.0}
            \left[\begin{array}{c} a \\ s \end{array}\right] \; = \;
            \left[\begin{array}{c} 0 \\ -2 \end{array}\right], \quad
            \left[\begin{array}{c} \sqrt{3} \\ 1 \end{array}\right], \quad
            \left[\begin{array}{c} -\sqrt{3} \\ 1 \end{array}\right]
        \end{equation}
        correspond to the limits $q^2=p^2=0$, $p^2=k^2=0$ and $q^2=k^2=0$, respectively.
        Since the three momenta are independent and the variables $p^2$, $q^2$ and $k^2$ can take any value $\in \mathds{R}_+$,
        all points within the triangle contribute to the spacelike region.\footnote{For a three-point function one obtains a similar doublet, namely a triangle in the variables $p_1^2$, $p_2^2$ and $p_3^2$.
        However, since in that case only two momenta are independent, the phase space is restricted to the interior of a unit circle embedded in the triangle (indicated in Fig.~\ref{fig:phasespace}), cf.~Ref.~\cite{Eichmann:2014xya}. \label{footnote-1}}
        The center of the triangle is the symmetric point where $p^2=q^2=k^2$.

        In practice it is convenient to parametrize the doublet by a radius and an angle:
        \renewcommand{\arraystretch}{1.0}
         \begin{equation}\label{doublet-parametrization}
             \mathcal{D}_0 = \mS_0\left[ \begin{array}{c} a \\ s \end{array}\right] =  \mS_0\,r \left[ \begin{array}{r} \sin\varphi \\ -\cos\varphi \end{array}\right],
         \end{equation}
         with $\varphi \in (0,2\pi)$, chosen such that $\varphi=0$ corresponds to the lower corner of the triangle.
         Note that the range or $r$ is a function of the angle $\varphi$ as spelled out below, Eq.~(\ref{rphi}).
        When expressed in $r$ and $\varphi$, the Mandelstam variables are cyclically related by a rotation $\varphi \rightarrow \varphi \pm 2\pi/3$:
        \begin{equation}
        \begin{split}
            p^2 &= \frac{4\mathcal{S}_0}{3}\left( 1+r\cos\big( \varphi+\tfrac{2\pi}{3}\big)\right), \\
            q^2 &= \frac{4\mathcal{S}_0}{3}\left( 1+r\cos\big( \varphi-\tfrac{2\pi}{3}\big)\right), \\
            k^2 &= \frac{4\mathcal{S}_0}{3}\left( 1+r\cos\varphi\right). \\
        \end{split}
        \end{equation}
        The sum of the cosines vanishes, so that $p^2+q^2+k^2=4\mathcal{S}_0$ from Eq.~\eqref{S0-def} is satisfied.
         The permutation-group invariants in Eq.~\eqref{perm-group-invariants-LI} become
         \begin{equation}
             \mathcal{S}_1 = \mS_0^2 \,r^2\,, \qquad
             \mathcal{S}_2= \mS_0^3 \,r^3 \cos 3\varphi\,.
         \end{equation}

        The transformation of the doublet under any permutation of momenta can be reconstructed from the Cayley diagram in Fig.~\ref{fig:cayley} and the
        representation matrices $\mathsf{M}_{12}$, $\mathsf{M}_{1234}$ in Eq.~\eqref{doublet-rep-matrices}. Permutations act on the angle $\varphi$ only:
        $P_{12}$ induces a reflection $\varphi\rightarrow-\varphi$ and $P_{1234}$ a  rotation and reflection $\varphi \rightarrow -\varphi - 2\pi/3$.
        Therefore, permutations within the $t$, $s$ or $u$ channels reflect the triangle
        along any of the three axes (dashed lines) in Fig.~\ref{fig:phasespace}; permutations between the channels rotate it by an angle $\pm 2\pi/3$.
        The maximum value of the radius $r$ (the boundary of the triangle) defines the boundary of the spacelike region:
        \begin{equation}
            r_\text{max}(\varphi) = \frac{1}{\sin\left[\varphi + \tfrac{\pi}{6}\,\big(1-4\,\lfloor\tfrac{3\varphi}{2\pi}\rfloor\big)\right]}\,.\label{rphi}
        \end{equation}
        This is again a permutation-group invariant function with $1 \leq r_\text{max}(\varphi) \leq 2$.
        The variable $\hat{r}=r/r_\text{max}(\varphi)$ takes values in the interval $[0,1]$ only.

         The Mandelstam triangle is convenient for discussing two-photon intermediate states that appear in the LbL amplitude.
         While the space-like region is free of singularities, it is still influenced by the time-like singularity structure.
         For example, it is well known that the pion poles in the two-photon channel
         produce a large (and presumably even the largest) QCD contribution to the LbL amplitude~\cite{Jegerlehner:2009ry}.
         Bose symmetry entails that two photons can only produce meson quantum numbers with even charge-conjugation parity:
         $J^{PC}=0^{-+}$ (pseudoscalar), $0^{++}$ (scalar), $1^{++}$ (axialvector), $1^{-+}$ (exotic vector), $2^{++}$ (tensor), etc.
         Such poles will occur at $p^2=-m^2$, $q^2=-m^2$ or $k^2=-m^2$, where $m$ is the respective pole mass (which is complex if the state carries a width).
        In terms of the radius this translates to
        \begin{equation}\label{timelike-poles}
            \hat{r} = 1 + \frac{3m^2}{4\mathcal{S}_0} > 1\,.
        \end{equation}
        Hence, each real pole constitutes another triangle that encompasses the one in Fig.~\ref{fig:phasespace}. They
        should have an impact on the spacelike region and induce a rise in the dressing functions towards the boundary $\hat{r}=1$.
        Because $\mS_0$ appears in the denominator of Eq.~\eqref{timelike-poles}, the sensitivity to timelike physics will be generally stronger at \textit{larger} values of $\mS_0$,
        i.e., in the UV region, because the poles approach the boundary of the triangle, whereas for \mbox{$\mS_0\rightarrow 0$} they move infinitely far away.
        A similar behavior is known from the three-gluon vertex, where the angular dependence is sizeable in the UV but negligible in the infrared~\cite{Eichmann:2014xya}.

\subsection{Triplet} \label{sec:phase-space-triplet}

         Whereas the Mandelstam triangle encodes the two-photon intermediate states, the triplet variables $\{u,v,w\}$ are
         related to the photon virtualities $x_i=p_i^2$, $i=1\dots 4$, and thereby encode the vector-meson poles.
         To see this, one derives the following relations between the $x_i$, $\mathcal{S}_0$ and $\omega_i$ from Eq.~\eqref{momenta-inverse}:
         \begin{equation}
         \begin{split}
             x_1 &= \mathcal{S}_0 + \tfrac{1}{2}\,( \omega_1-\omega_2-\omega_3)\,, \\
             x_2 &= \mathcal{S}_0 + \tfrac{1}{2}\,(-\omega_1+\omega_2-\omega_3)\,, \\
             x_3 &= \mathcal{S}_0 + \tfrac{1}{2}\,(-\omega_1-\omega_2+\omega_3)\,, \\
             x_4 &= \mathcal{S}_0 + \tfrac{1}{2}\,( \omega_1+\omega_2+\omega_3)\,.
         \end{split}
         \end{equation}
        In terms of the $x_i$ the singlet takes the form
         \begin{equation}
            \mathcal{S}_0=\frac{x_1+x_2+x_3+x_4}{4}
         \end{equation}
         and the triplet is given by
        \begin{equation}\label{LI-doublet-triplet-2}
           \mathcal{T}_0 = \frac{1}{4}\left[ \begin{array}{c} x_1+x_2+x_3-3x_4 \\ -\sqrt{2}\,(x_1+x_2-2x_3) \\ \sqrt{6}\,(x_1-x_2) \end{array}\right].
        \end{equation}

          \begin{figure}[t]
                    \begin{center}
                    \includegraphics[scale=0.25]{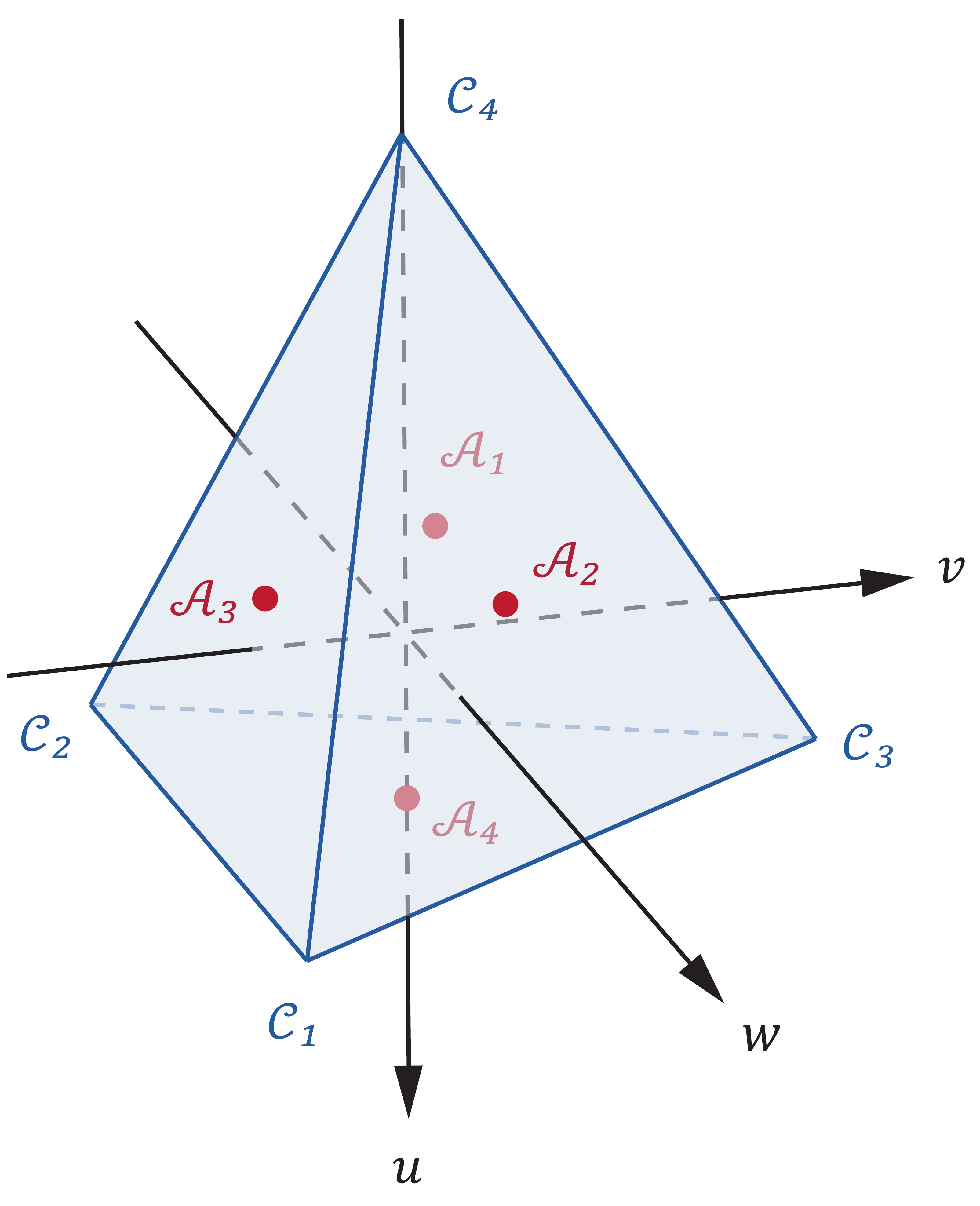}
                    \caption{ Triplet phase space for the photon four-point function in the $(u,v,w)$ space.
                                            }\label{fig:phasespace-tetrahedron}
                    \end{center}
        \end{figure}

        If we allow the $x_i$ to take any spacelike value $\in \mathds{R}_+$,
        the resulting phase space forms a tetrahedron which is shown in Fig.~\ref{fig:phasespace-tetrahedron}. Its four corners $\mathcal{T}_0=\mS_0\,\mathcal{C}_i$, $i=1\dots 4$,
        are defined by the kinematic limits where $x_i = 4\mathcal{S}_0$ and the other three virtualities $x_{j\neq i}$ vanish:
        \renewcommand{\arraystretch}{1.0}
        \begin{equation}
            \mathcal{C}_i = \left[\begin{array}{c} 1 \\ -\sqrt{2} \\ \sqrt{6} \end{array}\right], \quad
                            \left[\begin{array}{c} 1 \\ -\sqrt{2} \\ -\sqrt{6} \end{array}\right], \quad
                            \left[\begin{array}{c} 1 \\ 2\sqrt{2} \\ 0 \end{array}\right], \quad
                            \left[\begin{array}{c} -3 \\ 0 \\ 0 \end{array}\right].
        \end{equation}
        The soft kinematic limits where only one of the photons is real ($x_i=0$) define the four faces of the tetrahedron;
        i.e., they are the planes spanned by the points $\mathcal{C}_{j\neq i}$. These are the points which are relevant
        for $(g-2)_\mu$, since the external source of the magnetic field is represented by an on-shell photon \cite{Jegerlehner:2009ry}.
        The center points of the four faces $\mathcal{A}_i = -\mathcal{C}_i/3$ lie on the opposite sides
        of the corners $\mathcal{C}_i$. The center of the tetrahedron is the symmetric point where all $\omega_i$ vanish and consequently all $x_i=\mathcal{S}_0$ are the same.

          \begin{figure*}[t]
                    \begin{center}
                    \includegraphics[scale=0.27]{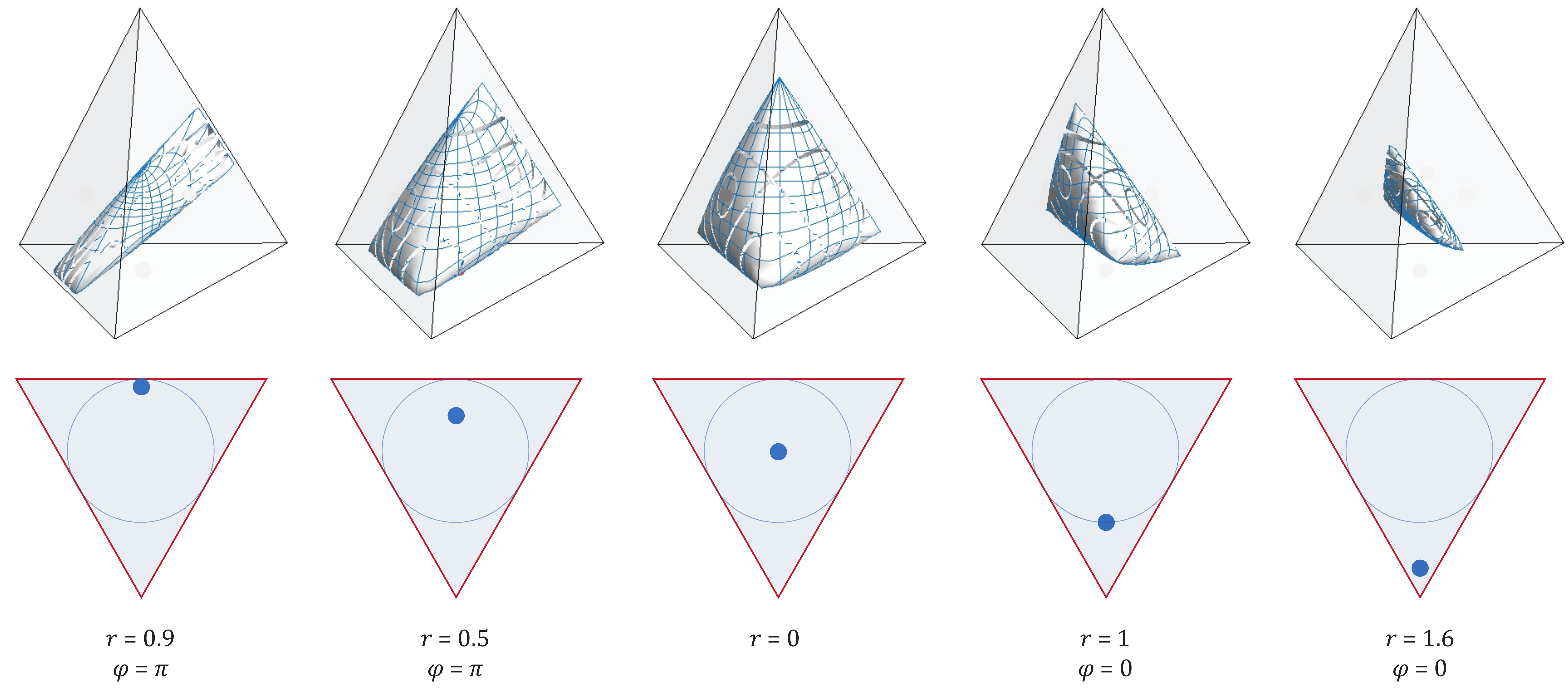}
                    \caption{ Different kinematic points in the doublet triangle (bottom row) and the corresponding spacelike region within the triplet tetrahedron (top row).
                              The alignments are the same as in Figs.~\ref{fig:phasespace} and~\ref{fig:phasespace-tetrahedron}. A rotation $\varphi\rightarrow\varphi \pm \tfrac{2\pi}{3}$
                              will also rotate the tetrahedron.
                              The figure in the center ($r=0$) contains both the central limit ($R=0$) and the limit of three equal photon momenta ($R=2$, the corners of the spacelike volume).
                                            }\label{fig:phasespace-triplet}
                    \end{center}
        \end{figure*}

        Similarly to Eq.~\eqref{doublet-parametrization}, we parametrize the triplet in spherical coordinates,
        \renewcommand{\arraystretch}{1.0}
         \begin{equation}\label{triplet-parametrization}
             \mT_0 = \mS_0 \left[ \begin{array}{c} u \\ v \\ w \end{array}\right] = \mS_0\,R\left[ \begin{array}{c}  -\cos\theta \\ \quad\sin\theta\,\cos\phi \\ -\sin\theta\,\sin\phi \end{array}\right], \quad
             \renewcommand{\arraystretch}{1.2}
             \begin{array}{rl} \theta & \in(0,\pi)\,, \\ \phi &\in(0,2\pi)\,.\end{array}
         \end{equation}
        For the $\omega_i$ this choice entails
        \begin{equation}
        \begin{split}
            \omega_1 &= \frac{2\mathcal{S}_0}{3}\,R\left( \cos\theta+\sqrt{2}\,\sin\theta\,\cos\big( \phi+\tfrac{2\pi}{3}\big)\right), \\
            \omega_2 &= \frac{2\mathcal{S}_0}{3}\,R\left( \cos\theta+\sqrt{2}\,\sin\theta\,\cos\big( \phi-\tfrac{2\pi}{3}\big)\right), \\
            \omega_3 &= \frac{2\mathcal{S}_0}{3}\,R\left( \cos\theta+\sqrt{2}\,\sin\theta\,\cos\phi\right), \\
        \end{split}
        \end{equation}
        which are again cyclically related by a rotation $\phi \rightarrow \phi \pm 2\pi/3$.
         The remaining invariants from Eq.~\eqref{perm-group-invariants-LI} then take the form
         \begin{align}
            \mathcal{S}_3 &= \mS_0^2\,R^2\,, \\
            \mathcal{S}_4 &=\mS_0^3\,r R^2\,\big[ \sin^2\theta\,\cos(2\phi+\varphi)  +\sqrt{2}\,\sin 2\theta\,\cos(\phi-\varphi)\big], \nonumber \\
            \mathcal{S}_5 &= \mS_0^3\,R^3\,\big(\cos 3\theta+ \cos^3 \theta + \sqrt{2}\,\sin^3\theta\,\cos 3\phi\,\big). \nonumber
         \end{align}

        The representation matrix $\mathsf{H}_{12}$ for the permutation $P_{12}$ induces only a reflection $\phi\to-\phi$,
        whereas $\mathsf{H}_{1234}$ is more complicated to express in these variables because it exchanges the faces of the tetrahedron.
        Still, going to the $s$-channel ($P_{12}\, P_{23}$)
        or $u$-channel ($P_{23}\, P_{12}$) is simple because it only induces a rotation $\phi\rightarrow\phi \pm 2\pi/3$.
        This is quite helpful in practical calculations of the four-point function
        when the dressing functions are expressed in the variables $\mathcal{S}_0$, $r$, $\varphi$, $R$, $\theta$, $\phi$.
        Once the sum of all $t-$channel diagrams has been obtained,
        the remaining permutations in Eq.~\eqref{3-permutations}  amount to nothing more than a simultaneous rotation of the angles $\varphi$ and $\phi$ by $\pm 2\pi/3$ each.

        A complication comes from the fact that not the whole interior of the tetrahedron contributes to the spacelike region.
        Only three of the momenta $p_i$ are independent, which leads to a restriction on the phase space.\footnote{This
        is similar to the restriction for the doublet phase space in a three-point function, see footnote~\ref{footnote-1}.}
        For given doublet variables $r$ and $\varphi$, the actual triplet domain is a complicated geometric object contained within the tetrahedron.
        It is visualized in Fig.~\ref{fig:phasespace-triplet} for five different kinematic configurations.
        Its surface is defined by the permutation-group invariant condition\footnote{Expressed by the variables in Eq.~\eqref{phasespace-hyperspherical-co} this corresponds to either $y^2=1$, $z^2=1$ or ${z'}^2=1$.}
        \begin{equation}\label{boundary-1}
            p^2 q^2 k^2 = \omega_1^2 \,p^2 + \omega_2^2 \,q^2 + \omega_3^2 \,k^2 -2\,\omega_1\,\omega_2\,\omega_3\,,
        \end{equation}
        which leads to a cubic equation for the radius $R$. Its solution is:
        \begin{equation}\label{Rmax}
        \begin{split}
            R_\text{max}(r,\varphi,\theta,\phi) = \frac{B}{A}\left[ 1+2\,\text{sgn}(X)\,\cos \frac{\Phi}{3}  \right], \\
            \Phi = \arccos|X| + 2\pi\,\big[ \Theta(X)\,\Theta(-A) - \Theta(A) \big].
        \end{split}
        \end{equation}
        Here, $\Theta$ is the unit step function and the remaining auxiliary quantities are the invariants
        \begin{equation}
            A = \frac{\mathcal{S}_5}{2\mS_0^3\,R^3}\,, \qquad
            B = 1+\frac{\mathcal{S}_4}{2\mS_0^3\,R^2}
        \end{equation}
        with domains $A \in [-1,1]$, $B \in[0,3]$ and
        \begin{equation}
            X = 1-\frac{2A^2 \,D}{B^3}\,,  \qquad
            D = 1-\frac{3\mS_0\,\mathcal{S}_1-\mathcal{S}_2}{4\mS_0^3}
        \end{equation}
        with domain 
        $D \in [0,1]$.
        Consequently, $R_\text{max}$ is also invariant under permutations.
        It is restricted to the interval $0 \leq R_\text{max} \leq 2$ and never reaches the corners of the tetrahedron with radius $R=3$.
        In other words, three photon virtualities can never vanish simultaneously in the domain that is integrated over in the $g-2$ calculation.
        The variable $\hat{R}=R/R_\text{max}(r,\varphi,\theta,\phi)$ is again permutation-group invariant
        and its domain is the interval $[0,1]$.

        Since the triplet phase space encodes the relations between the photon virtualities,
        additional momentum dependencies will come from
        vector-meson poles at timelike virtualities $x_i = -m_\rho^2$. Microscopically, each photon in the hadronic part of the LbL amplitude
        couples to a quark via a dressed quark-photon vertex, which automatically contains all vector-meson poles.
        The sequence of vector-meson poles corresponds to a sequence of tetrahedra
        that encompass the one in Fig.~\ref{fig:phasespace-tetrahedron}. In analogy to the doublet, the dressing functions
        in the spacelike domain should therefore rise in magnitude towards the spacelike boundary.
        Fig.~\ref{fig:phasespace-triplet} shows that the sensitivity to vector-meson poles should be weakest in the corners of the doublet triangle
        (where the triplet volume shrinks to the point $R=0$, rightmost figure) and strongest at the centers of its sides
        (where the triplet extends to the edges of the tetrahedron at $R=2$, leftmost figure).
        Generally, while the spacelike interior of the triangle and the tetrahedron is free of singularities,
        one- and two-photon singularities will therefore influence the behavior of the dressing functions from the timelike domain.

         \subsection{Special momentum configurations}

         The main practical goal of the permutation-group analysis was to facilitate the discussion of the phase space.
         Once a symmetric tensor basis is employed,  the dressing functions can only depend on the symmetric combinations in Eq.~\eqref{perm-group-invariants-LI}.
         If the basis is minimal, it fully absorbs the kinematic part of the momentum dependence.
         Hence, the momentum evolution of the form factors should be mainly governed by the scaling variable $\mS_0$ with lowest mass dimension: $f_i(\mS_0 \dots \mS_5) \approx f_i(\mS_0)$, and therefore become simple.
         However, angular dependencies induced by timelike singularities will become relevant towards the boundaries of the triangle and tetrahedron.

         We conclude this section by discussing some special momentum configurations,
         some of which have been also studied in the context of the four-gluon vertex~\cite{Binosi:2014kka}.
         Expressed in our permutation-group variables, they correspond to the following limits:
         \medskip

         (i) \, \textbf{Uniform soft limit}  ($\mathcal{S}_0=0$): all momenta vanish simultaneously, $p=q=k=0$, and consequently also all Lorentz invariants are zero.

\begin{figure}[t]
         \includegraphics[scale=0.14]{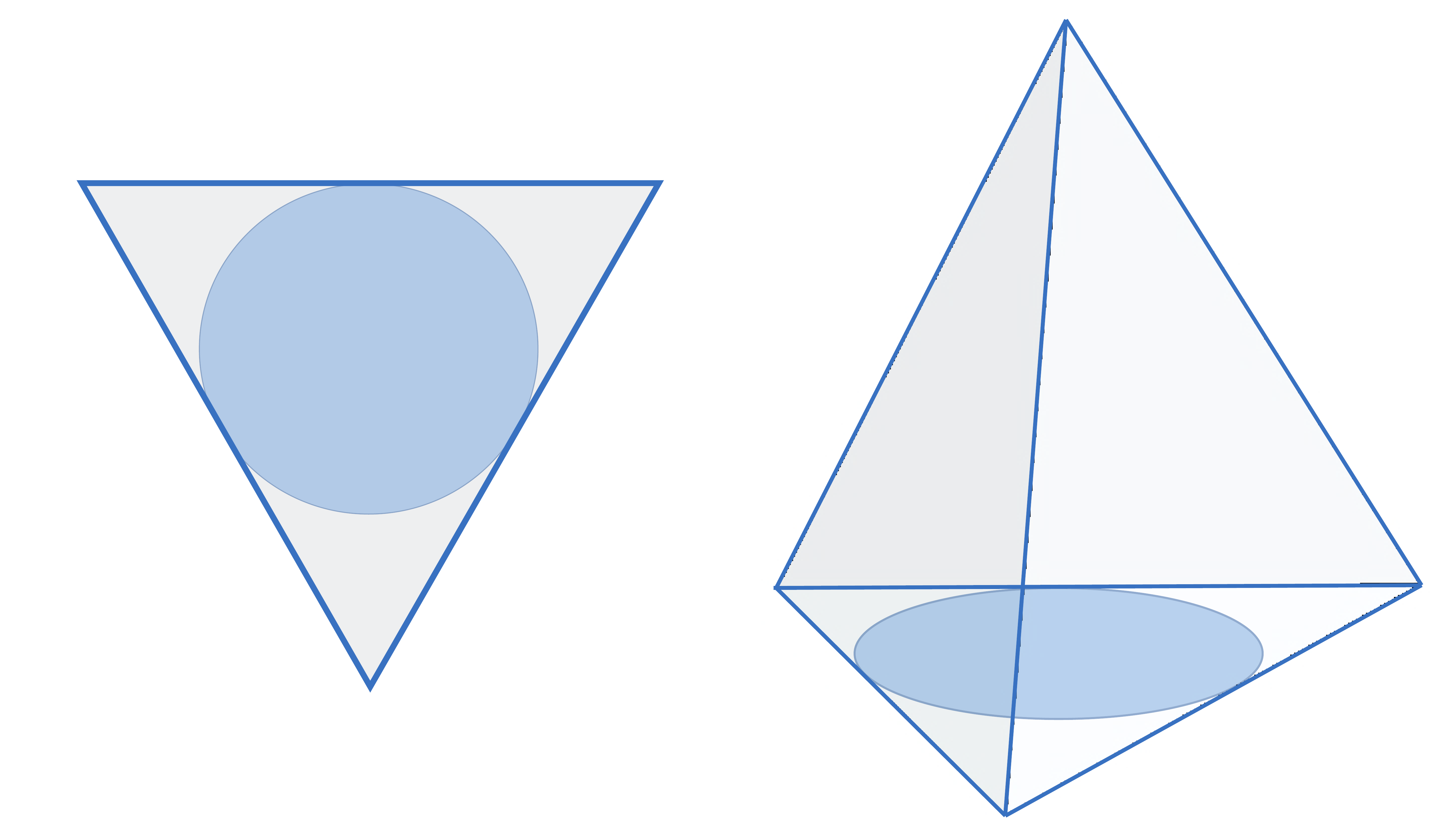}
         \caption{ Phase space that is relevant for $g-2$.}\label{fig:phasespace2}
       \end{figure}

         \medskip
         (ii) \, The \textbf{soft-photon limit} is the relevant limit for the $g-2$ problem:
                one external momentum vanishes, e.g. $p_4=0$ and therefore $p + q + k = 0$.
               The variables $p^2$, $q^2$ and $k^2$ are still independent but
               \begin{equation}
               \begin{split}
                  \omega_1 &= \tfrac{1}{2}\,(p^2-q^2-k^2), \\
                  \omega_2 &= -\tfrac{1}{2}\,(p^2-q^2+k^2),  \\
                  \omega_3 &= -\tfrac{1}{2}\,(p^2+q^2-k^2)\,.
               \end{split}
               \end{equation}
               In terms of doublet and triplet variables,
               $\mS_0$, $a$ and $s$ remain independent
               whereas
               \begin{equation*}
                  u=1, \quad v=-\sqrt{2}s, \quad w=-\sqrt{2}a\,.
               \end{equation*}
               This is the lower face of the tetrahedron, whose remaining variables $v$ and $w$
               are now proportional to those in the Mandelstam triangle $s$ and $a$.
               The triplet variables from Eq.~\eqref{triplet-parametrization} become
               \begin{equation*}
                  R^2 = 1+2r^2, \quad \cos\theta=-\frac{1}{\sqrt{1+2r^2}}\,, \quad \phi=\varphi\,.
               \end{equation*}
               Since $R=R_\text{max}$ and therefore $\hat{R}=1$, the combination with the upper limit for $R_\text{max}$ from Eq.~\eqref{Rmax} restricts the Mandelstam triangle
               to the unit circle: \mbox{$r\leq 1$}. The domain is visualized in Fig.~\ref{fig:phasespace2}.

         \medskip
         (iii) \, \textbf{Central limit:} the Mandelstam momenta have the same length and are orthogonal to each other:
               \begin{equation}
                   p^2 = q^2 = k^2 = \tfrac{4}{3}\,\mathcal{S}_0, \qquad
                   \omega_1=\omega_2=\omega_3=0\,.
               \end{equation}
               Therefore only $\mS_0\neq 0$, whereas $a=s=u=v=w=0$.
               This is the center of the triangle and the tetrahedron.
               The doublet and triplet radii vanish, $r=R=0$, and all photon virtualities are equal: $x_i = \mathcal{S}_0$.

         \medskip
         (iv) \, \textbf{Three equal momenta:} $p_1=p_2=p_3=-p_4/3$,
                hence all Mandelstam momenta are identical ($p=q=k$) as well as the Lorentz-invariants:
               \begin{equation}
                   p^2 = q^2 = k^2 = \omega_1 = \omega_2 = \omega_3 = \tfrac{4}{3}\,\mathcal{S}_0\,.
               \end{equation}
               It entails $a=s=v=w=0$ but $u=-2$;
               the photon virtualities are $ x_1=x_2=x_3=\tfrac{1}{3}\,\mathcal{S}_0$ and $x_4 = 3\mathcal{S}_0$.

         \medskip
         (v) \, \textbf{Two equal momentum pairs:} for example, $p_1=p_2=-p_3=-p_4$.
               As a consequence, $p=q=0$ and therefore $a=u=v=w=0$ and $s=-2$,
               which is the limiting case for the right panel in Fig.~\ref{fig:phasespace-triplet}.

         \medskip
         (vi) \, \textbf{Two opposite momenta:} e.g. $p_1=-p_2$ and therefore $p_3=-p_4$, so
               one of the Mandelstam momenta vanishes ($k=0$).
               There are now three independent variables: $p^2$, $q^2$ and $\omega_3=p\cdot q$, with $k^2=\omega_1=\omega_2=0$.
               Equivalently, $s=1$, $w=0$ and $v=-\sqrt{2}u$, which is the limiting case for the left panel in Fig.~\ref{fig:phasespace-triplet}.

         \medskip
         (vii) \, \textbf{Two equal momenta:} e.g. $p_1=p_2$ and therefore also $p=q$.
               Also here three independent variables remain: $k^2$, $p^2=q^2=\omega_3$,  and $p\cdot k=\omega_1=\omega_2$,
               which corresponds to $a=w=0$, $v=(u+s+2)/\sqrt{2}$.

\section{Tensor basis}\label{sec:tensorbasis}

        Having discussed the phase space, we will now proceed with the tensor basis construction for the photon four-point function.
        It should respect the symmetry properties of the permutation group $S_4$ and the transversality and analyticity constraints.
      Gauge invariance entails that the LbL amplitude is transverse, and analyticity implies that it must be at least quartic in the photon momenta.
      The form factors in such a basis should be free of kinematic singularities but also free of kinematic dependencies and kinematic zeros.
      If the tensor basis is made of permutation-group singlets,
      the form factors must be symmetric and can only depend on the singlet variables in Eq.~\eqref{perm-group-invariants-LI},
      so that their momentum dependence should become simple.

        In Sec.~\ref{sec:2photon} we worked out the simpler case of a scalar two-photon current which will serve as a template in what follows.
        However, we will encounter additional subtleties that do not appear in the two-photon example. In this respect, a closer analogue
        is nucleon Compton scattering, whose tensor basis was worked out by Bardeen and Tung~\cite{Bardeen:1969aw} and Tarrach~\cite{Tarrach:1975tu}.
        In that case the procedure is as follows:
        (i) Write down all possible 34 tensor structures that are compatible with Lorentz covariance.
            It turns out that two of them are redundant and can be eliminated, thus leaving a 32-dimensional type-I basis.
        (ii) Apply gauge invariance to arrive at a transverse basis with 18 elements.
            Tarrach noted that the resulting basis (Eq.~(12) in Ref.~\cite{Tarrach:1975tu}) was not minimal, i.e., the basis elements were still kinematically dependent,
            and suggested to replace three tensors by alternative ones in a certain kinematic limit (Eq.~(13) therein).
            It was later realized by the Mainz group~\cite{Drechsel:1997xv} that this would not have been necessary had one started directly
            with the 18-dimensional basis made from Tarrach's three alternative tensors, because that basis is minimal.
            The reason behind this becomes only clear when the basis elements are cast
            into permutation-group \textit{singlets}, because the corresponding three new tensors have lower powers in the photon momenta than the original ones~\cite{new:GE+GR}.\footnote{What
            complicates the power counting in Compton scattering is that
         the photon momentum powers are no longer equal to the mass dimension:
         the amplitude also depends on the average nucleon momentum, and the contraction with nucleon spinors can additionally raise these powers due to Gordon identities.}

      We will make similar observations also for the photon four-point function. In Sec.~\ref{sec:basis-counting} we construct a generic \mbox{type-I} basis with 138 elements, but
      it will turn out that two of them are redundant and can be eliminated. In Sec.~\ref{sec:basis-gaugeinv-II}, after applying the constraints from gauge invariance, we will arrive at 41 transverse tensor elements.

      The permutation-group technique will be an essential ingredient in determining the minimality of such bases. The criterion is that
      the \textit{singlet} basis elements should depend on the lowest possible powers of photon momenta,
      which translates into the lowest mass dimensions.
      In the Compton scattering example, invariance under charge conjugation and photon crossing
      leads to an $S_2\times S_2$ symmetry, and constructing a singlet basis is simple
      because it is sufficient to multiply the \mbox{(anti-)}symmetrized basis elements with appropriate momentum prefactors.
      For the photon-four point function the situation is more difficult because of the more complicated structure of the group $S_4$:
      one has to cast the tensor basis elements into irreducible representations of $S_4$, combine them into singlets, and ensure that the resulting basis elements have the lowest possible mass dimension.

      In any case, we proceed under the assumption that it \textit{should} be possible to find a 41-dimensional minimal basis
      without the need for swapping basis elements in different kinematic limits. Since the problem is quite formidable we have not yet succeeded in this goal,
      but we will describe the necessary steps in the following.

\subsection{Type-I basis} \label{sec:basis-counting}

        To begin with, we prove that the LbL amplitude (and also the four-gluon vertex) has indeed 136 independent Lorentz tensors and not 138.
        The system depends on three independent momenta $q_i$ ($i=1,2,3$);
        for example, we can work with the three Mandelstam momenta in Eq.~\eqref{Mandelstam-momenta}.
        A naive counting of all possible combinations of Kronecker deltas and four-momenta yields 138 elements:
        \begin{itemize}
        \item $\delta^{\mu\nu} \delta^{\rho\sigma}$ $\Rightarrow$ 3 permutations $\Rightarrow$ 3 elements,
        \item $\delta^{\mu\nu}\,q_i^\rho \, q_j^\sigma$ $\Rightarrow$ $3^2 \times 6 \,\text{permutations}= 54$ elements,
        \item $q_i^\mu \, q_j^\nu \, q_k^\rho \, q_l^\sigma$ $\Rightarrow$ $3^4 = 81$ elements.
        \end{itemize}
        The list is, however, redundant: for four- and higher $n$-point amplitudes
        the dimensionality of spacetime restricts the number of independent basis elements.
        This can be understood from a simple argument.
        Suppose we orthogonalize the three independent momenta to obtain three unit vectors $n_i$, $i=1,2,3$ that are transverse to each other.
        From three vectors one can construct an axialvector $v^\mu = \varepsilon^{\mu\alpha\beta\gamma}\,n_1^\alpha n_2^\beta n_3^\gamma$
        with opposite parity.
        A complete, orthonormal tensor basis follows from collecting
        all possible combinations of
        \begin{alignat}{6}\label{vrsd-basis}
        &\bullet \;\; && v^\mu\,v^\nu\,v^\rho\,v^\sigma \quad               && \Rightarrow \quad &&  1 \;\text{element}\,,  \nonumber \\
        &\bullet \;\; && v^\mu\,v^\nu\,n_i^\rho \, n_j^\sigma \quad         && \Rightarrow \quad &&  3^2 \times 6 = 54 \;\text{elements}   \\
        &\bullet \;\; && n_i^\mu \, n_j^\nu \, n_k^\rho \, n_l^\sigma \quad && \Rightarrow \quad &&  3^4 = 81 \; \text{elements},  \nonumber
        \end{alignat}
        where $v$ must appear in pairs to ensure the correct parity.
        This yields $136$ tensor structures instead of $138$.
      The list is already complete since the Kronecker delta can be written as the linear combination
      \begin{equation}\label{dd-vv}
         \delta^{\mu\nu} = v^\mu\,v^\nu + \sum_{i=1}^3 n_i^\mu\,n_i^\nu\,.
      \end{equation}
      In a frame where $v$, $n_1$, $n_2$ and $n_3$ are the unit vectors in $\mathds{R}_4$ this is obviously true;
      since the equation is Lorentz-covariant it holds in any frame. Therefore, the element $\delta^{\mu\nu} \delta^{\rho\sigma}$ and its permutations
      are already contained in the basis~\eqref{vrsd-basis} and do not generate additional structures.

        We emphasize that this reduction is a consequence of dimensionality and has nothing to do with gauge invariance (which we will exploit in Sec.~\ref{sec:basis-gaugeinv-II})
        or Bose symmetry (which does not reduce the number of independent tensors).
        One can repeat the exercise for higher $n-$point functions: also in that case at most three of the $(n-1)$ independent momenta
        can appear in their bases, which greatly reduces their dimensions.

      The same argument also gives the correct number of transverse elements.
      When applying transverse projectors, no elements with
      the same label and index survive: $p_1^\mu$ is longitudinal, and so are $p_2^\nu$, $p_3^\rho$ and $p_4^\sigma$.
      At least for counting purposes,
      removing these four objects from the basis is formally equivalent to crossing off any of the vectors $n_i$ from the list above, so that $i=1,2$ only.
      This leads to $1 + 2^2\times 6 + 2^4 = 41$ transverse elements.

      We note that the basis in Eq.~\eqref{vrsd-basis}
      is already orthonormal because all the vectors and axialvectors it contains are normalized and orthogonal to each other.
      Hence, it defines the simplest possible complete tensor basis for a vector four-point function.
      On the other hand, neither Bose symmetry nor gauge invariance or analyticity are implemented at this point.
      In the aforementioned Lorentz frame all basis elements are non-singular because they consist of pure numbers,
      but the corresponding (Lorentz-invariant) dressing functions will exhibit kinematic dependencies and zeros.

     Therefore, we should first construct the analogue of Eq.~\eqref{scalar-current-1}: a 136-dimensional type-I tensor basis that is free of kinematic singularities and made of permutation-group singlets.
     It will serve as our starting point for working out the transversality constraints.
     The simplest construction principle is to use the Mandelstam momenta $p$, $q$ and $k$ together with the permutation technique from Sec.~\ref{sec:multiplets}, i.e.,
     to write down the maximum number of seed elements and work out their permutations.
     The outcome of this procedure is collected in Table~\ref{tab:generic-basis}. For example, the permutations
     of the seed element $\delta^{\mu\nu} \delta^{\rho\sigma}$ are
      \begin{equation}
         \delta^{\mu\nu} \delta^{\rho\sigma}, \qquad
         \delta^{\nu\rho} \delta^{\mu\sigma}, \qquad
         \delta^{\rho\mu} \delta^{\nu\sigma}.
      \end{equation}
      With Eqs.~(\ref{perm-001}--\ref{perm-triplets}) they can be arranged into a singlet $\mathcal{S}$ and a doublet $\mathcal{D}_1$
      whose structure is analogous to those in Eq.~\eqref{color-dd}; all other multiplets vanish.
      Likewise, the seed $\delta^{\mu\nu} \,k^\rho k^\sigma$ generates the six permutations
      \begin{equation} \renewcommand{\arraystretch}{1.2}
         \begin{array}{l}
         \delta^{\mu\nu}\,k^\rho k^\sigma, \\
         \delta^{\rho\sigma}\,k^\mu k^\nu,
         \end{array}\qquad
         \begin{array}{l}
         \delta^{\nu\rho}\,p^\mu p^\sigma, \\
         \delta^{\mu\sigma}\,p^\nu p^\rho,
         \end{array}\qquad
         \begin{array}{l}
         \delta^{\rho\mu}\,q^\nu q^\sigma, \\
         \delta^{\nu\sigma}\,q^\rho q^\mu
         \end{array}
      \end{equation}
      which produce a singlet, a doublet $\mathcal{D}_1$ and a triplet $\mathcal{T}_1^+$.
      The remaining Lorentz tensors with one Kronecker delta and two identical momenta come from
      the seed $\delta^{\mu\nu} p^\rho p^\sigma$ (or equivalently $\delta^{\mu\nu} q^\rho q^\sigma$) which has 12 permutations:
      \begin{equation} \renewcommand{\arraystretch}{1.2}
         \begin{array}{l}
         \delta^{\mu\nu}\,p^\rho p^\sigma, \\
         \delta^{\rho\sigma}\,p^\mu p^\nu, \\[2mm]
         \delta^{\mu\nu}\,q^\rho q^\sigma, \\
         \delta^{\rho\sigma}\,q^\mu q^\nu,
         \end{array}\qquad
         \begin{array}{l}
         \delta^{\nu\rho}\,q^\mu q^\sigma, \\
         \delta^{\mu\sigma}\,q^\nu q^\rho, \\[2mm]
         \delta^{\nu\rho}\,k^\mu k^\sigma, \\
         \delta^{\mu\sigma}\,k^\nu k^\rho,
         \end{array}\qquad
         \begin{array}{l}
         \delta^{\rho\mu}\,k^\nu k^\sigma, \\
         \delta^{\nu\sigma}\,k^\rho k^\mu  \\[2mm]
         \delta^{\rho\mu}\,p^\nu p^\sigma, \\
         \delta^{\nu\sigma}\,p^\rho p^\mu.
         \end{array}
      \end{equation}
      One proceeds along these lines until the list is complete.
      The  resulting 136 Lorentz tensors in Table~\ref{tab:generic-basis} are arranged
      with increasing mass dimension $n$: there are 3, 54 and 79 elements for $n=0,2,4$, respectively.
      Because each seed produces only one symmetric singlet and there are 11 singlets in total,
      this is also the mini\-mum number of independent tensor elements in the four-point function: all 136 tensors
      can be reconstructed from those eleven through permutations.

    \renewcommand{\arraystretch}{1.5}

        \begin{table}[t]

                \begin{center}
                \begin{tabular}{  @{\;\;} l @{\;\;\;\;}     @{\;\;\;}l@{\;\;\;}   @{\;\;\;}l@{\;\;\;}   @{\;\;\;\;}l@{\;\;}      }

                           \hline\noalign{\smallskip}

                    $n$  & Seed         &  \#       & Multiplet type        \\   \noalign{\smallskip}\hline\hline\noalign{\smallskip}

                    $0$               &$\delta^{\mu\nu} \delta^{\rho\sigma}$             &  $3$     & $\mathcal{S}$, $\mathcal{D}_1$    \smallskip \\
                                      \noalign{\smallskip}\hline\noalign{\smallskip}

                    $2$               & $\delta^{\mu\nu}\,k^\rho\,k^\sigma$               &  $6$     & $\mathcal{S}$, $\mathcal{D}_1$, $\mathcal{T}_1^+$     \\
                                      & $\delta^{\mu\nu}\,p^\rho\,p^\sigma$               &  $12$    & $\mathcal{S}$, $\mathcal{D}_1$, $\mathcal{D}_2$, $\mathcal{T}_1^\pm$, $\mathcal{A}$     \\
                                      & $\delta^{\mu\nu}\,p^\rho\,q^\sigma$               &  $12$    & $\mathcal{S}$, $\mathcal{D}_1$, $\mathcal{T}_1^+$, $\mathcal{T}_2^\pm$     \\
                                      & $\delta^{\mu\nu}\,p^\rho\,k^\sigma$               &  $24$    & $\mathcal{S}$, $\mathcal{D}_1$, $\mathcal{D}_2$, $\mathcal{T}_1^\pm$, $\mathcal{T}_2^\pm$, $\mathcal{T}_3^\pm$, $\mathcal{A}$ \smallskip  \\
                                      \noalign{\smallskip}\hline\noalign{\smallskip}

                    $4$               & $p^\mu\,p^\nu\,p^\rho\,p^\sigma$             &  $3$     & $\mathcal{S}$, $\mathcal{D}_1$    \\
                                      & $p^\mu\,p^\nu\,q^\rho\,q^\sigma$             &  $6$    & $\mathcal{S}$, $\mathcal{D}_1$, $\mathcal{T}_1^-$   \\
                                      & $p^\mu\,p^\nu\,k^\rho\,k^\sigma$             &  $10$    & $\mathcal{S}$, ($\mathcal{D}_1$,) $\mathcal{D}_2$, $\mathcal{T}_1^\pm$, $\mathcal{A}$     \\
                                      & $p^\mu\,q^\nu\,k^\rho\,k^\sigma$             &  $12$    & $\mathcal{S}$, $\mathcal{D}_1$, $\mathcal{T}_1^+$, $\mathcal{T}_2^\pm$  \\
                                      & $p^\mu\,p^\nu\,p^\rho\,k^\sigma$             &  $24$    & $\mathcal{S}$, $\mathcal{D}_1$, $\mathcal{D}_2$, $\mathcal{T}_1^\pm$, $\mathcal{T}_2^\pm$, $\mathcal{T}_3^\pm$, $\mathcal{A}$     \\
                                      & $p^\mu\,p^\nu\,q^\rho\,k^\sigma$             &  $24$    & $\mathcal{S}$, $\mathcal{D}_1$, $\mathcal{D}_2$, $\mathcal{T}_1^\pm$, $\mathcal{T}_2^\pm$, $\mathcal{T}_3^\pm$, $\mathcal{A}$     \smallskip \\
                                      \noalign{\smallskip}\hline\hline\noalign{\smallskip}

                \end{tabular}
                \end{center}

               \caption{ 136-dimensional tensor basis for the vector four-point function, where gauge invariance is not yet implemented. The doublet in brackets
               is linearly dependent due to the spacetime restriction discussed in the text; its inclusion would lead to 138 instead of 136 tensor structures.}
               \label{tab:generic-basis}

        \end{table}

         \renewcommand{\arraystretch}{1.0}

      In principle, for $n=4$ there would be 81 independent elements but here the spacetime restriction discussed above
      comes into effect: the basis saturates with 136 elements and adding two more produces linear dependencies.
      Which ones to remove is not arbitrary because  unfortunate choices
      can produce kinematic singularities in the dressing functions already at this stage.
      If we label the eleven seed elements in Table~\eqref{tab:generic-basis} by $1 \dots 11$ from top to bottom,
      then one can show that the basis element $\mD_1(8)$ --- which is the one in brackets in the table ---
      is a linear combination of the multiplets
      \begin{equation} \renewcommand{\arraystretch}{1.2}
      \begin{split}
         &  \mS(2)-\mS(3), \; \mS(7)-\mS(8), \\
         &  \mD_1(1), \; \mD_1(2), \; \mD_1(3), \; \mD_2(3), \; \mD_1(7), \; \mD_2(8), \\
         &  \mT_1^+(4)-\mT_2^+(5), \; \mT_1^+(9) - \mT_3^+(11), \\
         &  \mT_2^-(5), \; \mT_3^-(11), \\
         &  \mA(3), \; \mA(8).
      \end{split}
      \end{equation}
      The coefficients are rather lengthy but they respect the doublet construction rules in Eq.~\eqref{perm-product-doublets}.
      It turns out that all coefficients share the denominator $\sim (r^2-16)$, where $r$ is the doublet radius defined in Eq.~\eqref{doublet-parametrization}.
      Hence, unless $r=4$ (which is never reached in practice because $r \leq 2$ in the spacelike domain) the element $\mD_1(8)$ is linearly dependent and can be removed.\footnote{This
      is not entirely satisfactory but sufficient for our present purposes.
      Ideally it should be possible to remove two elements in arbitrary kinematics, as it can be done for the Compton scattering amplitude~\cite{Tarrach:1975tu}.}

      In principle we still need to recast the tensor structures in Table~\ref{tab:generic-basis} into permutation-group singlets.
      Since the procedure is the same for the type-I basis and the transverse basis that we will derive next,
      we integrate the discussion into the following subsection.

\subsection{Transverse tensor basis} \label{sec:basis-gaugeinv-II}

       The remaining task is to work out the consequences of electromagnetic gauge invariance.
       We start from the expression~\eqref{4PA-decomposition-0} for the LbL amplitude,
         \begin{equation} \label{lbl-general-1}
             \mathcal{M}^{\mu\nu\rho\sigma}(p,q,k) = \sum_{i=1}^{136} f_i(\dots)\,\tau_i^{\mu\nu\rho\sigma}(p,q,k)\,,
         \end{equation}
       where the $\tau_i^{\mu\nu\rho\sigma}(p,q,k)$ are the basis elements from Table~\eqref{tab:generic-basis} or, alternatively,
       136 singlets constructed from them.
      The transversality conditions have the form
         \begin{equation}\label{transversality-00}
             p_1^\mu \, \mathcal{M}^{\mu\nu\rho\sigma} = 0, \quad \dots \quad
             p_4^\sigma \, \mathcal{M}^{\mu\nu\rho\sigma} = 0
         \end{equation}
         and reduce the basis to a subset of 41 transverse tensors.
         Transversality and analyticity require these tensors to be proportional to at least four powers in the photon momenta.
         Instead of Eq.~\eqref{transversality-00}, one can equivalently work out the Bose-symmetric condition
         \begin{equation}\label{transversality-1}
             T^{\mu\alpha}_{1}\,T^{\nu\beta}_{2}\,T^{\rho\gamma}_{3}\,T^{\sigma\delta}_{4}\,\mathcal{M}^{\alpha\beta\gamma\delta} \stackrel{!}{=} \mathcal{M}^{\mu\nu\rho\sigma}\,,
         \end{equation}
         where the $T_{i}^{\mu\nu}= \delta^{\mu\nu} - p_i^\mu \,p_i^\nu/p_i^2$ are the transverse projectors with respect to each photon momentum.
         This leads to relations between the dressing functions;
         if we denote the independent functions by $f_i$ and the dependent ones by $g_j$, they take the form
         \begin{equation}\label{95-eqs}
         \begin{split}
             g_1 &= g_1(f_1, \dots f_{41})\,, \\
                 &\vdots \\
             g_{95} &= g_{95}(f_1, \dots f_{41})\,.
         \end{split}
         \end{equation}
         They must be solved so that no kinematic singularities are introduced in the process, i.e., all $g_j$ remain regular.
         In analogy to Eq.~\eqref{scalar-current-transversality-relations} for the two-photon current example,
         one must choose the $g_j$ such that they carry no kinematic prefactors.
         In practice this is not always possible: there are equations where \textit{all} $g_j$ come with kinematic prefactors
         and one must divide by them, thereby introducing kinematic singularities.
         Since the $g_j$ are regular, some of them must vanish in these kinematic limits.
         Therefore, the division should be done such that only the \textit{minimal} number of $g_j$ picks up kinematical zeros.

         After reinserting Eqs.~\eqref{95-eqs} into the general expression~\eqref{lbl-general-1},
         the resulting amplitude will take the form
         \begin{equation}\label{amp-1}
             \mathcal{M}^{\mu\nu\rho\sigma} = \sum_{i=1}^{41} f_i\,\tau_{\perp i}^{\mu\nu\rho\sigma} + \sum_{j=1}^{95} g_j\,\tau_j^{\mu\nu\rho\sigma} ,
         \end{equation}
         which is the analogue of the two-photon current~\eqref{scalar-current-full}.
         The first term is the transverse part of the amplitude, with transverse tensors $\tau_{\perp i}^{\mu\nu\rho\sigma}$ that have mass dimension $4,6,8\dots$,
         and dressing functions $f_i$ that become constant in any kinematic limit.
         The second term constitutes the gauge part, which is neither longitudinal nor transverse.
         Here we have again added the $g_j$, which we eliminated in the first place; consequently, the gauge part
         must vanish if the amplitude is gauge invariant. In turn, if it does \textit{not} vanish gauge invariance must be violated ---
         either by a calculation that respects gauge invariance but is incomplete,
         or by an approach where gauge invariance is simply not built in.

         The fact that the $\tau_j$ remain with mass dimension $0, 2, 4, \dots$
         is also the reason why  violating gauge invariance can have severe consequences in practice.
         With another transverse projection of Eq.~\eqref{amp-1} everything collapses into the transverse part, in the same manner as in Eq.~\eqref{trans-proj-art-sing}.
         If the dressing functions $g_j$ are nonzero, they will introduce artificial singularities with momentum powers $-4, -2$, etc. into the $f_i$.
         In any case, the decomposition~\eqref{amp-1}
         provides a convenient \textit{filter} that allows one to quantify such gauge violations and, if possible,
         remove them to arrive at physically meaningful predictions.\footnote{We note that
         in the context of the LbL amplitude not even the constituent-quark loop is truly gauge invariant~\cite{Eichmann:2014ooa}.
         Instead, the gauge part is a constant, $(\delta^{\mu\nu}\,\delta^{\rho\sigma} + \delta^{\nu\rho}\,\delta^{\mu\sigma} + \delta^{\rho\mu}\,\delta^{\nu\sigma})/(24\pi^2)$,
         and drops out if the identity  $\mathcal{M}^{\mu\nu\rho\sigma} = -p_4^\lambda\,d\mathcal{M}^{\mu\nu\rho\lambda}/dp_4^\sigma$
         is employed as it is typically done in $g-2$ calculations.}

         While the procedure outlined here is at least in principle straightforward, it is almost impossible to perform by hand because of the sheer length of the expressions involved.
         Hence, we take the alternative route that we advertised in Sec.~\ref{sec:2photon}, which is the essence of Tarrach's procedure~\cite{Tarrach:1975tu}: construct tensors with lowest possible mass dimensions
         that are automatically free of kinematic singularities.
         The mass dimension must be even because the four-point function has positive parity.
         By working out the permutations of these tensors we can construct a linearly independent, complete basis made of 41 elements.

         To this end we employ
         \begin{equation}\label{t-and-epsilon}
         \begin{split}
             t^{\mu\nu}_{ij} &= p_i\cdot p_j\,\delta^{\mu\nu} - p_j^\mu \,p_i^\nu, \\
             \varepsilon^{\mu\nu}_{ij} &:= \varepsilon^{\mu\nu\alpha\beta}\,p_i^\alpha\,p_j^\beta
         \end{split}
         \end{equation}
         as the building blocks for the construction of such tensors~\cite{Eichmann:2012mp}.
         $t^{\mu\nu}_{ij}$ is transverse with respect to $p_i^\mu$ and $p_j^\nu$, and $\varepsilon^{\mu\nu}_{ij}$ is transverse to both momenta.
         The only two tensor structures with mass dimension four are then
      \begin{equation}\label{basis-seeds-tt-ee}
          \psi_1^{\mu\nu\rho\sigma} = t^{\mu\nu}_{12}\,t^{\rho\sigma}_{34} \quad \text{and} \quad \psi_2^{\mu\nu\rho\sigma} = \varepsilon^{\mu\nu}_{12}\,\varepsilon^{\rho\sigma}_{34}.
      \end{equation}
         They have a simple physical interpretation: $t^{\mu\nu}_{12}$ is the leading tensor
         of a scalar two-photon current with photon momenta $p_1$ and $p_2$ (which we now count as incoming), and $\varepsilon^{\mu\nu}_{12}$ is that of a pseudoscalar two-photon current (e.g. for the process $\pi\to\gamma\gamma$).
         Hence, if the LbL amplitude has scalar poles, they will appear in the form factor of $\psi_1$
         whereas the form factor of $\psi_2$ inherits the pion pole.

         Next, we employ these tensors as permutation-group seeds in analogy to the derivation of Table~\ref{tab:generic-basis}.
         We take $\psi_1$ and $\psi_2$ as seed elements and derive the multiplets according to Eqs.~(\ref{perm-multiplets}--\ref{perm-triplets}).
         It turns out that each of them generates a singlet $\mathcal{S}$ and a doublet of type $\mathcal{D}_1$;
         the other multiplets vanish.
       The only singlets of dimension $n=4$ are therefore the tensors
      \begin{equation}
      \begin{split}
          \mathcal{S}'(\psi_1) &= t^{\mu\nu}_{12}\,t^{\rho\sigma}_{34} + t^{\nu\rho}_{23}\,t^{\mu\sigma}_{14} + t^{\rho\mu}_{31}\,t^{\nu\sigma}_{24}, \\[1mm]
          \mathcal{S}'(\psi_2) &= \varepsilon^{\mu\nu}_{12}\,\varepsilon^{\rho\sigma}_{34} + \varepsilon^{\nu\rho}_{23}\,\varepsilon^{\mu\sigma}_{14} + \varepsilon^{\rho\mu}_{31}\,\varepsilon^{\nu\sigma}_{24}\,,
      \end{split}
      \end{equation}
      and their corresponding (fully symmetric) form factors should be expected to be the dominant ones.
        Here and in the following we denote the multiplets for the Lorentz tensors with primes to distinguish them from the momentum multiplets.

          \renewcommand{\arraystretch}{1.5}

        \begin{table*}[t]

                \begin{center}
                \begin{tabular}{  @{\;\;} l @{\;\;\;\;}     @{\;\;\;}l@{\;\;\;}  @{\quad}l@{\;\;\;}   @{\;\;\; }l@{\;\;\;}   @{\quad}c@{\;\;\;}  @{\;\;\;}c@{\;\;\;} @{\;\;\;}c@{\;\;\;} @{\;\;\;}c@{\;\;\;}  @{\;\;\;}c@{\;\;\;}  @{\;\;\;\;}c      }

                           \hline\noalign{\smallskip}

                    $n$               & Seed element                                                      & \#   & Multiplets  & $n=4$  & $n=6$  & $n=8$  & $n=10$   & $n=12$      \\
                                      \noalign{\smallskip}\hline\hline\noalign{\smallskip}

                   $4$                & $t^{\mu\nu}_{12}\, t^{\rho\sigma}_{34}$                          & $3$   & $\mathcal{S}$, $\mathcal{D}_1$    & $1$  & $1$  & $1$  &    &           \\

                                      & $\varepsilon^{\mu\nu}_{12}\, \varepsilon^{\rho\sigma}_{34}$      & $3$   & $\mathcal{S}$, $\mathcal{D}_1$    & $1$  & $1$  & $1$  &    &               \smallskip  \\
                                      \noalign{\smallskip}\hline\noalign{\smallskip}

                   $6$                & $\varepsilon^{\mu\lambda\alpha}_1 t^{\alpha\nu}_{22}\,\varepsilon^{\rho\lambda\beta}_3 t^{\beta\sigma}_{44}$
                                                                                                         & $12$   & $\mathcal{S}$, $\mathcal{D}_1$, $\mathcal{D}_2$, $\mathcal{T}_2^+$, $\mathcal{T}_2^-$, $\mA$  &      & $1$  & $3$  & $5$     &  $3$         \\

                                      & $t^{\mu\nu}_{12}\,t^{\rho\lambda}_{33}\,t^{\lambda\sigma}_{44}$ & $6$    & $\mathcal{S}$, $\mathcal{D}_1$, $\mathcal{T}_1^+$ &      & $1$  & $2$  & $3$   &        \\

                                      & $t^{\mu\nu}_{12}\,t^{\rho\lambda}_{31}\,t^{\lambda\sigma}_{24}$ & $7$    & $\mathcal{S}$, $\mathcal{T}_1^+$, $\mathcal{T}_1^-$  &      & $1$  & $1$  & $3$    &   $2$     \\

                                      & $\varepsilon^{\mu\nu}_{12}\,\varepsilon^{\rho\lambda}_{31}\,t^{\lambda\sigma}_{24}$
                                                                                                         & $7$   & $\mathcal{D}_2$, $\mathcal{T}_2^+$, \gray{$\mathcal{T}_1^-$, $\mathcal{T}_2^-$}  &      &      & $2$  &  $5$   &     \smallskip  \\

                                      \noalign{\smallskip}\hline\noalign{\smallskip}

                     $8$              & $t^{\mu\nu}_{12}\,t^{\rho\alpha}_{31}\,t^{\alpha\beta}_{12}\,t^{\beta\sigma}_{24}$
                                                                                                         & $3$   & $\mathcal{S}$, \gray{$\mathcal{D}_1$, $\mathcal{T}_1^+$}  &      &      & $1$  & $2$   &     \smallskip \\

                                      \noalign{\smallskip}\hline\hline\noalign{\smallskip}

                                     &  Total                                                           & $41$    &                    & $2$  & $5$  & $11$ & $18$ & $5$

                \end{tabular}
                \end{center}

               \caption{         41-dimensional tensor basis for the transverse part of the photon four-point function.
                                 $n$ denotes the mass dimension of the seed elements and $\#$ the number of the resulting singlets with mass dimension $n$.
                                 For the grayed terms  we keep only the lowest-dimensional singlets with $n=10$.  }
               \label{tab:41-basis}

        \end{table*}

         \renewcommand{\arraystretch}{1.0}

      To obtain the remaining basis elements, we define
      \begin{equation}\label{regular-projectors}
      \begin{split}
          t^{\mu\alpha\beta}_i &:= \delta^{\mu\beta}\,p_i^\alpha - \delta^{\mu\alpha}\,p_i^\beta\,,\\
          \varepsilon^{\mu\alpha\beta}_i &:= \varepsilon^{\mu\alpha\beta\lambda}\,p_i^\lambda\,.
      \end{split}
      \end{equation}
      These are the lowest-dimensional Lorentz tensors that are transverse without introducing
      kinematic singularities.
      $t^{\mu\alpha\beta}_i$ is transverse to the momentum $p_i^\mu$ and $\varepsilon^{\mu\alpha\beta}_i$ is transverse in all Lorentz indices.
      Both are antisymmetric in $\alpha$ and $\beta$.\footnote{Note
      that the electromagnetic field-strength tensor and its dual can be expressed in terms of these quantities:
      \begin{equation}
          F^{\mu\nu} \sim t_\p^{\alpha\mu\nu} A^\alpha, \quad
          \widetilde{F}^{\mu\nu} \sim \varepsilon_\p^{\alpha\mu\nu} A^\alpha.
      \end{equation}}
      The quantities in Eq.~\eqref{t-and-epsilon} are
      their momentum contractions:
      \begin{equation}
          t^{\mu\nu}_{ij} = t^{\mu\alpha\nu}_i p_j^\alpha\,, \qquad
          \varepsilon^{\mu\nu}_{ij} = \varepsilon^{\mu\alpha\nu}_i p_j^\alpha\,,
      \end{equation}
      and by contracting once more we can define
      \begin{equation}
      \begin{split}
          t^{\mu}_{ijk} &:= t^{\mu\alpha}_{ij}\,p^\alpha_k = p_i\cdot p_j\,p_k^\mu - p_i\cdot p_k\,p_j^\mu\,, \\
          \varepsilon^\mu_{ijk} &:= \varepsilon^{\mu\alpha}_{ij}\,p_k^\alpha = \varepsilon^{\mu\alpha\beta\gamma}\,p_i^\alpha\,p_j^\beta\,p_k^\gamma\,.
      \end{split}
      \end{equation}

      At dimension $n=6$ one can find many possible Lorentz tensors by taking suitable combinations of these quantities and their momentum contractions; however,
      only few of them are linearly independent. In particular, it turns out that the seed elements
      \begin{equation}
      \begin{split}
          \psi^{\mu\nu\rho\sigma}_3 &= \varepsilon^{\mu\lambda\alpha}_1 t^{\alpha\nu}_{22}\,\varepsilon^{\rho\lambda\beta}_3 t^{\beta\sigma}_{44}, \\
          \psi^{\mu\nu\rho\sigma}_4 &= t^{\mu\nu}_{12}\,t^{\rho\lambda}_{33}\,t^{\lambda\sigma}_{44}\,, \\
          \psi^{\mu\nu\rho\sigma}_5 &= t^{\mu\nu}_{12}\,t^{\rho\lambda}_{31}\,t^{\lambda\sigma}_{24} \,, \\
          \psi^{\mu\nu\rho\sigma}_6 &= \varepsilon^{\mu\nu}_{12}\,\varepsilon^{\rho\lambda}_{31}\,t^{\lambda\sigma}_{24}   \\
      \end{split}
      \end{equation}
      together with one element at $n=8$,
      \begin{equation}
          \psi^{\mu\nu\rho\sigma}_7 = t^{\mu\nu}_{12}\,t^{\rho\alpha}_{31}\,t^{\alpha\beta}_{12}\,t^{\beta\sigma}_{24}\,,
      \end{equation}
      are sufficient to generate a complete tensor basis with 41 elements.
      The multiplets that they produce are collected in Table~\ref{tab:41-basis}.
      We will discuss them in a moment, but let us first resolve the remaining issue.

      Ultimately we would like to cast the 41 tensor elements into permutation-group singlets, so the question is:
      how can one construct singlets from, for example, a doublet? According to Eq.~\eqref{perm-product-singlets}, the only possibility is to take dot products
      with other doublets. They will be made from the momentum multiplets $\mS_0$, $\mD_0$ and $\mT_0$ that we defined in Eqs.~(\ref{S0-def}--\ref{triplet-def}):
      $\mS_0$ has mass dimension two whereas $\mD_0$ and $\mT_0$ have dimension four.
      The second row in Table~\ref{tab:momentum-multiplets-0} collects all possible doublets at dimension four and six.
      Combined with $\mD_0$, these are
      \begin{equation}
         \mD_0\,, \quad \mD_0\ast\mD_0\,, \quad \mT_0 \ast\mT_0\,, \quad \mD_0 \ast (\mT_0\ast\mT_0)
      \end{equation}
      apart from further trivial multiplications with singlets.
      Now take for example the doublet $\mD' = \mD_1'(\psi_1)$, which is obtained from the tensor structure $\psi_1$. It has two independent components, and by dotting it into two independent doublets from the list above
      we can generate two singlets. Restricting ourselves to the lowest-dimensional possible combinations, these are
      \begin{equation}\label{alpha-beta}
      \begin{split}
         \mS_1' &= \mD_0 \cdot \mD', \\
         \mS_2' &= (\alpha\,\mD_0\ast\mD_0 + \beta\,\mT_0\ast\mT_0)\cdot\mD',
      \end{split}
      \end{equation}
      where $\alpha,\beta$ are constants. Hence, in the process of constructing singlets we have raised the dimension by two and four, respectively:
      a doublet with dimension $n$ generates a singlet at $n+2$ and another singlet at $n+4$.

      To construct singlets from a triplet, we have to dot it into three independent triplets.
      The available options from Table~\ref{tab:momentum-multiplets-0} are
      \begin{equation}\label{mom-multiplets-triplets-1}
         \mT_0\,, \quad \mT_0\vee\mD_0\,, \quad \mT_0\vee\mT_0
      \end{equation}
      which have dimension two, four and four, respectively.
      Therefore, a triplet of dimension $n$ generates a singlet at $n+2$ and two singlets at $n+4$.
      Similarly, the fourth row in Table~\ref{tab:momentum-multiplets-0} shows that an antitriplet of dimension $n$ produces
      a singlet at $n+4$ and two singlets at $n+6$, and the last row entails that an antisinglet of dimension $n$ leads to a singlet at $n+6$.

      In this regard, Table~\ref{tab:41-basis} should be read as follows.
      We start from the seven independent seed elements $\psi_1 \dots \psi_7$ defined above, with dimensions  $n=4,6,8$.
      With their help we can generate 41 linearly independent tensors because each seed generates a number of multiplets.
      Combining them with the momentum multiplets in the way described above, we generate further singlets whose dimension has raised:
      the singlets have dimension $n=4,6,8,10,12$.

      Working out all multiplets by hand would be rather tedious, but it is easy to implement in \texttt{Mathematica}.
      We start from a large number of seed elements (essentially all conceivable tensor structures at $n=4,6,8$ etc.)
      and let \texttt{Mathematica} generate the multiplets. We then add them up until all singlets at $n=4$ are found, proceed to $n=6$, etc.
      At each step we check for linear independence, i.e., whether the condition
         \begin{equation}
             \text{if} \quad \sum_{i=1}^{N} f_i\,\tau_i^{\mu\nu\rho\sigma}(p,q,k) = 0 \quad \Rightarrow \quad f_i = 0
         \end{equation}
         is still satisfied, until $N=41$ is reached (or $N=136$ in the case of Table~\ref{tab:generic-basis}).
         In that way one also confirms directly that there cannot be more than 41 linearly independent transverse elements (or more than 136 elements in general).

         Of course it is possible to construct many equivalent bases by this procedure, but they share some common features.
         First, the maximum number of singlets for a given mass dimension $n$ (the last row in Table~\ref{tab:41-basis}), ordered with increasing $n$, is fixed:
         we found at most two singlets with $n=4$, five singlets with $n=6$, etc.
         Second, we never found fewer than five singlets with $n=12$.
      By contrast, the same procedure applied to Table~\ref{tab:generic-basis} would
      produce singlets with $n=10$ at most (they originate from the antitriplet and antisinglet seeds with $n=4$ whose mass dimension is raised by 6.)
      The appearance of $n=12$ singlets in the transverse basis therefore suggests that not all of them are related in a simple way
      to the type-I basis without kinematic prefactors, as it was the case in the two-photon current example~\eqref{scalar-current-temp},
      and that divisions must have been necessary in the solution process of Eqs.~\eqref{95-eqs}.
      Barring oversights, we are therefore led to believe that Table~\ref{tab:41-basis} can indeed serve as a minimal basis for the LbL amplitude.

      There is, however, a remaining problem. The construction of singlets with lowest mass dimension from a given multiplet is not unique,
      as one can infer from the parameters $\alpha$, $\beta$ in Eq.~\eqref{alpha-beta}.
      There are two momentum doublets at $n=4$ and both of them are equally suitable for constructing a singlet.
      Choosing one over the other can result, once again, in kinematic singularities.
      Similarly, there are three antitriplets at $n=6$ and
      two antisinglets at $n=6$. In fact, only the triplet case is unique because there are three momentum triplets up to $n=4$ (those in Eq.~\eqref{mom-multiplets-triplets-1}).
      One might conclude that it is simply impossible to construct a 41-dimensional transverse basis made of singlets, thus effectively leading to a redundant basis.
      On the other hand, one can argue that the solution of the system of equations~\eqref{95-eqs}, which we circumvented so far,
      should be unambiguous and determine these coefficients in the process. (We mean `unambiguous' in the sense that linear combinations of singlets with the same mass dimension are still allowed.)
      Ultimately it might turn out to be unavoidable to solve Eqs.~\eqref{95-eqs} directly,
      because even with a 41-dimensional transverse tensor basis at hand one still needs to construct a gauge part that is kinematically safe and consistent with it.

      We were recently made aware\footnote{We thank Gilberto Colangelo for bringing this to our attention.} of a similar attempt in constructing a transverse basis for the LbL amplitude~\cite{Stoffer:2014rka}.
      Expressed in our language, the seed elements (Eq.~(3.14) of Part III therein) have the form
      \begin{equation} \renewcommand{\arraystretch}{1.4}
      \begin{split}
      &
      \begin{array}{rl}
         T_1 &\sim \varepsilon^{\mu\nu}_{12}\,\varepsilon^{\rho\sigma}_{34}, \\
         T_4 &\sim t^{\mu\nu}_{12}\,t^{\rho\sigma}_{34}, \\
      \end{array}\qquad
      \begin{array}{rl}
         T_7 &\sim t_{12}^{\mu\nu}\,t_{31}^{\rho\lambda}\,t_{14}^{\lambda\sigma}, \\
         T_{19} &\sim t_{12}^{\mu\nu}\,t_{31}^{\rho\lambda}\,t_{24}^{\lambda\sigma},
      \end{array} \\
      &\qquad
      \begin{array}{rl}
         T_{31} &\sim t_{12}^{\mu\nu}\,t_{312}^\rho\,t_{412}^\sigma, \\
         T_{37} &\sim t_{134}^\mu\,t^{\nu\alpha\beta}_2\,t^{\rho\alpha\lambda}_3\,t^{\sigma\beta\lambda}_4, \\
         T_{49} &\sim (t^{\mu\alpha}_{14}\,t^{\beta\nu}_{32}-t^{\mu\beta}_{13}\,t^{\alpha\nu}_{42})\,t^{\rho\alpha\lambda}_3\,t^{\sigma\beta\lambda}_4\,.
      \end{array}
      \end{split}
      \end{equation}
      The problem of minimality is not addressed, but in terms of counting mass dimensions these tensors are equivalent to the seeds in Table~\ref{tab:41-basis}:
      there are two seeds with $n=2$, four with $n=6$ and one with $n=8$.
      After working out the permutations, also the distribution of singlets is the same:
      we found an `optimal' arrangement where $41=2+5+11+18+5$, as in the last row in Table~\ref{tab:41-basis}.

      Finally, we should comment on the four-gluon vertex. In that case the construction of the transverse part is the same as in Table~\ref{tab:41-basis},
      except for the final construction of the singlets. The reason is of course the additional color structure which also produces multiplets (see Sec.~\ref{sec:color}),
      so there are more possibilities for constructing singlets which have lower mass dimensions. In addition, the four-gluon vertex has a nonzero gauge part,
      which makes a solution of Eqs.~\eqref{95-eqs} mandatory. On the other hand, the presence of the gauge part also simplifies the problem because
      a type-I basis such as that in Table~\ref{tab:generic-basis} (or even the one in Eq.~\eqref{vrsd-basis}) would suffice for its solution.
      The contraction with transverse gluon propagators (in Landau gauge) will remove the purely longitudinal elements, so that only the `survivors' remain.
      This is exemplified by a calculation of the three-gluon vertex,
      including its full tensor structure and full kinematics, with a type-I basis only~\cite{Eichmann:2014xya}. To reiterate, the problem of transversality
      and analyticity is only truly relevant when gauge-\textit{invariant} amplitudes are considered.

\section{Summary and conclusions}\label{sec:summary}

	In this work we discussed in some detail the application of the permutation group $S_4$ to
	four-point functions of particles with four external gauge bosons. In particular, we explored
	the case of light-by-light scattering that has important applications in the calculation of
	hadronic contributions of the anomalous magnetic moment of the muon. The main problem was to
	identify the transverse, gauge invariant components and represent these in terms of tensor
	structures that are free of kinematical singularities. To this end we made judicious use of
	the permutation group. We introduced an efficient notation for constructing the $S_4$ multiplets,
	and we applied it to organize the kinematic phase space and to construct appropriate tensor bases.
	The multiplet analysis provides us with a straightforward way to find a complete basis of
	136 linearly independent elements. Implementing gauge invariance, we constructed a transverse
	basis of 41 elements that has the required analyticity properties. The remaining problem is
	to cast this basis into one made of permutation-group singlets and combine it with a consistent
	gauge part. This will have to be addressed in future work. Nevertheless, we do hope that our
	general framework serves to establish common grounds for communication among the several groups
	studying the light-by-light scattering amplitude in different approaches.

    \section*{Acknowledgements}

    We thank Gilberto Colangelo, Peter Stoffer and Richard Williams for helpful interactions.
	This work was supported by the German Science Foundation DFG under project number TR-16,
	by the German Federal Ministry of Education and Research BMBF under project number 06GI7121,
	by the Helmholtz International Center for FAIR within the LOEWE program of the State of Hesse,
	and by the Helmholtzzentrum GSI.

      \bibliographystyle{apsrev4-1-mod}
      \bibliography{lit}

\begin{thebibliography}{40}%
\makeatletter
\providecommand \@ifxundefined [1]{%
 \@ifx{#1\undefined}
}%
\providecommand \@ifnum [1]{%
 \ifnum #1\expandafter \@firstoftwo
 \else \expandafter \@secondoftwo
 \fi
}%
\providecommand \@ifx [1]{%
 \ifx #1\expandafter \@firstoftwo
 \else \expandafter \@secondoftwo
 \fi
}%
\providecommand \natexlab [1]{#1}%
\providecommand \enquote  [1]{``#1''}%
\providecommand \bibnamefont  [1]{#1}%
\providecommand \bibfnamefont [1]{#1}%
\providecommand \citenamefont [1]{#1}%
\providecommand \href@noop [0]{\@secondoftwo}%
\providecommand \href [0]{\begingroup \@sanitize@url \@href}%
\providecommand \@href[1]{\@@startlink{#1}\@@href}%
\providecommand \@@href[1]{\endgroup#1\@@endlink}%
\providecommand \@sanitize@url [0]{\catcode `\\12\catcode `\$12\catcode
  `\&12\catcode `\#12\catcode `\^12\catcode `\_12\catcode `\%12\relax}%
\providecommand \@@startlink[1]{}%
\providecommand \@@endlink[0]{}%
\providecommand \url  [0]{\begingroup\@sanitize@url \@url }%
\providecommand \@url [1]{\endgroup\@href {#1}{\urlprefix }}%
\providecommand \urlprefix  [0]{URL }%
\providecommand \Eprint [0]{\href }%
\providecommand \doibase [0]{http://dx.doi.org/}%
\providecommand \selectlanguage [0]{\@gobble}%
\providecommand \bibinfo  [0]{\@secondoftwo}%
\providecommand \bibfield  [0]{\@secondoftwo}%
\providecommand \translation [1]{[#1]}%
\providecommand \BibitemOpen [0]{}%
\providecommand \bibitemStop [0]{}%
\providecommand \bibitemNoStop [0]{.\EOS\space}%
\providecommand \EOS [0]{\spacefactor3000\relax}%
\providecommand \BibitemShut  [1]{\csname bibitem#1\endcsname}%
\let\auto@bib@innerbib\@empty
\bibitem [{\citenamefont {Jegerlehner}\ and\ \citenamefont
  {Nyffeler}(2009)}]{Jegerlehner:2009ry}%
  \BibitemOpen
  \bibfield  {author} {\bibinfo {author} {\bibfnamefont {F.}~\bibnamefont
  {Jegerlehner}}\ and\ \bibinfo {author} {\bibfnamefont {A.}~\bibnamefont
  {Nyffeler}},\ }\href {\doibase 10.1016/j.physrep.2009.04.003} {\bibfield
  {journal} {\bibinfo  {journal} {Phys. Rept.}\ }\textbf {\bibinfo {volume}
  {477}},\ \bibinfo {pages} {1} (\bibinfo {year} {2009})}\BibitemShut {NoStop}%
\bibitem [{\citenamefont {Ball}\ and\ \citenamefont
  {Chiu}(1980{\natexlab{a}})}]{Ball:1980ax}%
  \BibitemOpen
  \bibfield  {author} {\bibinfo {author} {\bibfnamefont {J.~S.}\ \bibnamefont
  {Ball}}\ and\ \bibinfo {author} {\bibfnamefont {T.-W.}\ \bibnamefont
  {Chiu}},\ }\href {\doibase 10.1103/PhysRevD.22.2550,
  10.1103/PhysRevD.23.3085} {\bibfield  {journal} {\bibinfo  {journal}
  {Phys.Rev.}\ }\textbf {\bibinfo {volume} {D22}},\ \bibinfo {pages} {2550}
  (\bibinfo {year} {1980}{\natexlab{a}})}\BibitemShut {NoStop}%
\bibitem [{\citenamefont {Eichmann}\ \emph {et~al.}(2010)\citenamefont
  {Eichmann}, \citenamefont {Alkofer}, \citenamefont {Krassnigg},\ and\
  \citenamefont {Nicmorus}}]{Eichmann:2009qa}%
  \BibitemOpen
  \bibfield  {author} {\bibinfo {author} {\bibfnamefont {G.}~\bibnamefont
  {Eichmann}}, \bibinfo {author} {\bibfnamefont {R.}~\bibnamefont {Alkofer}},
  \bibinfo {author} {\bibfnamefont {A.}~\bibnamefont {Krassnigg}}, \ and\
  \bibinfo {author} {\bibfnamefont {D.}~\bibnamefont {Nicmorus}},\ }\href
  {\doibase 10.1103/PhysRevLett.104.201601} {\bibfield  {journal} {\bibinfo
  {journal} {Phys. Rev. Lett.}\ }\textbf {\bibinfo {volume} {104}},\ \bibinfo
  {pages} {201601} (\bibinfo {year} {2010})}\BibitemShut {NoStop}%
\bibitem [{\citenamefont {Sanchis-Alepuz}\ \emph {et~al.}(2011)\citenamefont
  {Sanchis-Alepuz}, \citenamefont {Eichmann}, \citenamefont {Villalba-Chavez},\
  and\ \citenamefont {Alkofer}}]{SanchisAlepuz:2011jn}%
  \BibitemOpen
  \bibfield  {author} {\bibinfo {author} {\bibfnamefont {H.}~\bibnamefont
  {Sanchis-Alepuz}}, \bibinfo {author} {\bibfnamefont {G.}~\bibnamefont
  {Eichmann}}, \bibinfo {author} {\bibfnamefont {S.}~\bibnamefont
  {Villalba-Chavez}}, \ and\ \bibinfo {author} {\bibfnamefont {R.}~\bibnamefont
  {Alkofer}},\ }\href {\doibase 10.1103/PhysRevD.84.096003} {\bibfield
  {journal} {\bibinfo  {journal} {Phys.Rev.}\ }\textbf {\bibinfo {volume}
  {D84}},\ \bibinfo {pages} {096003} (\bibinfo {year} {2011})}\BibitemShut
  {NoStop}%
\bibitem [{\citenamefont {Eichmann}\ \emph {et~al.}()\citenamefont {Eichmann},
  \citenamefont {Fischer},\ and\ \citenamefont {Heupel}}]{tetraquark}%
  \BibitemOpen
  \bibfield  {author} {\bibinfo {author} {\bibfnamefont {G.}~\bibnamefont
  {Eichmann}}, \bibinfo {author} {\bibfnamefont {C.~S.}\ \bibnamefont
  {Fischer}}, \ and\ \bibinfo {author} {\bibfnamefont {W.}~\bibnamefont
  {Heupel}},\ }\href@noop {} {\bibinfo  {journal} {{in preparation}}\
  }\BibitemShut {NoStop}%
\bibitem [{\citenamefont {Blum}\ \emph {et~al.}(2014)\citenamefont {Blum},
  \citenamefont {Huber}, \citenamefont {Mitter},\ and\ \citenamefont {von
  Smekal}}]{Blum:2014gna}%
  \BibitemOpen
\bibfield  {journal} {  }\bibfield  {author} {\bibinfo {author} {\bibfnamefont
  {A.}~\bibnamefont {Blum}}, \bibinfo {author} {\bibfnamefont {M.~Q.}\
  \bibnamefont {Huber}}, \bibinfo {author} {\bibfnamefont {M.}~\bibnamefont
  {Mitter}}, \ and\ \bibinfo {author} {\bibfnamefont {L.}~\bibnamefont {von
  Smekal}},\ }\href {\doibase 10.1103/PhysRevD.89.061703} {\bibfield  {journal}
  {\bibinfo  {journal} {Phys.Rev.}\ }\textbf {\bibinfo {volume} {D89}},\
  \bibinfo {pages} {061703} (\bibinfo {year} {2014})}\BibitemShut {NoStop}%
\bibitem [{\citenamefont {Eichmann}\ \emph
  {et~al.}(2014{\natexlab{a}})\citenamefont {Eichmann}, \citenamefont
  {Williams}, \citenamefont {Alkofer},\ and\ \citenamefont
  {Vujinovic}}]{Eichmann:2014xya}%
  \BibitemOpen
  \bibfield  {author} {\bibinfo {author} {\bibfnamefont {G.}~\bibnamefont
  {Eichmann}}, \bibinfo {author} {\bibfnamefont {R.}~\bibnamefont {Williams}},
  \bibinfo {author} {\bibfnamefont {R.}~\bibnamefont {Alkofer}}, \ and\
  \bibinfo {author} {\bibfnamefont {M.}~\bibnamefont {Vujinovic}},\ }\href
  {\doibase 10.1103/PhysRevD.89.105014} {\bibfield  {journal} {\bibinfo
  {journal} {Phys. Rev.}\ }\textbf {\bibinfo {volume} {D89}},\ \bibinfo {pages}
  {105014} (\bibinfo {year} {2014}{\natexlab{a}})}\BibitemShut {NoStop}%
\bibitem [{\citenamefont {Braun}\ \emph {et~al.}(2014)\citenamefont {Braun},
  \citenamefont {Fister}, \citenamefont {Pawlowski},\ and\ \citenamefont
  {Rennecke}}]{Braun:2014ata}%
  \BibitemOpen
  \bibfield  {author} {\bibinfo {author} {\bibfnamefont {J.}~\bibnamefont
  {Braun}}, \bibinfo {author} {\bibfnamefont {L.}~\bibnamefont {Fister}},
  \bibinfo {author} {\bibfnamefont {J.~M.}\ \bibnamefont {Pawlowski}}, \ and\
  \bibinfo {author} {\bibfnamefont {F.}~\bibnamefont {Rennecke}},\ }\href@noop
  {} {\ }\Eprint {http://arxiv.org/abs/1412.1045} {1412.1045 [hep-ph]}
  \BibitemShut {NoStop}%
\bibitem [{\citenamefont {Aguilar}\ \emph {et~al.}(2014)\citenamefont
  {Aguilar}, \citenamefont {Binosi}, \citenamefont {Ibanez},\ and\
  \citenamefont {Papavassiliou}}]{Aguilar:2014lha}%
  \BibitemOpen
  \bibfield  {author} {\bibinfo {author} {\bibfnamefont {A.}~\bibnamefont
  {Aguilar}}, \bibinfo {author} {\bibfnamefont {D.}~\bibnamefont {Binosi}},
  \bibinfo {author} {\bibfnamefont {D.}~\bibnamefont {Ibanez}}, \ and\ \bibinfo
  {author} {\bibfnamefont {J.}~\bibnamefont {Papavassiliou}},\ }\href {\doibase
  10.1103/PhysRevD.90.065027} {\bibfield  {journal} {\bibinfo  {journal}
  {Phys.Rev.}\ }\textbf {\bibinfo {volume} {D90}},\ \bibinfo {pages} {065027}
  (\bibinfo {year} {2014})}\BibitemShut {NoStop}%
\bibitem [{\citenamefont {Kellermann}\ and\ \citenamefont
  {Fischer}(2008)}]{Kellermann:2008iw}%
  \BibitemOpen
  \bibfield  {author} {\bibinfo {author} {\bibfnamefont {C.}~\bibnamefont
  {Kellermann}}\ and\ \bibinfo {author} {\bibfnamefont {C.~S.}\ \bibnamefont
  {Fischer}},\ }\href {\doibase 10.1103/PhysRevD.78.025015} {\bibfield
  {journal} {\bibinfo  {journal} {Phys.Rev.}\ }\textbf {\bibinfo {volume}
  {D78}},\ \bibinfo {pages} {025015} (\bibinfo {year} {2008})}\BibitemShut
  {NoStop}%
\bibitem [{\citenamefont {Cyrol}\ \emph {et~al.}(2015)\citenamefont {Cyrol},
  \citenamefont {Huber},\ and\ \citenamefont {von Smekal}}]{Cyrol:2014kca}%
  \BibitemOpen
  \bibfield  {author} {\bibinfo {author} {\bibfnamefont {A.~K.}\ \bibnamefont
  {Cyrol}}, \bibinfo {author} {\bibfnamefont {M.~Q.}\ \bibnamefont {Huber}}, \
  and\ \bibinfo {author} {\bibfnamefont {L.}~\bibnamefont {von Smekal}},\
  }\href {\doibase 10.1140/epjc/s10052-015-3312-1} {\bibfield  {journal}
  {\bibinfo  {journal} {Eur.Phys.J.}\ }\textbf {\bibinfo {volume} {C75}},\
  \bibinfo {pages} {102} (\bibinfo {year} {2015})}\BibitemShut {NoStop}%
\bibitem [{\citenamefont {Binosi}\ \emph {et~al.}(2014)\citenamefont {Binosi},
  \citenamefont {Ibanez},\ and\ \citenamefont
  {Papavassiliou}}]{Binosi:2014kka}%
  \BibitemOpen
  \bibfield  {author} {\bibinfo {author} {\bibfnamefont {D.}~\bibnamefont
  {Binosi}}, \bibinfo {author} {\bibfnamefont {D.}~\bibnamefont {Ibanez}}, \
  and\ \bibinfo {author} {\bibfnamefont {J.}~\bibnamefont {Papavassiliou}},\
  }\href {\doibase 10.1007/JHEP09(2014)059} {\bibfield  {journal} {\bibinfo
  {journal} {JHEP}\ }\textbf {\bibinfo {volume} {1409}},\ \bibinfo {pages}
  {059} (\bibinfo {year} {2014})}\BibitemShut {NoStop}%
\bibitem [{\citenamefont {Lee~Roberts}(2011)}]{LeeRoberts:2011zz}%
  \BibitemOpen
  \bibfield  {author} {\bibinfo {author} {\bibfnamefont {B.}~\bibnamefont
  {Lee~Roberts}} (\bibinfo {collaboration} {Fermilab P989}),\ }\href {\doibase
  10.1016/j.nuclphysbps.2011.06.038} {\bibfield  {journal} {\bibinfo  {journal}
  {Nucl.Phys.Proc.Suppl.}\ }\textbf {\bibinfo {volume} {218}},\ \bibinfo
  {pages} {237} (\bibinfo {year} {2011})}\BibitemShut {NoStop}%
\bibitem [{\citenamefont {Iinuma}(2011)}]{Iinuma:2011zz}%
  \BibitemOpen
  \bibfield  {author} {\bibinfo {author} {\bibfnamefont {H.}~\bibnamefont
  {Iinuma}} (\bibinfo {collaboration} {J-PARC New g-2/EDM experiment}),\ }\href
  {\doibase 10.1088/1742-6596/295/1/012032} {\bibfield  {journal} {\bibinfo
  {journal} {J.Phys.Conf.Ser.}\ }\textbf {\bibinfo {volume} {295}},\ \bibinfo
  {pages} {012032} (\bibinfo {year} {2011})}\BibitemShut {NoStop}%
\bibitem [{\citenamefont {Goecke}\ \emph {et~al.}(2013)\citenamefont {Goecke},
  \citenamefont {Fischer},\ and\ \citenamefont {Williams}}]{Goecke:2012qm}%
  \BibitemOpen
  \bibfield  {author} {\bibinfo {author} {\bibfnamefont {T.}~\bibnamefont
  {Goecke}}, \bibinfo {author} {\bibfnamefont {C.~S.}\ \bibnamefont {Fischer}},
  \ and\ \bibinfo {author} {\bibfnamefont {R.}~\bibnamefont {Williams}},\
  }\href {\doibase 10.1103/PhysRevD.87.034013} {\bibfield  {journal} {\bibinfo
  {journal} {Phys. Rev.}\ }\textbf {\bibinfo {volume} {D87}},\ \bibinfo {pages}
  {034013} (\bibinfo {year} {2013})}\BibitemShut {NoStop}%
\bibitem [{\citenamefont {Pauk}\ and\ \citenamefont
  {Vanderhaeghen}(2014)}]{Pauk:2014rfa}%
  \BibitemOpen
  \bibfield  {author} {\bibinfo {author} {\bibfnamefont {V.}~\bibnamefont
  {Pauk}}\ and\ \bibinfo {author} {\bibfnamefont {M.}~\bibnamefont
  {Vanderhaeghen}},\ }\href {\doibase 10.1103/PhysRevD.90.113012} {\bibfield
  {journal} {\bibinfo  {journal} {Phys.Rev.}\ }\textbf {\bibinfo {volume}
  {D90}},\ \bibinfo {pages} {113012} (\bibinfo {year} {2014})}\BibitemShut
  {NoStop}%
\bibitem [{\citenamefont {Blum}\ \emph {et~al.}(2015)\citenamefont {Blum},
  \citenamefont {Chowdhury}, \citenamefont {Hayakawa},\ and\ \citenamefont
  {Izubuchi}}]{Blum:2014oka}%
  \BibitemOpen
  \bibfield  {author} {\bibinfo {author} {\bibfnamefont {T.}~\bibnamefont
  {Blum}}, \bibinfo {author} {\bibfnamefont {S.}~\bibnamefont {Chowdhury}},
  \bibinfo {author} {\bibfnamefont {M.}~\bibnamefont {Hayakawa}}, \ and\
  \bibinfo {author} {\bibfnamefont {T.}~\bibnamefont {Izubuchi}},\ }\href
  {\doibase 10.1103/PhysRevLett.114.012001} {\bibfield  {journal} {\bibinfo
  {journal} {Phys.Rev.Lett.}\ }\textbf {\bibinfo {volume} {114}},\ \bibinfo
  {pages} {012001} (\bibinfo {year} {2015})}\BibitemShut {NoStop}%
\bibitem [{\citenamefont {Dorokhov}\ \emph {et~al.}(2014)\citenamefont
  {Dorokhov}, \citenamefont {Radzhabov},\ and\ \citenamefont
  {Zhevlakov}}]{Dorokhov:2014iva}%
  \BibitemOpen
  \bibfield  {author} {\bibinfo {author} {\bibfnamefont {A.~E.}\ \bibnamefont
  {Dorokhov}}, \bibinfo {author} {\bibfnamefont {A.}~\bibnamefont {Radzhabov}},
  \ and\ \bibinfo {author} {\bibfnamefont {A.}~\bibnamefont {Zhevlakov}},\
  }\href {\doibase 10.1134/S0021364014140045} {\bibfield  {journal} {\bibinfo
  {journal} {JETP Lett.}\ }\textbf {\bibinfo {volume} {100}},\ \bibinfo {pages}
  {133} (\bibinfo {year} {2014})}\BibitemShut {NoStop}%
\bibitem [{\citenamefont {Colangelo}\ \emph
  {et~al.}(2014{\natexlab{a}})\citenamefont {Colangelo}, \citenamefont
  {Hoferichter}, \citenamefont {Procura},\ and\ \citenamefont
  {Stoffer}}]{Colangelo:2014dfa}%
  \BibitemOpen
  \bibfield  {author} {\bibinfo {author} {\bibfnamefont {G.}~\bibnamefont
  {Colangelo}}, \bibinfo {author} {\bibfnamefont {M.}~\bibnamefont
  {Hoferichter}}, \bibinfo {author} {\bibfnamefont {M.}~\bibnamefont
  {Procura}}, \ and\ \bibinfo {author} {\bibfnamefont {P.}~\bibnamefont
  {Stoffer}},\ }\href {\doibase 10.1007/JHEP09(2014)091} {\bibfield  {journal}
  {\bibinfo  {journal} {JHEP}\ }\textbf {\bibinfo {volume} {1409}},\ \bibinfo
  {pages} {091} (\bibinfo {year} {2014}{\natexlab{a}})}\BibitemShut {NoStop}%
\bibitem [{\citenamefont {Colangelo}\ \emph
  {et~al.}(2014{\natexlab{b}})\citenamefont {Colangelo}, \citenamefont
  {Hoferichter}, \citenamefont {Kubis}, \citenamefont {Procura},\ and\
  \citenamefont {Stoffer}}]{Colangelo:2014pva}%
  \BibitemOpen
  \bibfield  {author} {\bibinfo {author} {\bibfnamefont {G.}~\bibnamefont
  {Colangelo}}, \bibinfo {author} {\bibfnamefont {M.}~\bibnamefont
  {Hoferichter}}, \bibinfo {author} {\bibfnamefont {B.}~\bibnamefont {Kubis}},
  \bibinfo {author} {\bibfnamefont {M.}~\bibnamefont {Procura}}, \ and\
  \bibinfo {author} {\bibfnamefont {P.}~\bibnamefont {Stoffer}},\ }\href
  {\doibase 10.1016/j.physletb.2014.09.021} {\bibfield  {journal} {\bibinfo
  {journal} {Phys.Lett.}\ }\textbf {\bibinfo {volume} {B738}},\ \bibinfo
  {pages} {6} (\bibinfo {year} {2014}{\natexlab{b}})}\BibitemShut {NoStop}%
\bibitem [{\citenamefont {Eichmann}\ \emph
  {et~al.}(2014{\natexlab{b}})\citenamefont {Eichmann}, \citenamefont
  {Fischer}, \citenamefont {Heupel},\ and\ \citenamefont
  {Williams}}]{Eichmann:2014ooa}%
  \BibitemOpen
  \bibfield  {author} {\bibinfo {author} {\bibfnamefont {G.}~\bibnamefont
  {Eichmann}}, \bibinfo {author} {\bibfnamefont {C.~S.}\ \bibnamefont
  {Fischer}}, \bibinfo {author} {\bibfnamefont {W.}~\bibnamefont {Heupel}}, \
  and\ \bibinfo {author} {\bibfnamefont {R.}~\bibnamefont {Williams}},\
  }\href@noop {} {\ }\Eprint {http://arxiv.org/abs/1411.7876} {1411.7876
  [hep-ph]} \BibitemShut {NoStop}%
\bibitem [{\citenamefont {Benayoun}\ \emph {et~al.}(2014)\citenamefont
  {Benayoun}, \citenamefont {Bijnens}, \citenamefont {Blum}, \citenamefont
  {Caprini}, \citenamefont {Colangelo} \emph {et~al.}}]{Benayoun:2014tra}%
  \BibitemOpen
  \bibfield  {author} {\bibinfo {author} {\bibfnamefont {M.}~\bibnamefont
  {Benayoun}}, \bibinfo {author} {\bibfnamefont {J.}~\bibnamefont {Bijnens}},
  \bibinfo {author} {\bibfnamefont {T.}~\bibnamefont {Blum}}, \bibinfo {author}
  {\bibfnamefont {I.}~\bibnamefont {Caprini}}, \bibinfo {author} {\bibfnamefont
  {G.}~\bibnamefont {Colangelo}},  \emph {et~al.},\ }\href@noop {} {\ }\Eprint
  {http://arxiv.org/abs/1407.4021} {1407.4021 [hep-ph]} \BibitemShut {NoStop}%
\bibitem [{\citenamefont {Bardeen}\ and\ \citenamefont
  {Tung}(1968)}]{Bardeen:1969aw}%
  \BibitemOpen
  \bibfield  {author} {\bibinfo {author} {\bibfnamefont {W.~A.}\ \bibnamefont
  {Bardeen}}\ and\ \bibinfo {author} {\bibfnamefont {W.}~\bibnamefont {Tung}},\
  }\href {\doibase 10.1103/PhysRevD.4.3229, 10.1103/PhysRev.173.1423}
  {\bibfield  {journal} {\bibinfo  {journal} {Phys. Rev.}\ }\textbf {\bibinfo
  {volume} {173}},\ \bibinfo {pages} {1423} (\bibinfo {year}
  {1968})}\BibitemShut {NoStop}%
\bibitem [{\citenamefont {Tarrach}(1975)}]{Tarrach:1975tu}%
  \BibitemOpen
  \bibfield  {author} {\bibinfo {author} {\bibfnamefont {R.}~\bibnamefont
  {Tarrach}},\ }\href {\doibase 10.1007/BF02894857} {\bibfield  {journal}
  {\bibinfo  {journal} {Nuovo Cim.}\ }\textbf {\bibinfo {volume} {A28}},\
  \bibinfo {pages} {409} (\bibinfo {year} {1975})}\BibitemShut {NoStop}%
\bibitem [{\citenamefont {Drechsel}\ \emph {et~al.}(1997)\citenamefont
  {Drechsel}, \citenamefont {Knochlein}, \citenamefont {Metz},\ and\
  \citenamefont {Scherer}}]{Drechsel:1996ag}%
  \BibitemOpen
  \bibfield  {author} {\bibinfo {author} {\bibfnamefont {D.}~\bibnamefont
  {Drechsel}}, \bibinfo {author} {\bibfnamefont {G.}~\bibnamefont {Knochlein}},
  \bibinfo {author} {\bibfnamefont {A.}~\bibnamefont {Metz}}, \ and\ \bibinfo
  {author} {\bibfnamefont {S.}~\bibnamefont {Scherer}},\ }\href {\doibase
  10.1103/PhysRevC.55.424} {\bibfield  {journal} {\bibinfo  {journal} {Phys.
  Rev.}\ }\textbf {\bibinfo {volume} {C55}},\ \bibinfo {pages} {424} (\bibinfo
  {year} {1997})}\BibitemShut {NoStop}%
\bibitem [{\citenamefont {Drechsel}\ \emph {et~al.}(1998)\citenamefont
  {Drechsel}, \citenamefont {Knochlein}, \citenamefont {Korchin}, \citenamefont
  {Metz},\ and\ \citenamefont {Scherer}}]{Drechsel:1997xv}%
  \BibitemOpen
  \bibfield  {author} {\bibinfo {author} {\bibfnamefont {D.}~\bibnamefont
  {Drechsel}}, \bibinfo {author} {\bibfnamefont {G.}~\bibnamefont {Knochlein}},
  \bibinfo {author} {\bibfnamefont {A.~Y.}\ \bibnamefont {Korchin}}, \bibinfo
  {author} {\bibfnamefont {A.}~\bibnamefont {Metz}}, \ and\ \bibinfo {author}
  {\bibfnamefont {S.}~\bibnamefont {Scherer}},\ }\href {\doibase
  10.1103/PhysRevC.57.941} {\bibfield  {journal} {\bibinfo  {journal} {Phys.
  Rev.}\ }\textbf {\bibinfo {volume} {C57}},\ \bibinfo {pages} {941} (\bibinfo
  {year} {1998})}\BibitemShut {NoStop}%
\bibitem [{\citenamefont {L'vov}\ \emph {et~al.}(2001)\citenamefont {L'vov},
  \citenamefont {Scherer}, \citenamefont {Pasquini}, \citenamefont {Unkmeir},\
  and\ \citenamefont {Drechsel}}]{L'vov:2001fz}%
  \BibitemOpen
  \bibfield  {author} {\bibinfo {author} {\bibfnamefont {A.}~\bibnamefont
  {L'vov}}, \bibinfo {author} {\bibfnamefont {S.}~\bibnamefont {Scherer}},
  \bibinfo {author} {\bibfnamefont {B.}~\bibnamefont {Pasquini}}, \bibinfo
  {author} {\bibfnamefont {C.}~\bibnamefont {Unkmeir}}, \ and\ \bibinfo
  {author} {\bibfnamefont {D.}~\bibnamefont {Drechsel}},\ }\href {\doibase
  10.1103/PhysRevC.64.015203} {\bibfield  {journal} {\bibinfo  {journal} {Phys.
  Rev.}\ }\textbf {\bibinfo {volume} {C64}},\ \bibinfo {pages} {015203}
  (\bibinfo {year} {2001})}\BibitemShut {NoStop}%
\bibitem [{\citenamefont {Drechsel}\ \emph {et~al.}(2003)\citenamefont
  {Drechsel}, \citenamefont {Pasquini},\ and\ \citenamefont
  {Vanderhaeghen}}]{Drechsel:2002ar}%
  \BibitemOpen
  \bibfield  {author} {\bibinfo {author} {\bibfnamefont {D.}~\bibnamefont
  {Drechsel}}, \bibinfo {author} {\bibfnamefont {B.}~\bibnamefont {Pasquini}},
  \ and\ \bibinfo {author} {\bibfnamefont {M.}~\bibnamefont {Vanderhaeghen}},\
  }\href {\doibase 10.1016/S0370-1573(02)00636-1} {\bibfield  {journal}
  {\bibinfo  {journal} {Phys. Rept.}\ }\textbf {\bibinfo {volume} {378}},\
  \bibinfo {pages} {99} (\bibinfo {year} {2003})}\BibitemShut {NoStop}%
\bibitem [{\citenamefont {Gorchtein}(2010)}]{Gorchtein:2009wz}%
  \BibitemOpen
  \bibfield  {author} {\bibinfo {author} {\bibfnamefont {M.}~\bibnamefont
  {Gorchtein}},\ }\href {\doibase 10.1103/PhysRevC.81.015206} {\bibfield
  {journal} {\bibinfo  {journal} {Phys. Rev.}\ }\textbf {\bibinfo {volume}
  {C81}},\ \bibinfo {pages} {015206} (\bibinfo {year} {2010})}\BibitemShut
  {NoStop}%
\bibitem [{\citenamefont {Meyers}\ and\ \citenamefont
  {Swanson}(2013)}]{Meyers:2012ka}%
  \BibitemOpen
  \bibfield  {author} {\bibinfo {author} {\bibfnamefont {J.}~\bibnamefont
  {Meyers}}\ and\ \bibinfo {author} {\bibfnamefont {E.~S.}\ \bibnamefont
  {Swanson}},\ }\href {\doibase 10.1103/PhysRevD.87.036009} {\bibfield
  {journal} {\bibinfo  {journal} {Phys.Rev.}\ }\textbf {\bibinfo {volume}
  {D87}},\ \bibinfo {pages} {036009} (\bibinfo {year} {2013})}\BibitemShut
  {NoStop}%
\bibitem [{\citenamefont {Sanchis-Alepuz}\ \emph {et~al.}(2015)\citenamefont
  {Sanchis-Alepuz}, \citenamefont {Fischer}, \citenamefont {Kellermann},\ and\
  \citenamefont {von Smekal}}]{Sanchis-Alepuz:2015hma}%
  \BibitemOpen
  \bibfield  {author} {\bibinfo {author} {\bibfnamefont {H.}~\bibnamefont
  {Sanchis-Alepuz}}, \bibinfo {author} {\bibfnamefont {C.~S.}\ \bibnamefont
  {Fischer}}, \bibinfo {author} {\bibfnamefont {C.}~\bibnamefont {Kellermann}},
  \ and\ \bibinfo {author} {\bibfnamefont {L.}~\bibnamefont {von Smekal}},\
  }\href@noop {} {\ }\Eprint {http://arxiv.org/abs/1503.06051} {1503.06051
  [hep-ph]} \BibitemShut {NoStop}%
\bibitem [{\citenamefont {Ball}\ and\ \citenamefont
  {Chiu}(1980{\natexlab{b}})}]{Ball:1980ay}%
  \BibitemOpen
  \bibfield  {author} {\bibinfo {author} {\bibfnamefont {J.~S.}\ \bibnamefont
  {Ball}}\ and\ \bibinfo {author} {\bibfnamefont {T.-W.}\ \bibnamefont
  {Chiu}},\ }\href {\doibase 10.1103/PhysRevD.22.2542} {\bibfield  {journal}
  {\bibinfo  {journal} {Phys. Rev.}\ }\textbf {\bibinfo {volume} {D22}},\
  \bibinfo {pages} {2542} (\bibinfo {year} {1980}{\natexlab{b}})}\BibitemShut
  {NoStop}%
\bibitem [{\citenamefont {Eichmann}\ and\ \citenamefont
  {Fischer}(2013)}]{Eichmann:2012mp}%
  \BibitemOpen
  \bibfield  {author} {\bibinfo {author} {\bibfnamefont {G.}~\bibnamefont
  {Eichmann}}\ and\ \bibinfo {author} {\bibfnamefont {C.~S.}\ \bibnamefont
  {Fischer}},\ }\href {\doibase 10.1103/PhysRevD.87.036006} {\bibfield
  {journal} {\bibinfo  {journal} {Phys. Rev.}\ }\textbf {\bibinfo {volume}
  {D87}},\ \bibinfo {pages} {036006} (\bibinfo {year} {2013})}\BibitemShut
  {NoStop}%
\bibitem [{\citenamefont {Carimalo}(1993)}]{Carimalo:1992ia}%
  \BibitemOpen
  \bibfield  {author} {\bibinfo {author} {\bibfnamefont {C.}~\bibnamefont
  {Carimalo}},\ }\href {\doibase 10.1063/1.530334} {\bibfield  {journal}
  {\bibinfo  {journal} {J. Math. Phys.}\ }\textbf {\bibinfo {volume} {34}},\
  \bibinfo {pages} {4930} (\bibinfo {year} {1993})}\BibitemShut {NoStop}%
\bibitem [{\citenamefont {Eichmann}(2011)}]{Eichmann:2011vu}%
  \BibitemOpen
  \bibfield  {author} {\bibinfo {author} {\bibfnamefont {G.}~\bibnamefont
  {Eichmann}},\ }\href {\doibase 10.1103/PhysRevD.84.014014} {\bibfield
  {journal} {\bibinfo  {journal} {Phys. Rev.}\ }\textbf {\bibinfo {volume}
  {D84}},\ \bibinfo {pages} {014014} (\bibinfo {year} {2011})}\BibitemShut
  {NoStop}%
\bibitem [{\citenamefont {van Beveren}(1998)}]{vanBeveren:1998}%
  \BibitemOpen
  \bibfield  {author} {\bibinfo {author} {\bibfnamefont {E.}~\bibnamefont {van
  Beveren}},\ }\href@noop {} {\bibfield  {journal} {\bibinfo  {journal}
  {Lecture notes, Universidade de Coimbra}\ } (\bibinfo {year}
  {1998})}\BibitemShut {NoStop}%
\bibitem [{\citenamefont {\mbox{Wolfram Research, Inc.}}(2012)}]{mma}%
  \BibitemOpen
  \bibfield  {author} {\bibinfo {author} {\bibnamefont {\mbox{Wolfram Research,
  Inc.}}},\ }\href@noop {} {\bibfield  {journal} {\bibinfo  {journal}
  {Mathematica, Version 9.0, Champaign, IL}\ } (\bibinfo {year}
  {2012})}\BibitemShut {NoStop}%
\bibitem [{\citenamefont {Pascual}\ and\ \citenamefont
  {Tarrach}(1980)}]{Pascual:1980yu}%
  \BibitemOpen
  \bibfield  {author} {\bibinfo {author} {\bibfnamefont {P.}~\bibnamefont
  {Pascual}}\ and\ \bibinfo {author} {\bibfnamefont {R.}~\bibnamefont
  {Tarrach}},\ }\href {\doibase 10.1016/0550-3213(80)90193-5} {\bibfield
  {journal} {\bibinfo  {journal} {Nucl.Phys.}\ }\textbf {\bibinfo {volume}
  {B174}},\ \bibinfo {pages} {123} (\bibinfo {year} {1980})}\BibitemShut
  {NoStop}%
\bibitem [{\citenamefont {Eichmann}\ and\ \citenamefont
  {Ramalho}()}]{new:GE+GR}%
  \BibitemOpen
  \bibfield  {author} {\bibinfo {author} {\bibfnamefont {G.}~\bibnamefont
  {Eichmann}}\ and\ \bibinfo {author} {\bibfnamefont {G.}~\bibnamefont
  {Ramalho}},\ }\href@noop {} {\bibinfo  {journal} {in preparation}\
  }\BibitemShut {NoStop}%
\bibitem [{\citenamefont {Stoffer}(2014)}]{Stoffer:2014rka}%
  \BibitemOpen
\bibfield  {journal} {  }\bibfield  {author} {\bibinfo {author} {\bibfnamefont
  {P.}~\bibnamefont {Stoffer}},\ }\href@noop {} {\ }\Eprint
  {http://arxiv.org/abs/1412.5171} {1412.5171 [hep-ph]} \BibitemShut {NoStop}%
\end{thebibliography}%

\end{document}